\documentclass[aps,prd,showpacs,letterpaper,onecolumn,longbibliography,superscriptaddress,notitlepage,nofootinbib]{revtex4-1}%
\usepackage[caption=false]{subfig}
\usepackage{graphicx}
\usepackage{bm}
\usepackage{epsf}
\usepackage{rotating}
\usepackage{epsfig,graphics,rotate,color}
\usepackage{wrapfig}
\usepackage{amssymb}
\usepackage{amsmath}
\usepackage{amsfonts}
\usepackage{booktabs}
\usepackage{gensymb}
\usepackage{placeins}
\usepackage{diagbox}
\usepackage[colorlinks=true,citecolor=blue,linkcolor=blue, allcolors=blue]{hyperref}
\usepackage{listings}
\lstdefinestyle{mystyle}{
    basicstyle=\ttfamily,
    breakatwhitespace=false,         
    breaklines=true,                 
    captionpos=b,                    
    keepspaces=true,                 
    showspaces=false,                
    showstringspaces=false,
    showtabs=false,                  
    tabsize=4
}
\AtBeginDocument{
\heavyrulewidth=.08em
\lightrulewidth=.05em
\cmidrulewidth=.03em
\belowrulesep=.65ex
\belowbottomsep=0pt
\aboverulesep=.4ex
\abovetopsep=0pt
\cmidrulesep=\doublerulesep
\cmidrulekern=.5em
\defaultaddspace=.5em
}

\newcommand{\Emf}{E_{\mu}^{\rm f}}
\newcommand{\Emi}{E_{\mu}^{\rm i}}
\newcommand{\ECR}{E_{\rm CR}}
\newcommand{\Prob}{{\cal P}}
\newcommand{\Ppuncor}{\Prob_{\rm pass}^{\rm uncor}}
\newcommand{\Ppcor}{\Prob_{\rm pass}^{\rm cor}}
\newcommand{\Pzmproto}{\Prob_{0\text{-}\mu}^{\rm shower}}
\newcommand{\Pzmsib}{\Prob_{0\text{-}\mu}^{\rm sib}}
\newcommand{\MCEq}{\texttt{MCE{\scriptsize Q}}}
\newcommand{\MMC}{\texttt{MMC}}
\newcommand{\PROPOSAL}{\texttt{PROPOSAL}}
\newcommand{\CORSIKA}{\texttt{CORSIKA}}
\newcommand{\Python}{\texttt{Python}}
\newcommand{\nuveto}{{\LARGE $\nu$}\texttt{eto}}
\newcommand{\barparen}[2]{\overset{\scriptscriptstyle (-)}{#1}_{\!\!#2}}

\begin{document}

\hfill{\tt IFIC/18-24}
\vskip 0.2in

\title{Unified atmospheric neutrino passing fractions for large-scale neutrino telescopes}

\author{Carlos A.~Arg\"uelles}
\email{caad@mit.edu}
\thanks{ORCID: \href{https://orcid.org/0000-0003-4186-4182}{0000-0003-4186-4182}}
\affiliation{Department of Physics, Massachusetts Institute of Technology, Cambridge, MA 02139, USA}
\author{Sergio Palomares-Ruiz}
\email{sergiopr@ific.uv.es}
\thanks{ORCID: \href{https://orcid.org/0000-0001-9049-2288}{0000-0001-9049-2288}}
\affiliation{Instituto de F\'{\i}sica Corpuscular (IFIC), CSIC-Universitat de Val\`{e}ncia, Apartado de Correos 22085, E-46071 Valencia, Spain}
\author{Austin Schneider} 
\email{aschneider@icecube.wisc.edu}
\thanks{ORCID: \href{https://orcid.org/0000-0002-0895-3477}{0000-0002-0895-3477}}
\affiliation{Department of Physics and Wisconsin IceCube Particle Astrophysics Center, University of Wisconsin, Madison, WI 53706, USA}
\author{Logan Wille}
\email{lwille@icecube.wisc.edu}
\affiliation{Department of Physics and Wisconsin IceCube Particle Astrophysics Center, University of Wisconsin, Madison, WI 53706, USA}
\author{Tianlu Yuan}
\email{tyuan@icecube.wisc.edu}
\thanks{ORCID: \href{http://orcid.org/0000-0002-7041-5872}{0000-0002-7041-5872}}
\affiliation{Department of Physics and Wisconsin IceCube Particle Astrophysics Center, University of Wisconsin, Madison, WI 53706, USA}

\begin{abstract}
The atmospheric neutrino passing fraction, or self-veto, is defined as the probability for an atmospheric neutrino not to be accompanied by a detectable muon from the same cosmic-ray air shower. Building upon previous work, we propose a redefinition of the passing fractions by unifying the treatment for muon and electron neutrinos. Several approximations have also been removed. This enables performing detailed estimations of the uncertainties in the passing fractions from several inputs: muon losses, cosmic-ray spectrum, hadronic-interaction models and atmosphere-density profiles. We also study the passing fractions under variations of the detector configuration: depth, surrounding medium and muon veto trigger probability. The calculation exhibits excellent agreement with passing fractions obtained from Monte Carlo simulations. Finally, we provide a general software framework to implement this veto technique for all large-scale neutrino observatories.
\end{abstract}

\date{\today}

\maketitle

\section{Introduction}
\label{sec:intro}

The detection of the first high-energy neutrinos of extraterrestrial origin~\cite{Aartsen:2013bka} marked the beginning of neutrino astronomy. As we move from this first observation to precision physics using these astrophysical neutrino events, it becomes crucial to improve our understanding of atmospheric neutrinos, which together with atmospheric muons, represent the main backgrounds. Currently, the IceCube neutrino telescope has detected high-energy astrophysical neutrinos at a significance of $> 6 \sigma$ using high-energy starting events (HESE)~\cite{Aartsen:2013jdh, Aartsen:2014gkd, Aartsen:2015zva, Aartsen:2017mau}. This is further complemented by the observation of a compatible astrophysical flux using through-going muon neutrino events~\cite{Aartsen:2015rwa, Aartsen:2016xlq}. Quantifying uncertainties on the atmospheric neutrino background becomes even more relevant for the medium energy starting event (MESE) sample~\cite{Aartsen:2014muf}, which extends to lower energies than the HESE sample. Recently, the ANTARES neutrino telescope has also carried out a search for the cosmic diffuse neutrino flux using nine years of up-going track- and shower-like events~\cite{Albert:2017nsd}. A small excess of high-energy events over the expected atmospheric background was identified, with a significance of $1.6 \sigma$ and fully compatible with IceCube observations.

A major background in searches for astrophysical neutrinos comes from atmospheric muons. For this reason, neutrino telescopes primarily considered neutrinos coming from the direction passing through the Earth. The Earth is a very efficient shield of the high atmospheric muon flux, almost eliminating this background. This strategy, however, limits the solid angle of observation to half of the sky. Using also cascades~\cite{Beacom:2004jb} and down-going events increases the statistics significantly. To use down-going neutrinos in astrophysical neutrino searches, an efficient method of vetoing atmospheric muons is required. Although ANTARES is too small to include an outer muon veto, larger detectors such as IceCube~\cite{Achterberg:2006md} have used part of their instrumented volumes as active vetos for down-going muon, and KM3NeT~\cite{Adrian-Martinez:2016fdl} and Baikal-GVD~\cite{Baikal2017} telescopes may. These muons are produced together with down-going atmospheric neutrinos and may reach the detector at the same time and from the same direction. A neutrino signal in coincidence with a passing muon could significantly reduce the expected atmospheric neutrino background. This idea was first applied by Sch\"onert, Gaisser, Resconi, and Schulz (SGRS)~\cite{Schonert:2008is} to the case of muons produced from the decay of the same parent as the muon neutrino. This was later extended for both electron and muon neutrinos by Gaisser, Jero, Karle, and van Santen (GJKvS)~\cite{Gaisser:2014bja}, who considered muons produced from other parents in the same cosmic-ray shower.

\begin{figure}
\centering
    \subfloat{
    \includegraphics[width=0.5\linewidth]{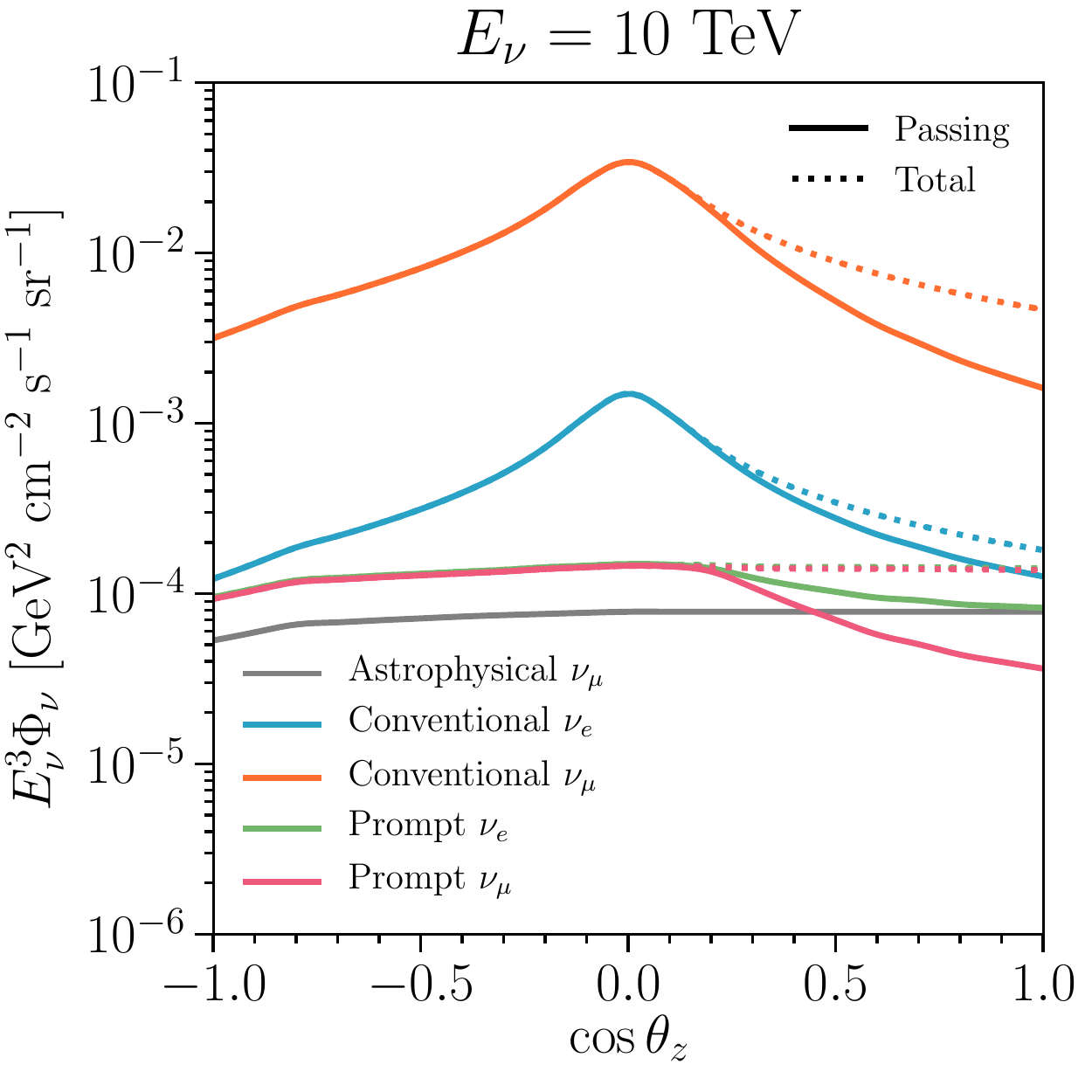}
    }
    \subfloat{
    \includegraphics[width=0.5\linewidth]{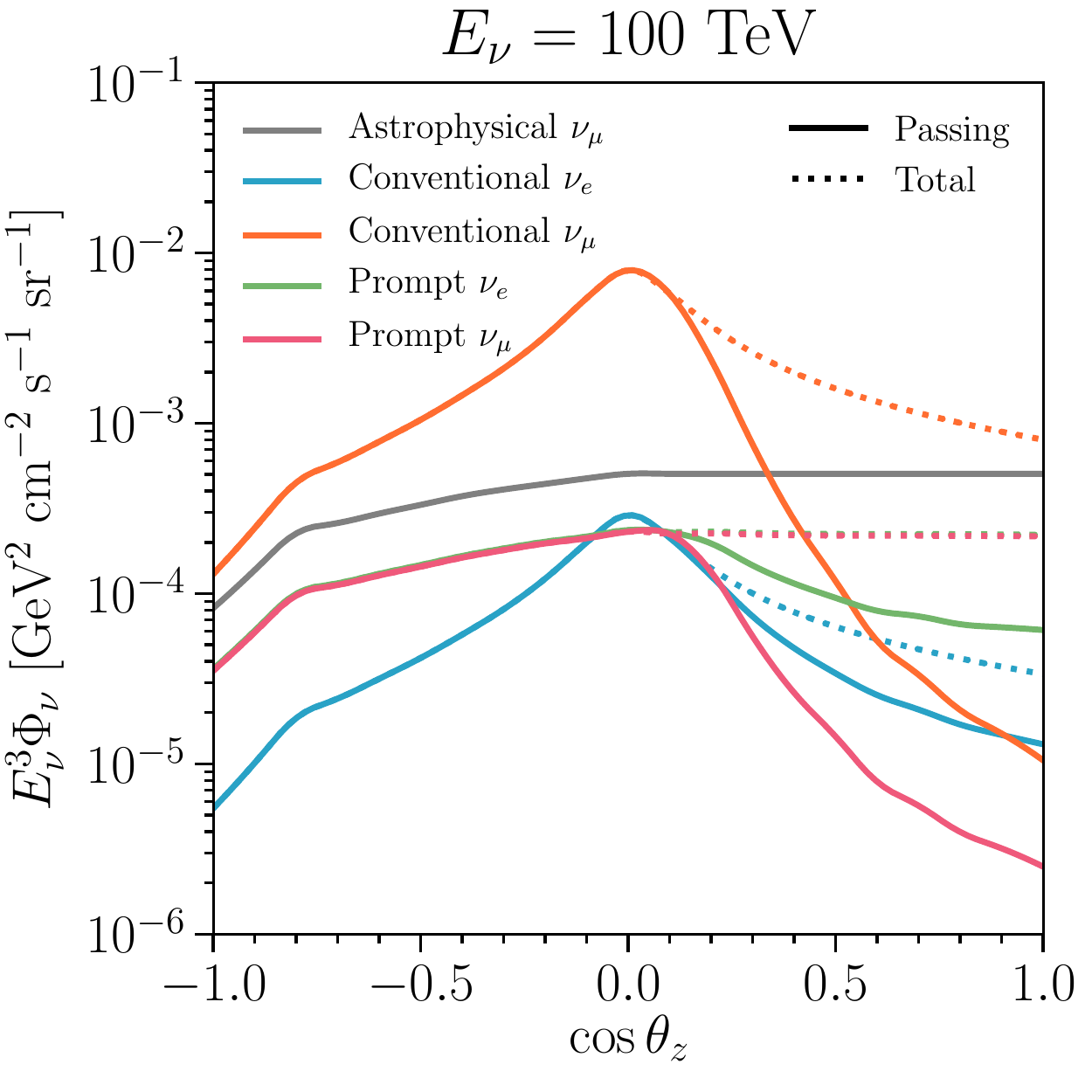}
    }
\caption{\textbf{\textit{Atmospheric neutrino passing fluxes,}} as a function of the cosine of the zenith angle $\theta_z$. We compare passing fluxes after applying the rejection factor described in this work (solid) with the fluxes before entering the detector (dotted). The results are shown, from top to bottom in the left panel, for conventional $\nu_\mu$ (orange), conventional $\nu_e$ (blue), prompt $\nu_e$ (green) and prompt $\nu_\mu$ (magenta) fluxes. The astrophysical $\nu_\mu$ (gray) $E_\nu^{-2.19}$ flux~\cite{Aartsen:2017mau} lies beneath the others and is not subject to the atmospheric rejection factor. Note that absorption in the Earth is included for $\cos \theta_z < 0$. \textit{Left panel:} $E_\nu = 10$~TeV. \textit{Right panel:} $E_\nu = 100$~TeV. Note the conventional $\nu_e$ flux drops below the other channels at $\cos \theta_z = -1$ while the astrophysical $\nu_\mu$ flux rises to just below the conventional $\nu_\mu$ channel.}
\label{fig:fluxes}
\end{figure}

Indeed, an outer layer veto is used by some IceCube analyses to significantly reduce the atmospheric muon and neutrino backgrounds. Atmospheric muons are rejected if they deposit enough light on the veto layer before entering the fiducial volume of the analysis. Atmospheric neutrinos are indirectly rejected if they are accompanied by muons, produced in the same air shower, that trigger the veto. These accompanying muons reduce the probability that neutrinos that pass the veto are of atmospheric origin. This is in fact what drives the bound on the atmospheric prompt neutrino contribution in HESE~\cite{Aartsen:2014gkd} and MESE~\cite{Aartsen:2014muf} analyses. At high energies, veto suppression in the Southern hemisphere turns out to be larger than that of Earth absorption in the Northern one (see Fig.~\ref{fig:fluxes}), in contradiction with the observed zenith distribution. The previous works did not study the systematic uncertainties of these probabilities and used several simplifying approximations. The objective of this work is to present a more precise calculation that also allows for including systematic uncertainties and detector characteristics. To summarize, Fig.~\ref{fig:fluxes} shows the effect of the suppression of the atmospheric neutrino fluxes, as a function of the cosine of the zenith angle $\theta_z$, for two energies (left 10~TeV and right 100~TeV). Note that neutrino absorption in the Earth is also included for $\cos \theta_z <0$ using the \texttt{$\nu$FATE} code~\cite{Vincent:2017svp}. The ratio of the passing to the total flux is called the passing fraction.

The atmospheric neutrino passing fraction depends on the primary shower development and on the muon energy losses in the medium. This probability is conditional on the neutrino flavor, energy, zenith angle, and parent particle. In this work, we are especially concerned with uncertainties in the muon energy losses, the primary cosmic-ray spectrum, the hadronic-interaction model, and the atmosphere-density model. In order to study these uncertainties, we created a new framework that relies on the Matrix Cascade Equation (\MCEq) package~\cite{MCEq, Fedynitch:2015zma, Fedynitch:2018vfe}. Naturally, our calculation is designed so that the primary cosmic-ray spectrum, the hadronic-interaction model, and the atmosphere-density model can be changed. In addition, we include accurate treatment of muon energy losses in different media and the possibility to tune the detector's characterization: depth, medium, and veto's response to incident muons. Moreover, to validate our calculation of the atmospheric neutrino passing fractions, we have compared our results with those obtained from a detailed simulation using the COsmic Ray SImulations for KAscade (\CORSIKA{}) program~\cite{Heck:1998vt, Heck:2018} and we find them in extremely good agreement. 

The paper is organized as follows. Section~\ref{sec:shower-muons} describes the main ingredients that enter our computation of the passing fractions. In section~\ref{sec:shower_physics} we discuss different inputs related to air shower development. In section~\ref{sec:preach_plight} we define the probability for a muon to reach the detector and trigger the veto at some depth below the surface. In section~\ref{sec:pf}, we build upon previous work~\cite{Schonert:2008is, Gaisser:2014bja} and define the passing fractions for atmospheric electron (section~\ref{sec:pfnue}), muon (section~\ref{sec:pfnumu}), and tau (section~\ref{sec:pfnutau}) neutrinos. We also evaluate the effect of previous approximations on the passing fraction and perform a direct comparison to those results in section~\ref{sec:improvements}. Different sources of systematic uncertainty are discussed in section~\ref{sec:systematics}: muon energy losses in section~\ref{sec:muonlosses}, primary cosmic-ray spectrum in section~\ref{sec:CR}, hadronic-interaction model in section~\ref{sec:hadronic}, and atmosphere-density model in section~\ref{sec:atmosphere}. In section~\ref{sec:detector}, we illustrate the effects of changing parameters related to the detector configuration: the detector depth in section~\ref{sec:depth}, the surrounding medium in section~\ref{sec:medium}, and the veto's response to passing muons in section~\ref{sec:response}. Finally, in section~\ref{sec:conclusions} we draw our final conclusions. In appendix~\ref{app:code} we briefly describe the publicly available code to calculate atmospheric neutrino passing fractions and in appendix~\ref{app:tables} we provide, in tabulated form, passing fractions for our default settings.

\section{Cosmic-ray showers and penetrating muons}
\label{sec:shower-muons}

\subsection{Shower physics}
\label{sec:shower_physics}

Interactions of cosmic rays with nuclei of the Earth's atmosphere produce a flux secondary particles, which can eventually decay into muons and neutrinos that could reach detectors located at surface or underground. While the spectra of these atmospheric leptons peak around GeV energies, it extends to high energies with an approximate power-law dependence. The final lepton spectra at the Earth's surface depend on the energy distribution and composition of the primary cosmic-ray flux, as well as on the properties of the hadronic interactions that control their production, on the density profile, and composition of the atmosphere. We use the open-source numerical code \MCEq~\cite{MCEq, Fedynitch:2015zma, Fedynitch:2018vfe} to compute the different distributions of all secondary particles, as a function of energy, direction, and slant depth. Moreover, for comparisons and tests of our computations, we also show the results from a dedicated full Monte Carlo simulation with \CORSIKA~\cite{Heck:1998vt, Heck:2018}.

Except otherwise noted, our default model for the primary cosmic-ray spectrum is the three-population model from Gaisser~\cite{Gaisser:2011cc} (H3a), that follows Hillas's proposal~\cite{Hillas:2006ms}, with each population containing five groups of nuclei that cut off exponentially at a characteristic rigidity. This is the minimal assumption if the galactic-extragalactic transition occurs at the ankle. In section~\ref{sec:CR} we also consider other models for the primary cosmic-ray spectrum.

For the description of multiparticle production and extensive air shower development we use the recently released SIBYLL~2.3c~\cite{Riehn:2017mfm} as our default hadronic-interaction model, which also includes charmed particles production. In section~\ref{sec:hadronic} we also compare the results obtained with other hadronic-interaction models.

To characterize the density profile of the atmosphere, we use the MSIS-90-E model at the South Pole corresponding to July 1, 1997~\cite{Labitzke:1985, Hedin:1991}, as our default model, location, and period of the year. We comment below on the uncertainties on the characterization of the Earth's atmosphere, from model to model differences, from seasonal variations, and from different locations.

Given that the \CORSIKA{} simulation uses SIBYLL~2.3~\cite{Engel:2015dxa, Riehn:2015oba} as the hadronic-interaction model, instead of SIBYLL~2.3c, whenever we compare our results with those from the simulation, we also use SIBYLL~2.3. Our default cosmic-ray primary spectrum and atmosphere-density model are chosen to coincide used in the \CORSIKA{} simulation.

Atmospheric neutrinos are typically categorized as conventional or prompt based on their parent particle~\cite{Gaisser:2002jj}. Conventional atmospheric muon neutrinos are defined to be those that come from pion or kaon decay~\cite{Barr:2004br, Honda:2006qj, Honda:2011nf, Petrova:2012qf, Sinegovskaya:2014pia, Gaisser:2014pda, Honda:2015fha}. At energies above $\sim 100$~GeV the flux of conventional muon neutrinos transitions from predominantly those produced by pion decays to predominantly those produced by kaon decays. At those high energies, conventional electron neutrinos are produced almost entirely from kaon decays. In this work, conventional neutrinos are defined as those produced from the decays of $\pi^{\pm}$, $K^{\pm}$, $K^0_L$, $K^0_s$, and $\mu^{\pm}$. Prompt atmospheric neutrinos are produced from the decay of charmed and bottom hadrons, which have a very short lifetime and decay almost immediately after production~\cite{Gondolo:1995fq, Pasquali:1998ji, Pasquali:1998xf, Martin:2003us, Enberg:2008te, Bhattacharya:2015jpa, Garzelli:2015psa, Gauld:2015yia, Gauld:2015kvh, Bhattacharya:2016jce, Garzelli:2016xmx, Benzke:2017yjn, Sinegovsky:2017gkg}. This is also the reason their flux follows a power-law with a spectral index one unit harder than the conventional one. This flux begins to dominate at around $100$~TeV and the dominant contributions come from $D^{\pm}$, $D^0$, $\bar{D}^0$, $\Lambda_c^+$, and $D_s^{\pm}$, with the last two components being less important~\cite{Fedynitch:2015zma}. In this work, we consider prompt atmospheric neutrinos from the decays of charmed mesons, $D^{\pm}$, $D^{\pm}_s$, $D^0$, and $\bar{D}^0$. The dominant two-body decay channels are $\pi^{\pm} \rightarrow \mu^{\pm} +  \barparen{\nu}{\mu}$ ($99.99\%$) and $K^{\pm} \rightarrow \mu^{\pm} + \barparen{\nu}{\mu}$ ($63.56\%$). Almost all other atmospheric neutrinos are produced via $n$-body processes. As described in section~\ref{sec:pfnumu}, we include the $n$-body decay spectra, obtained using \texttt{PYTHIA~8.1}~\cite{Sjostrand:2008py}, of $K^0_L$ for conventional muon neutrinos and $D^+$, $D^0$, $D_s^+$, and their antiparticles for prompts.

\subsection{Penetrating muons and vetos in underground detectors}
\label{sec:preach_plight}

The concept of the atmospheric neutrino passing fraction is based on the coincidence, in time and direction, between a neutrino event and a muon produced in the same cosmic-ray shower. Thus, one of the key ingredients in the calculation of the atmospheric neutrino passing fractions is $\Prob_{\rm reach}\left(\Emf \, | \, \Emi , \, X_\mu\right)$, the probability for a muon, produced with energy $\Emi$ and traversing a slant depth $X_\mu$ (in g/cm$^2$) in a particular medium, to reach the detector with energy $\Emf$. Given that the slant depth of the atmosphere is very small, muon energy losses before entering the Earth are negligible, so only the path inside the Earth before reaching the detector is relevant. This is fully determined by $\theta_z$, the zenith angle in detector's coordinates of the incoming muon, whose direction is assumed to coincide with that of the neutrino. In Earth's surface coordinates, $\sin^2 \theta_z^s = \left(1 - d_{\rm det}/R_\oplus\right)^2 \, \sin^2 \theta_z$, where $d_{\rm det}$ is the depth of the detector and $R_\oplus$ is the Earth's radius.

In the original work where the passing fraction for atmospheric muon neutrinos was proposed~\cite{Schonert:2008is}, $\Prob_{\rm reach}$ was approximated by a Dirac delta function with the final muon energy $\Emf$ set by a simple average relation obtained from the Muon Monte Carlo (\MMC) code~\cite{Chirkin:2004hz}. Explicitly, 
$\Prob_{\rm reach}^{\rm SGRS} (\Emf \, | \, \Emi , \, X_\mu) = \delta \left(X_\mu - X_{\mu}^{\rm median}(\Emi, \Emf)\right) \left|dX_{\mu}^{\rm median}/d \Emf \right|$,
where $X_{\mu}^{\rm median}(\Emi, \Emf)$ is the median slant depth a muon with initial energy $\Emi$ propagates so that its final energy is $\Emf$. In this work, we take into account fluctuations from stochastic losses of muons propagating through the Earth toward the detector, and build probability distribution functions for $\Prob_{\rm reach}$ using \MMC~\cite{Chirkin:2004hz}.\footnote{Its successor code, \PROPOSAL, provides identical results~\cite{Koehne:2013gpa}.} The inclusion of stochastic losses results in shorter average propagation distances~\cite{Lipari:1991ut} than if only continuous losses are considered. Moreover, taking into account fluctuations is especially important when the tails of the survival probabilities are relevant for distances beyond the average range, and results in larger atmospheric neutrino passing fractions for more horizontal directions. Nevertheless, for more vertical trajectories, these tails become important for distances below the average range and tend to reduce the passing fractions. This is discussed in the next section.

Next, in order to account for the detector veto's response to muons, we introduce $\Prob_{\rm light}(\Emf)$, which is the probability for a muon entering the detector with energy $\Emf$ to \textit{light} the veto. This probability is both detector and event selection dependent and should be explicitly calculated for each analysis. In this paper, following previous work, we study toy $\Prob_{\rm light}(\Emf)$.  With our notation, the approach followed in Refs.~\cite{Schonert:2008is, Gaisser:2014bja} implies that $\Prob_{\rm light}$ is equivalent to a step-like function above some muon energy threshold, $E_{\mu}^{\rm th}$, that is, $\Prob_{\rm light}^{\rm SGRS}(\Emf) = \Theta\left(\Emf - E_{\mu}^{\rm th}\right)$. As in Ref.~\cite{Gaisser:2014bja}, we consider as our default $E_{\mu}^{\rm th} = 1$~TeV. Nevertheless, our implementation is completely general and allows one to use functions with a smooth transition around the nominal threshold energy. As an example, we also consider a sigmoid, defined as the cumulative distribution function of a Gaussian, $\Prob_{\rm light}(\Emf) = \Phi\left(\left(\Emf - E_{\mu}^{\rm th}\right)/\sigma_\mu\right)$ with $E_{\mu}^{\rm th} = 0.75$~TeV and $\sigma_\mu = 0.25$~TeV. Note that this trigger probability applies individually to each muon that reaches the detector. To fully account for the different energies in bundles of muons produced in the shower, $\Prob_{\rm light}$ can be redefined to dependent on the muon multiplicities and the bundle energy distribution. 

Once $\Prob_{\rm reach}$ is obtained and $\Prob_{\rm light}$ determine, what matters in practice is the fraction of muons that reach and trigger the detector's veto, which we define as
\begin{equation}
\label{eq:PrPl}
\Prob_{\rm det}\left(\Emi , \, X_\mu(\theta_z)\right) \equiv \int d\Emf \, \Prob_{\rm reach}\left(\Emf \, | \, \Emi , \, X_\mu(\theta_z)\right) \, \Prob_{\rm light} (\Emf) ~.
\end{equation}
Note that losses also depend on the medium and not only on the slant depth, whose dependence on $\theta_z$ we write explicitly. If the medium surrounding the detector is homogeneous, $X_\mu$ can be exchanged by the distance without loss of generality. Also notice that, if the range is approximated by its median value ($\Prob_{\rm reach}$ as a Dirac delta function), $\Prob_{\rm det}$ follows the same shape as $\Prob_{\rm light}$. If $\Prob_{\rm light}$ is taken to be a step-like function, it results in the $\Prob_{\rm det}$ effectively used in Refs.~\cite{Schonert:2008is, Gaisser:2014bja},
\begin{equation}
\label{eq:PrPlSGRS}
\Prob_{\rm det}^{\rm SGRS} \left(\Emi , \, X_\mu\right) = \int d\Emf \, \Prob_{\rm reach}^{\rm SGRS} (\Emf \, | \, \Emi , \, X_\mu) \, \Prob_{\rm light}^{\rm SGRS} (\Emf) = \Theta\left(X_{\mu}^{\rm median}(\Emi, E_{\mu}^{\rm th}) - X_{\mu}\right) ~.
\end{equation}
For clarity, from now on we will use $\theta_z$ instead of $X_\mu$ and will assume ice as the medium muons traverse before reaching the detector.
\begin{figure}
\centering
    \subfloat{
        \includegraphics[width=0.5\linewidth]{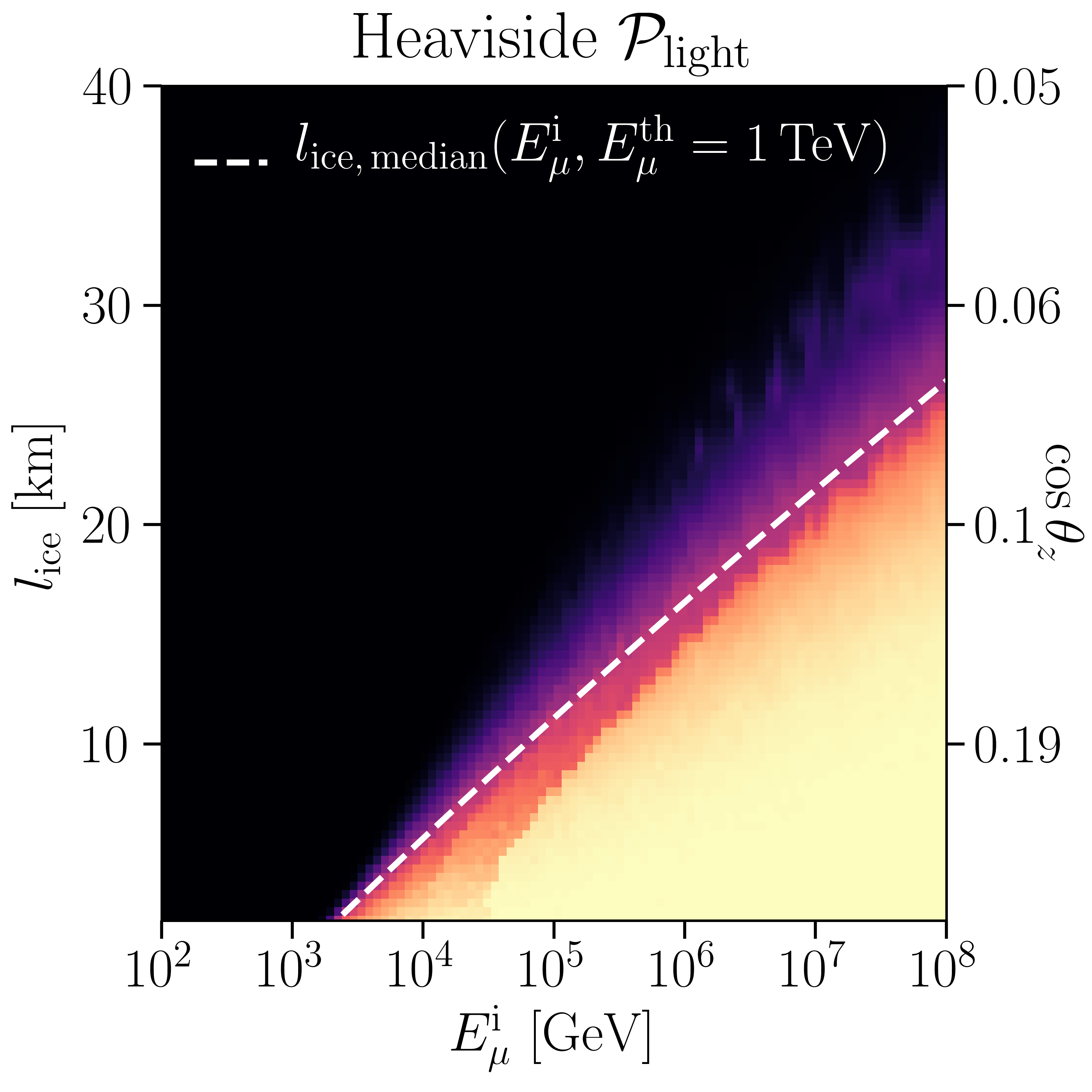}
    }
    \subfloat{
        \includegraphics[width=0.5\linewidth]{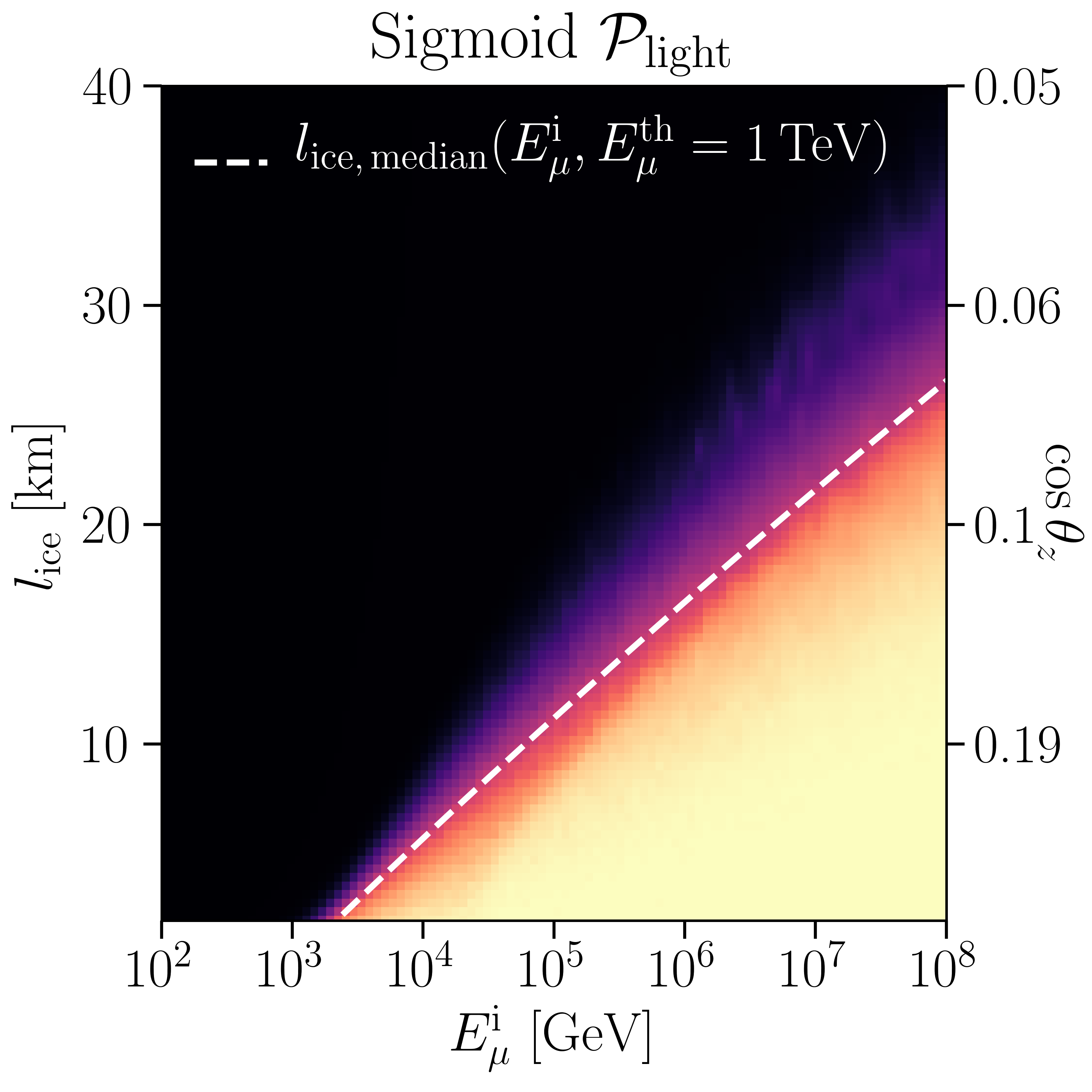}
    }\\[0.3ex]
    \subfloat{
        \includegraphics[width=0.5\linewidth]{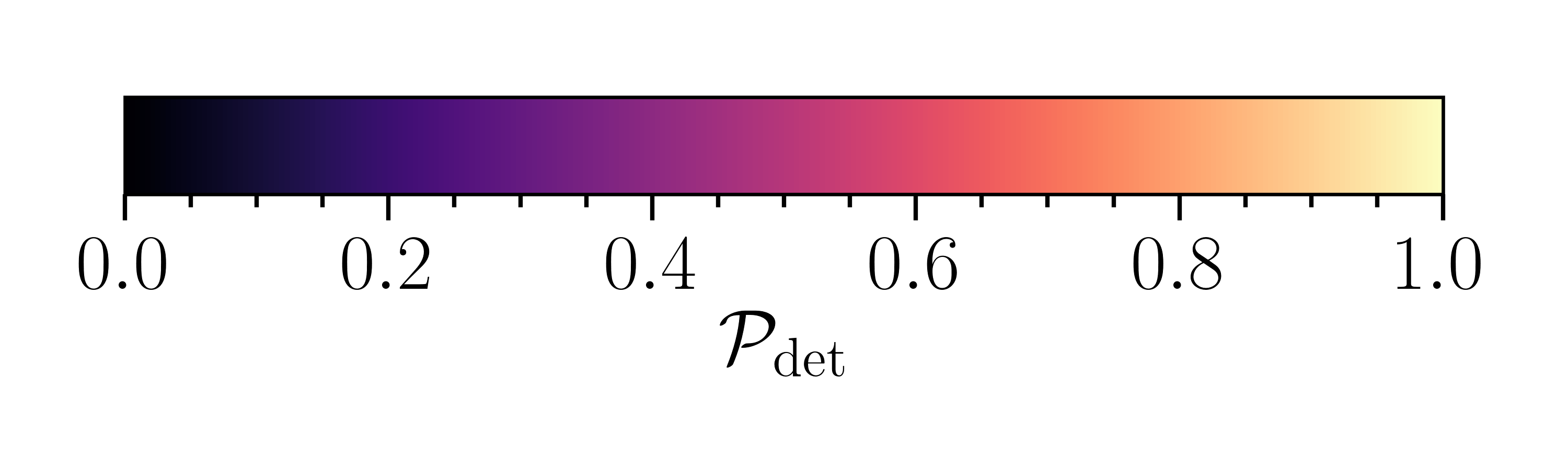}
    }
\caption{\textbf{\textit{Fraction of muons that reach and trigger the detector's veto, $\boldsymbol{\Prob_{\rm det}}$}}, as a function of the initial muon energy, $\Emi$, and the distance in ice, $l_{\rm ice}$. Also indicated is the equivalent $\cos \theta_z$ for a detector located at $d_{\rm det} = 1.95$~km. We show $\Prob_{\rm det}$ using the muon range distribution (color) and the $\Emi$ required for the muon to survive a distance $l_{\rm ice}$ with at least $\Emf = 1$~TeV 50\% of the time (dashed), from Ref.~\cite{Gaisser:2014bja}. \textit{Left panel:} Heaviside trigger probability, $\Prob_{\rm light}(\Emf) = \Theta(\Emf - E_{\mu}^{\rm th})$ with $E_{\mu}^{\rm th} = 1$~TeV. \textit{Right panel:} Sigmoid (defined as the cumulative distribution function of a Gaussian) trigger probability, $\Prob_{\rm light}(\Emf) = \Phi\left(\left(\Emf - E_{\mu}^{\rm th}\right)/\sigma_\mu\right)$ with $E_{\mu}^{\rm th} = 0.75$~TeV and $\sigma_\mu = 0.25$~TeV. The minimum distance is $(l_{\rm ice})_{\rm min} = 1.95$~km.}
\label{fig:pdet}
\end{figure}

In Fig.~\ref{fig:pdet} we show $\Prob_{\rm det}$ as a function of the initial muon energy, $\Emi$, and the distance from the Earth's surface to the detector in ice, $l_{\rm ice}$. Also indicated is the equivalent $\cos \theta_z$ for a detector at depth $d_{\rm det} = 1.95$~km, which we set as our default. The results are depicted for $\Prob_{\rm light}$ defined as a Heaviside function (left) or a sigmoid (right) and with the default setup in \MMC{} to calculate $\Prob_{\rm reach}$ (see section~\ref{sec:muonlosses}). The darker regions indicate a lower probability and the lighter ones a higher probability of a muon to trigger the veto. In both panels, we also indicate the initial energy $\Emi$ required for the muon to survive a distance $l_{\rm ice}$ with at least $\Emf = 1$~TeV 50\% of the time (dashed), as computed in Ref.~\cite{Gaisser:2014bja}; below that line $\Prob_{\rm det}^{\rm SGRS} = 1$ and above $\Prob_{\rm det}^{\rm SGRS} = 0$. In section~\ref{sec:depth} we study the dependence of the passing fractions on the depth of the detector and in section~\ref{sec:medium} on the medium surrounding the detector. The impact of different choices of $\Prob_{\rm light}$ are discussed in section~\ref{sec:response} and the effect of using the approximation with the median range, as compared to the complete treatment of muon energy losses, is studied in section~\ref{sec:pf}. As we will describe, the correct form of $\Prob_{\rm reach}$ introduces non-negligible corrections to the atmospheric neutrino passing fractions.

\section{Atmospheric neutrino passing fractions}
\label{sec:pf}

The atmospheric neutrino passing fraction is the probability for an atmospheric neutrino not to be accompanied by a detectable muon from the same air shower. We denote this probability by $\Prob_{\rm pass}$~\cite{Schonert:2008is, Gaisser:2014bja}, which is defined as
\begin{equation}
\Prob_{\rm pass} (E_\nu, \theta_z) = \frac{\phi_\nu^{\rm pass}(E_\nu, \theta_z)}{\phi_\nu(E_\nu, \theta_z)} ~,
\end{equation}
where $E_\nu$ is the neutrino energy, $\phi_\nu(E_\nu, \theta_z)$ is the atmospheric neutrino differential flux, and $\phi_\nu^{\rm pass}(E_\nu, \theta_z)$ is the differential flux of atmospheric neutrinos that are not accompanied by a muon from the same shower that triggers the veto.

For reference, the atmospheric neutrino passing fractions obtained using our default settings are tabulated in Tables~\ref{app:tables}. We define and discuss them below.

\subsection{Passing fraction for electron neutrinos}
\label{sec:pfnue}

\begin{figure}
\centering
    \subfloat{
    \includegraphics[width=0.5\linewidth]{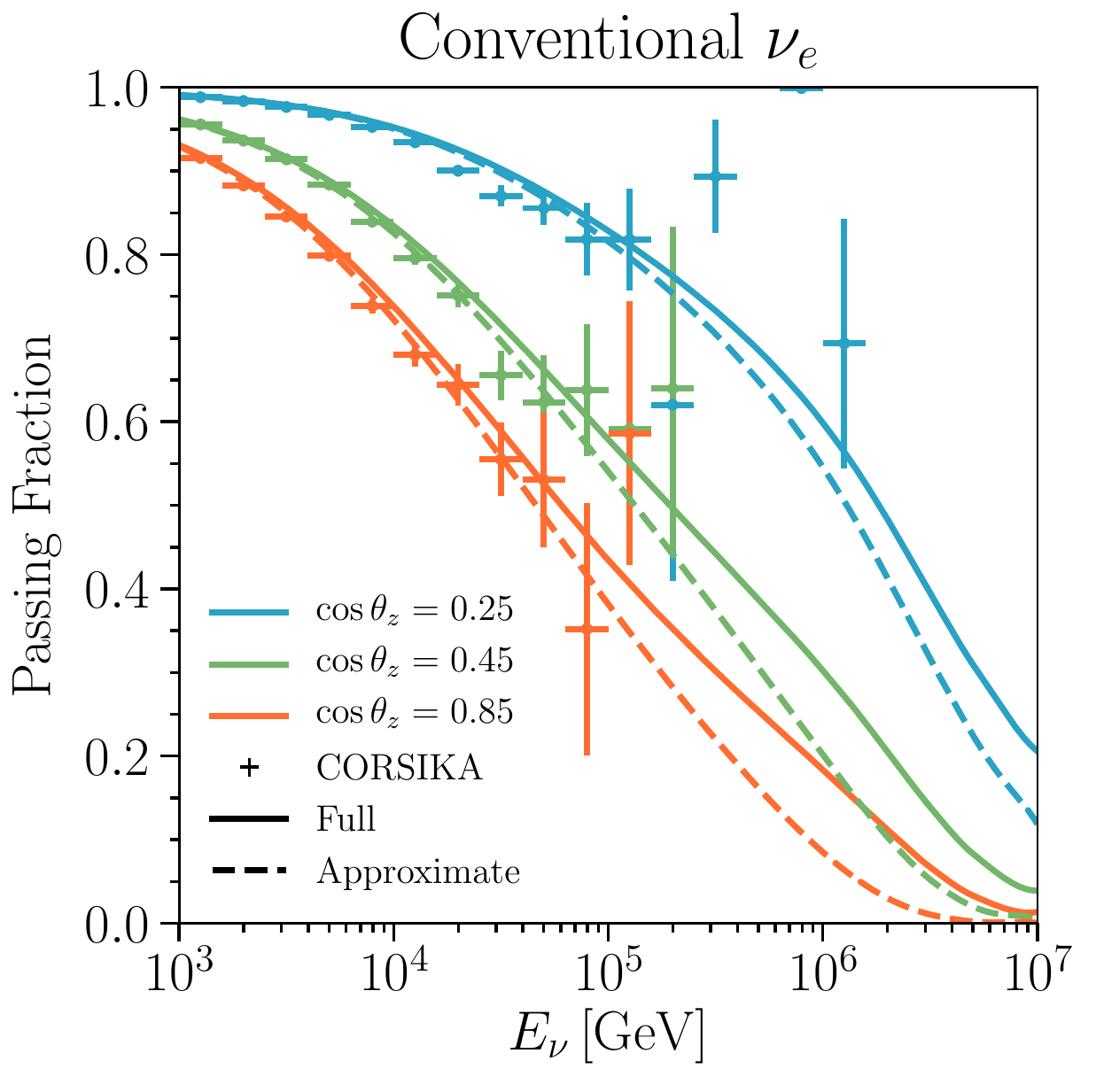}
    }
    \subfloat{
    \includegraphics[width=0.5\linewidth]{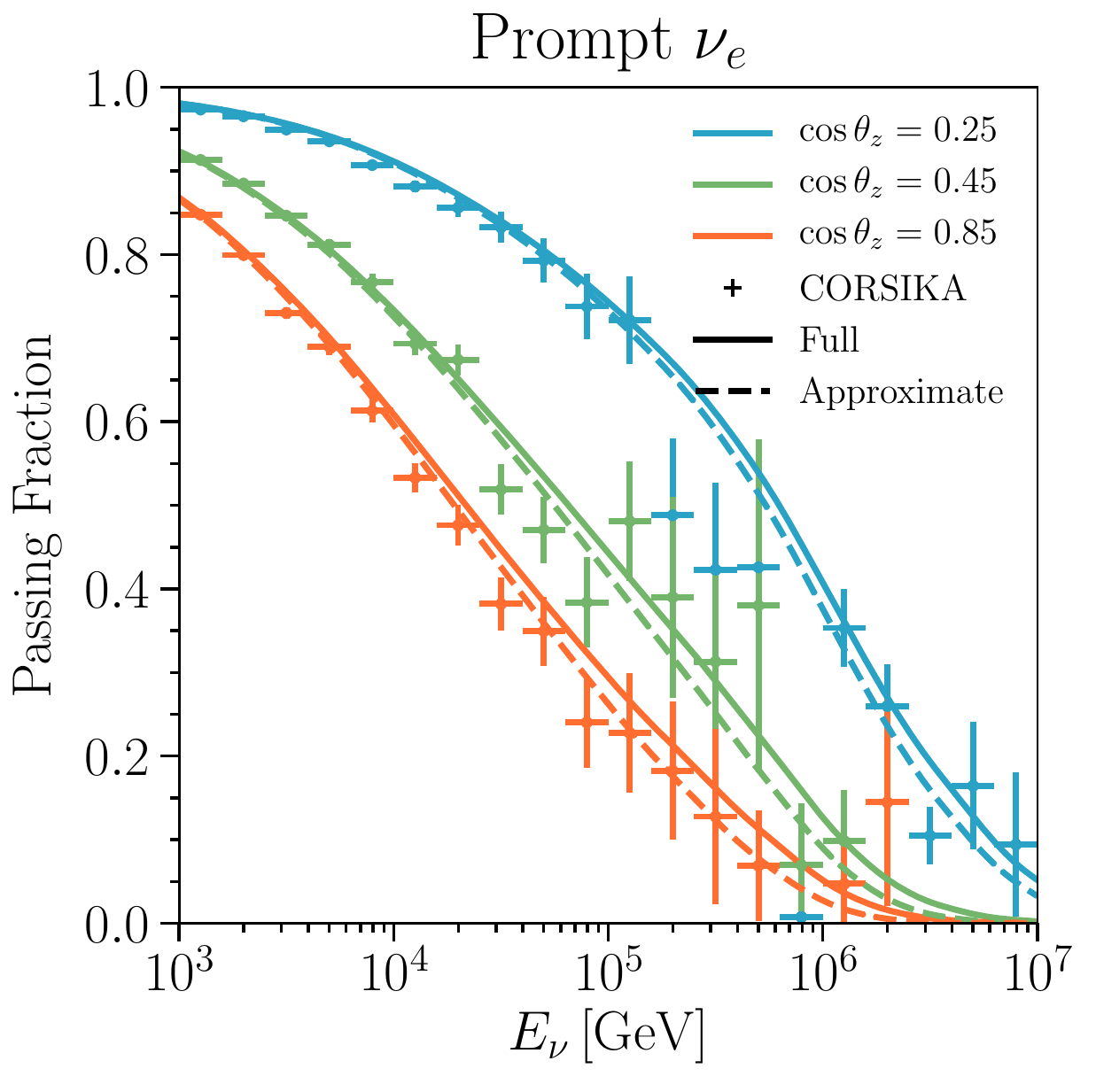}
    }
\caption{\textbf{\textit{Passing fractions: effect of approximations on the energy of the shower giving rise to uncorrelated muons.}} Results are shown for three values of $\cos\theta_z$ (from top to bottom): 0.25 (blue), 0.45 (green), and 0.85 (orange); with the approach of this work (solid), Eq.~(\ref{eq:PuncorGUE}), where the energy carried by the neutrino parent is subtracted from the rest of the shower which produces the muons to be triggered (i.e., $\ECR - E_p$), and without this subtraction (dashed), Eq.~(\ref{eq:PuncorGJKvS}), considering the cumulative muon yield from a shower with energy $\ECR$. Results from the \CORSIKA{} simulation are shown as crosses, with statistical error bars only. In all cases, the H3a primary cosmic-ray spectrum~\cite{Gaisser:2011cc}, the SIBYLL~2.3 hadronic-interaction model~\cite{Engel:2015dxa, Riehn:2015oba} and the MSIS-90-E atmosphere-density model at the South Pole on July 1, 1997~\cite{Labitzke:1985, Hedin:1991} are used, and $d_{\rm det} = 1.95$~km in ice and $\Prob_{\rm light}(\Emf) = \Theta(\Emf - 1\,{\rm TeV})$ are assumed. \textit{Left panel:} Conventional $\nu_e$ passing fraction. \textit{Right panel:} Prompt $\nu_e$ passing fraction.
}
\label{fig:nue_passing-double-counting}
\end{figure}

Atmospheric electron neutrinos (antineutrinos) are created together with positrons (electrons) at the parent's decay vertex and very rarely have a sibling muon. However, muons are produced in other branches of the shower. Following the approach of Ref.~\cite{Gaisser:2014bja}, we define the passing fraction for electron neutrinos using the average properties of muons in a prototypical shower produced by an individual cosmic-ray nucleus $A$ with energy $\ECR$. The atmospheric neutrino flux is given by,
\begin{equation}
\label{eq:nuphi}
\phi_\nu(E_\nu, \theta_z)  = \sum_A \int d\ECR \, \frac{dN_{A, \nu}}{dE_\nu}(\ECR, E_\nu, \theta_z) \, \phi_A(\ECR)~,
\end{equation}
where $\phi_A(\ECR)$ is the flux of primary cosmic-ray of nuclei $A$ and $dN_{A, \nu}/dE_\nu(\ECR, E_\nu, \theta_z)$ is the yield of neutrinos from a prototypical shower. The passing fraction can be obtained by weighting the total flux by the Poisson probability, $\Pzmproto$, that no accompanying muon triggers the veto~\cite{Gaisser:2014bja},
\begin{equation}
\label{eq:PuncorGJKvS}
\Prob_{\rm pass}^{\rm uncor, GJKvS}(E_\nu, \theta_z)  =  \frac{1}{\phi_\nu(E_\nu, \theta_z)} \sum_A \int d\ECR \, \frac{dN_{A, \nu}}{dE_\nu}(\ECR, E_\nu, \theta_z) \, \phi_A(\ECR) \, \Pzmproto \left(N_\mu = 0  ; \bar N_{A, \mu}^{\rm GJKvS}(\ECR, \theta_z)\right)~,
\end{equation}
where
\begin{equation}
\label{eq:pnomuproto}
\Pzmproto \left(N_\mu = 0  ; \bar N_{A, \mu}(\ECR, \theta_z)\right) = e^{- \bar N_{A, \mu}(\ECR, \theta_z)} ~,
\end{equation}
and $\bar N_{A, \mu}(\ECR, \theta_z)$ represents the expected number of muons in the prototypical shower that reach the detector and trigger the veto,
\begin{equation}
\label{eq:Nmu}
\bar N_{A, \mu}(\ECR,\theta_z) = \int d\Emi \, \frac{dN_{A, \mu}}{d\Emi}(\ECR, \Emi,  \theta_z) \, \Prob_{\rm det}\left(\Emi , \theta_z\right) ~,
\end{equation}
where $dN_{A, \mu}/d\Emi(\ECR, \Emi)$ is the energy spectrum of muons with energy $\Emi$ from a prototypical shower with energy $\ECR$ initiated from cosmic-ray nucleus $A$. If $\Prob_{\rm reach}$ is defined as a Dirac delta function following the dashed line in Fig.~\ref{fig:pdet} and $\Prob_{\rm light}$ is taken as a Heaviside function with the threshold at $\Emf = 1$~TeV, then this expression would correspond to the expected number of muons, $\bar N_{A, \mu}^{\rm GJKvS}$, as described in Ref.~\cite{Gaisser:2014bja}. 

\begin{figure}
\centering
    \subfloat{
        \includegraphics[width=0.5\linewidth]{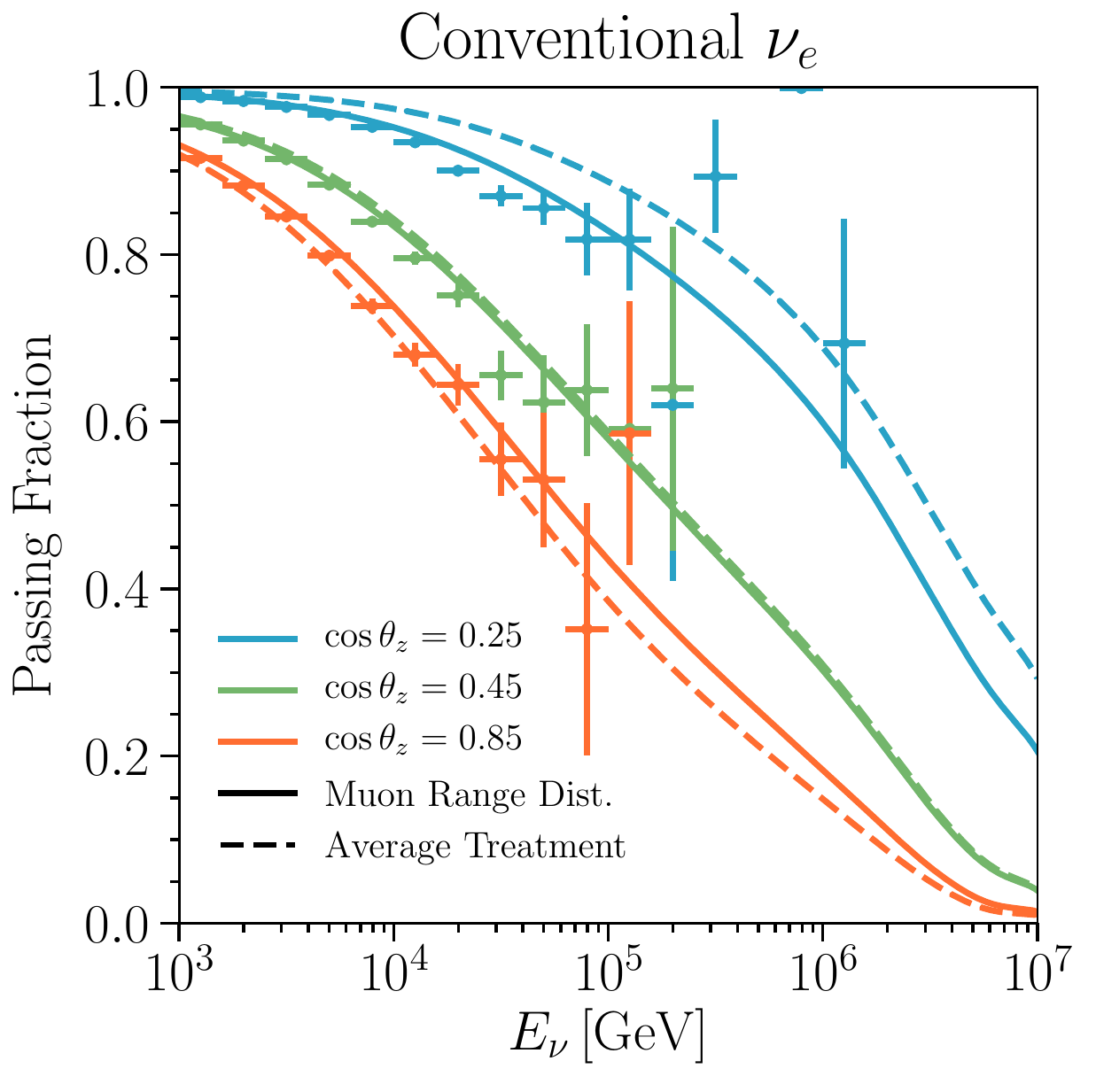}
    }
    \subfloat{
        \includegraphics[width=0.5\linewidth]{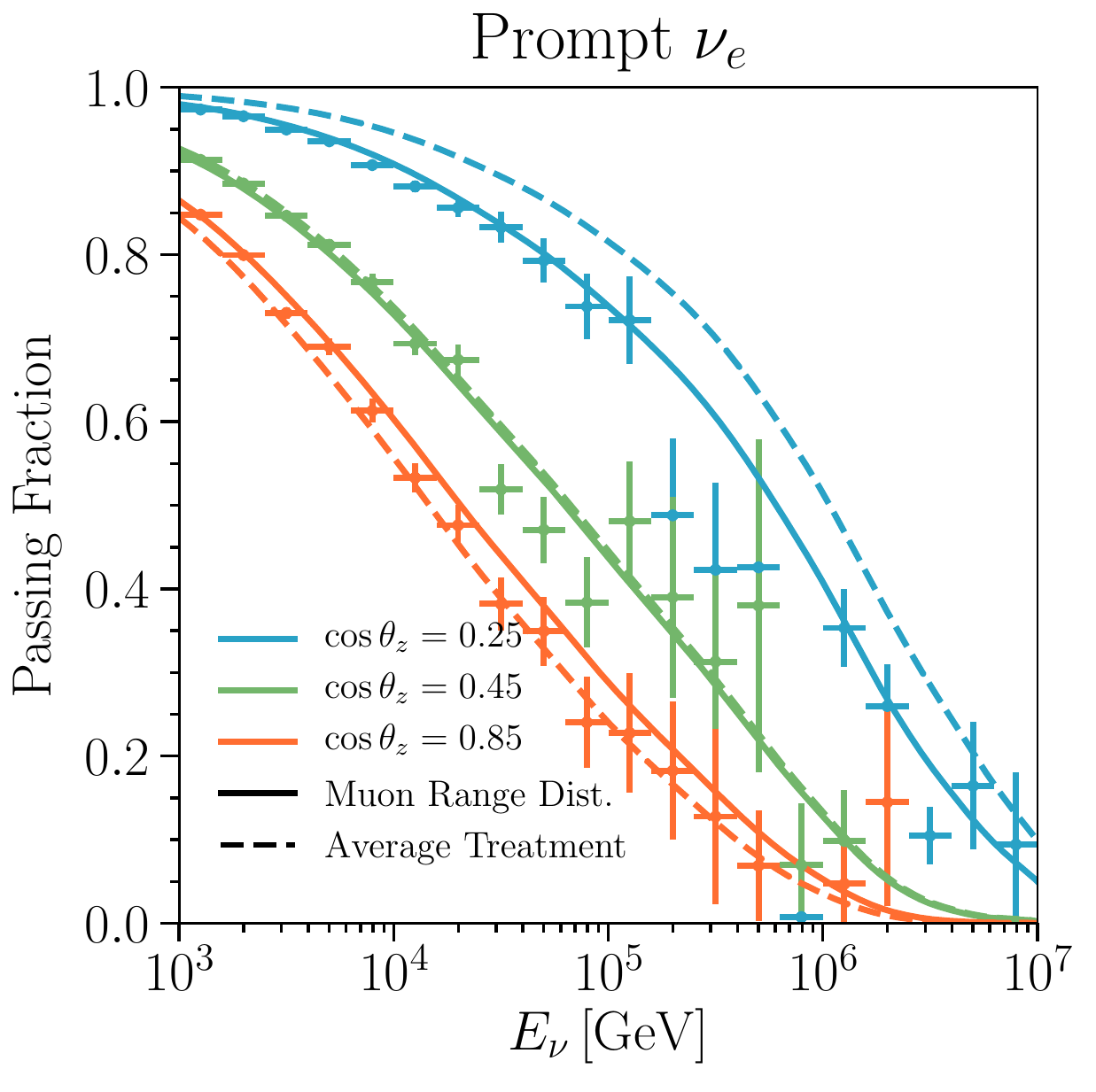}
    }    
\caption{\textbf{\textit{Passing fractions: effect of the treatment of muon losses in ice.}} Results are shown for three values of $\cos\theta_z$ (from top to bottom): 0.25 (blue), 0.45 (green), and 0.85 (orange); using the full muon range distribution (solid) or the median muon range (dashed), see Fig.~\ref{fig:pdet}. Results from the \CORSIKA{} simulation are shown as crosses, with statistical error bars only. In all cases, the H3a primary cosmic-ray spectrum~\cite{Gaisser:2011cc}, the SIBYLL~2.3 hadronic-interaction model~\cite{Engel:2015dxa, Riehn:2015oba} and the MSIS-90-E atmosphere-density model at the South Pole on July 1, 1997~\cite{Labitzke:1985, Hedin:1991} are used, and $d_{\rm det} = 1.95$~km in ice and $\Prob_{\rm light}(\Emf) = \Theta(\Emf - 1\,{\rm TeV})$ are assumed. \textit{Left panel:} Conventional $\nu_e$ passing fraction. \textit{Right panel:} Prompt $\nu_e$ passing fraction.
}
\label{fig:nue_passing-preach-effect}
\end{figure}

\begin{figure}
\centering
    \subfloat{
    \includegraphics[width=0.5\linewidth]{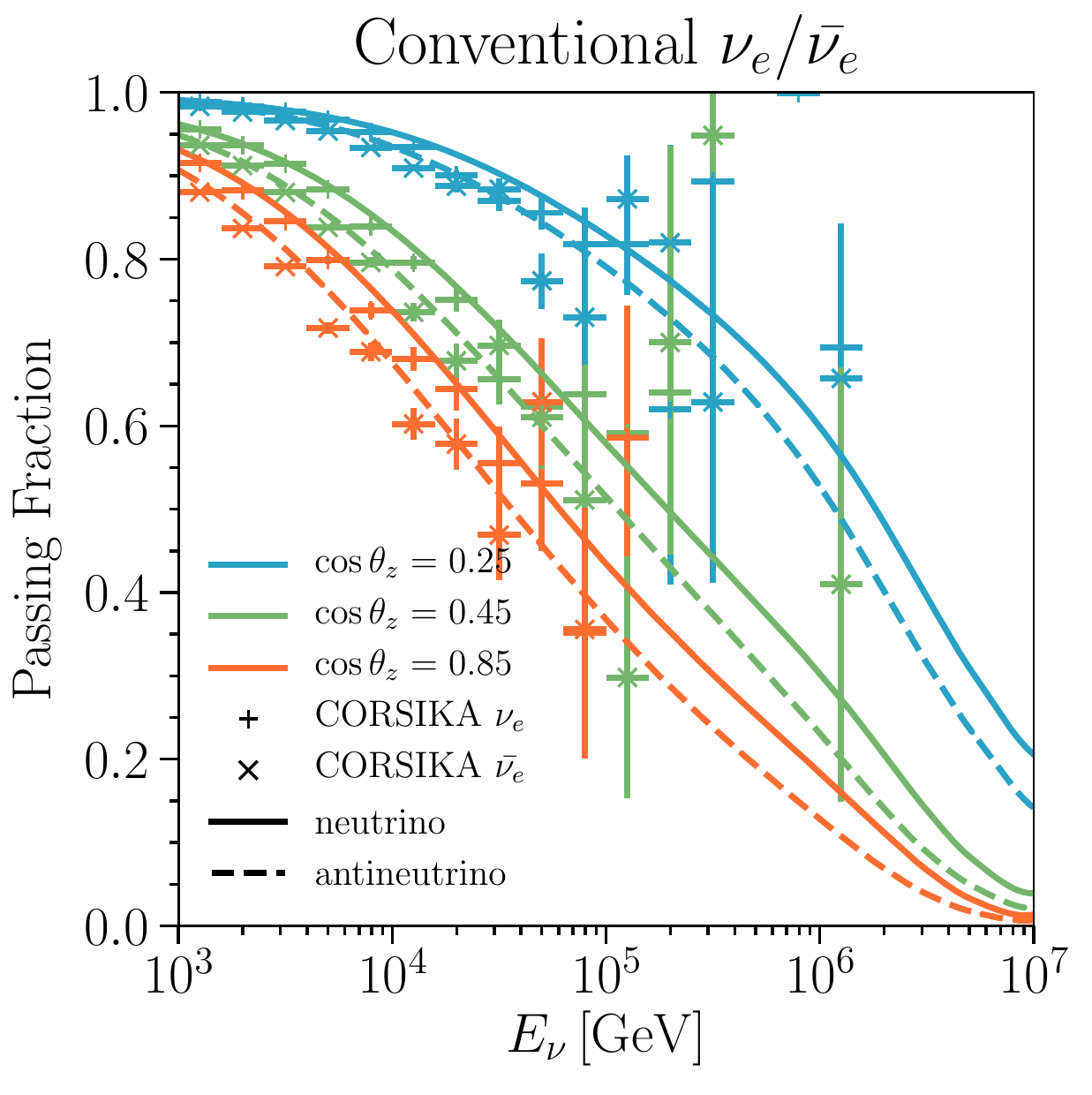}
    }
    \subfloat{
    \includegraphics[width=0.5\linewidth]{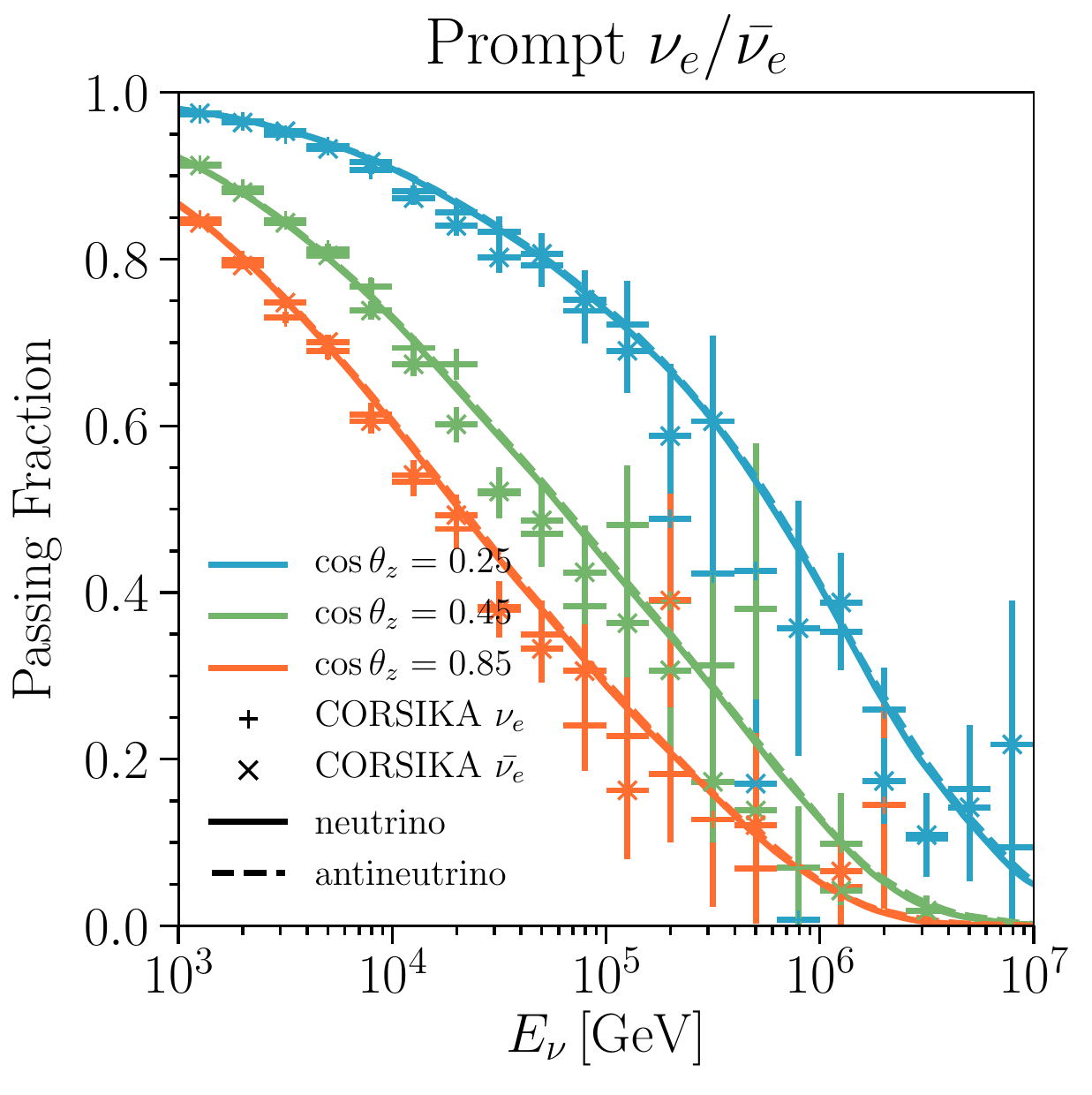}
    }
\caption{\textbf{\textit{Passing fractions: neutrinos versus antineutrinos.}} Results are shown for three values of $\cos\theta_z$ (from top to bottom): 0.25 (blue), 0.45 (green), and 0.85 (orange); for neutrinos (solid) and antineutrinos (dashed). Results from the \CORSIKA{} simulation for neutrinos ($+$) and antineutrinos ($\times$) are also shown, with statistical error bars only. In all cases, the H3a primary cosmic-ray spectrum~\cite{Gaisser:2011cc}, the SIBYLL~2.3 hadronic-interaction model~\cite{Engel:2015dxa, Riehn:2015oba} and the MSIS-90-E atmosphere-density model at the South Pole on July 1, 1997~\cite{Labitzke:1985, Hedin:1991} are used, and $d_{\rm det} = 1.95$~km in ice and $\Prob_{\rm light}(\Emf) = \Theta(\Emf - 1\,{\rm TeV})$ are assumed. \textit{Left panel:} Conventional $\nu_e$ passing fraction. \textit{Right panel:} Prompt $\nu_e$ passing fraction. Note that for prompts there are not differences between $\nu_e$ and $\bar\nu_e$. 
}
\label{fig:nu-e-neutrino-vs-antineutrino}
\end{figure}

As already recognized in Ref.~\cite{Gaisser:2014bja}, in Eq.~(\ref{eq:Nmu}) the actual energy of the shower producing the uncorrelated muons is overestimated. This is because the energy of the parent particle that produces the observed neutrino must be subtracted. If the parent particle $p$ that creates the observed neutrino has energy $E_p$, the remaining energy is $\ECR - E_p$.\footnote{Nephew muons from decays of the sibling mesons are suppressed because their production occurs deeper in the atmosphere where the air density is higher.} Therefore, we have to expand the neutrino yield in Eq.~(\ref{eq:PuncorGJKvS}), from a cosmic-ray shower initiated by nucleus $A$, in terms of the yields from the different parent particles in the shower,
\begin{equation}
\frac{dN_{A, \nu}}{dE_\nu}(\ECR, E_\nu, \theta_z) = \int dE_p \, \int \frac{dX}{\lambda_p(E_p, X)} \, \frac{dN_{p, \nu}}{dE_\nu}(E_p, E_\nu) \, \frac{dN_{A, p}}{dE_p}(\ECR, E_p, X) ~,
\end{equation}
where $dX/\lambda_p(E_p, X)$ is the probability of a parent $p$ with energy $E_p$ to decay between slant depth $X$ and $X + dX$, and $\lambda_p(E_p, X) = \rho(X) \, \tau_p \, E_p/m_p$ is the density, $\rho(X)$, times the decay length of parent particle $p$ with lifetime $\tau_p$, energy $E_p$, and mass $m_p$.\footnote{The path length $x$ and vertical height from the surface of the Earth $h$ are related as $h = x \, \cos\theta^*_z$, where $\theta^*_z$ is the curvature-corrected zenith angle at the coordinates of the neutrino production point at height $h$, and is given by $\sin \theta_z^* = \left(R_\oplus/(R_\oplus + h)\right) \, \sin \theta_z^s$.} The spectrum of neutrinos with energy $E_\nu$ from the decay of a parent particle $p$ with energy $E_p$ is given by $dN_{p, \nu}/dE_\nu(E_p,E_\nu)$. The spectrum of the parent particle $p$ at slant depth $X$, produced by a prototypical shower of energy $\ECR$, initiated from cosmic-ray nucleus $A$ is given by $dN_{A, p}/dE_p(\ECR, E_p, X)$.

In this way, the passing fraction for atmospheric electron neutrinos and antineutrinos can be rewritten,
\begin{align}
\label{eq:PuncorGUE}
\Ppuncor (E_\nu, \theta_z) & = \frac{1}{\phi_\nu(E_\nu, \theta_z)} \, \sum_A \sum_{p} \int dE_p \int \frac{dX}{\lambda_p(E_p, X)} \int d\ECR \nonumber \\
& \frac{dN_{p, \nu}}{dE_\nu}(E_p, E_\nu) \, \frac{dN_{A, p}}{dE_p}(\ECR, E_p, X) \, \phi_A(\ECR) \, \Pzmproto \left(N_\mu = 0  ; \bar N_{A, \mu}(\ECR - E_p, \theta_z)\right) ~.
\end{align}
If $\Pzmproto = 1$, then $\Ppuncor = 1$ by construction. Note that we have explicitly subtracted $E_p$ from the energy $\ECR$ of the cosmic-ray nuclei $A$ that generates the shower. This effect is shown in Fig.~\ref{fig:nue_passing-double-counting}. The passing fractions for the conventional (left) and prompt (right) atmospheric $\nu_e$ flux are depicted for three different zenith angles, with (solid) and without (dashed) the $E_p$ subtraction in Eq.~(\ref{eq:PuncorGUE}). We also show the results obtained from a dedicated Monte Carlo simulation with \CORSIKA~\cite{Heck:1998vt, Heck:2018} (crosses). The agreement with our results is excellent and we stress that there is no fit involved in our computations. 

The approximate calculation without $E_p$ subtraction (dashed) slightly under-predicts the passing fractions. This is because $\bar N_{A, \mu}(\ECR, \theta_z) > \bar N_{A, \mu}(\ECR - E_p, \theta_z)$. The absolute difference is less than $0.05$ below $10^5$~TeV for conventional neutrinos and for all energies for prompt neutrinos. For energies above $10^5$~TeV the differences for conventional neutrinos can be more important, in particular for more vertical directions for which the passing fractions are smaller.

Once we have Eq.~(\ref{eq:PuncorGUE}), we can evaluate the impact of $\Prob_{\rm reach}$, as described in the previous section. In Fig.~\ref{fig:nue_passing-preach-effect}, we illustrate the effect of taking the median muon range (dashed) instead of the full muon range distribution (solid). Both for the conventional (left) and the prompt (right) atmospheric $\nu_e$ fluxes, the median approximation overestimates the passing fraction for horizontal trajectories (right-blue), but underestimates it for more vertical directions (left-orange). This is also apparent from results presented in Ref.~\cite{Gaisser:2014bja} and it can be understood from Fig.~\ref{fig:pdet}. For small $\cos \theta_z$ muons have to traverse a larger portion of ice before reaching the detector. For propagation distances above $\sim 5$~km ($\cos \theta_z \lesssim 0.4$) with the Heaviside $\Prob_{\rm light}$, muons with lower $\Emi$ than that defining the median muon range have a small, but non-negligible, probability to trigger the veto. Thus, fewer muons reach the detector in the approximate case and consequently, the passing fractions are larger. On the contrary, for more vertical events, muons with lower $\Emi$ have a negligible probability to trigger the veto. This probability approaches one at higher $\Emi$, but not as sharply as the median approximation. Thus, when integrated, muons are more likely to trigger the veto in the approximate case and consequently, the passing fractions are smaller.

So far we have only shown results for $\nu_e$, but the passing fractions for the conventional $\bar\nu_e$ flux are different. In Fig.~\ref{fig:nu-e-neutrino-vs-antineutrino} we illustrate the differences between $\nu_e$ and $\bar\nu_e$ for the same angles considered in previous figures, along with results from the \CORSIKA{} simulation. Again, the agreement between the simulation and our modeling is excellent for antineutrinos. The conventional $\bar\nu_e$ passing fractions are lower than those of conventional $\nu_e$. This is due to positively charged mesons being preferentially produced, leading to a harder spectrum than the negatively charged ones. For the prompt flux, the passing fractions are almost identical as they are mainly produced from gluon fusion~\cite{Arguelles:2015wba}.

\subsection{Passing fraction for muon neutrinos}
\label{sec:pfnumu}

\begin{figure}
\centering
    \subfloat{
        \includegraphics[width=0.5\linewidth]{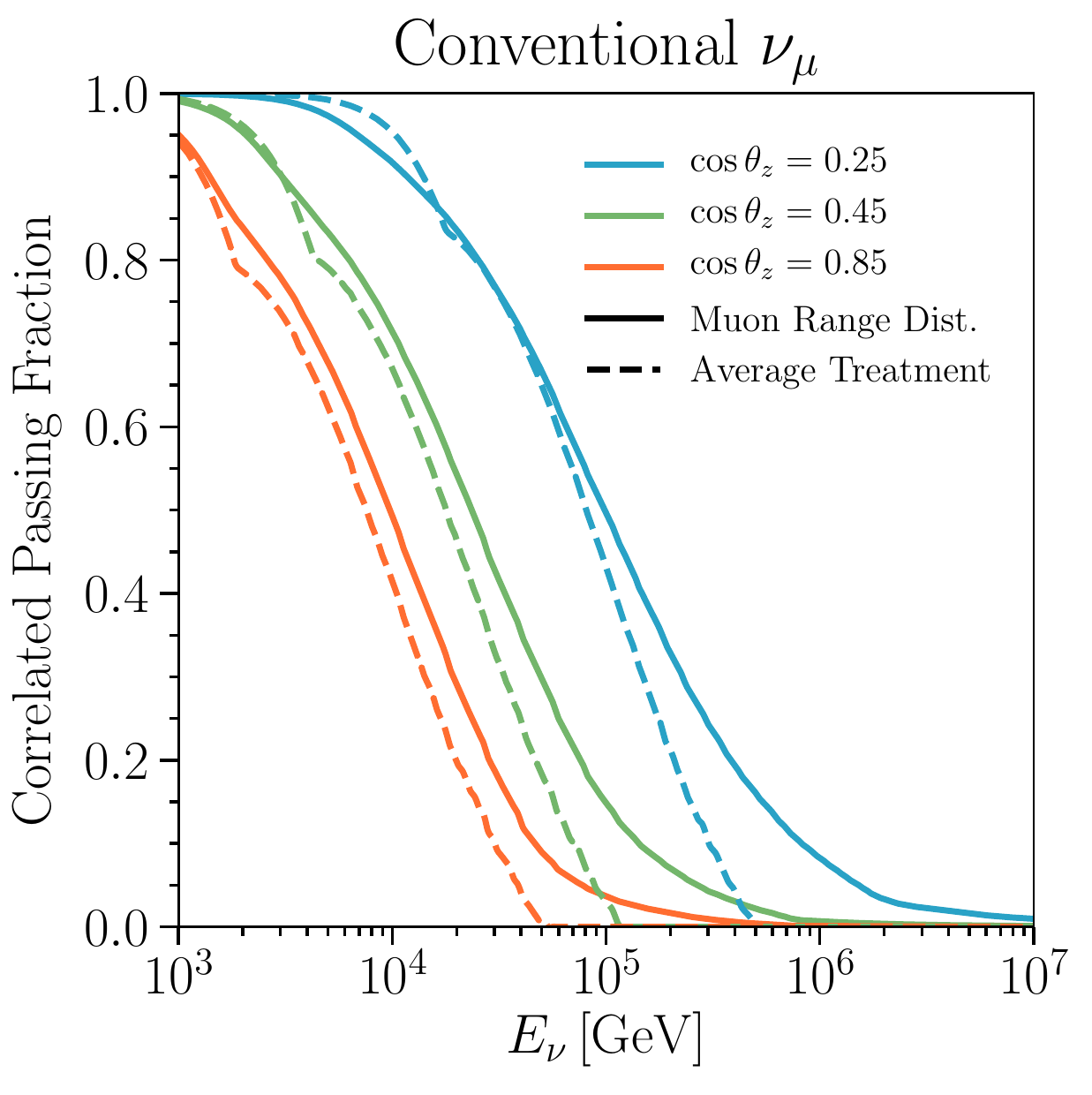}
    }
    \subfloat{
        \includegraphics[width=0.5\linewidth]{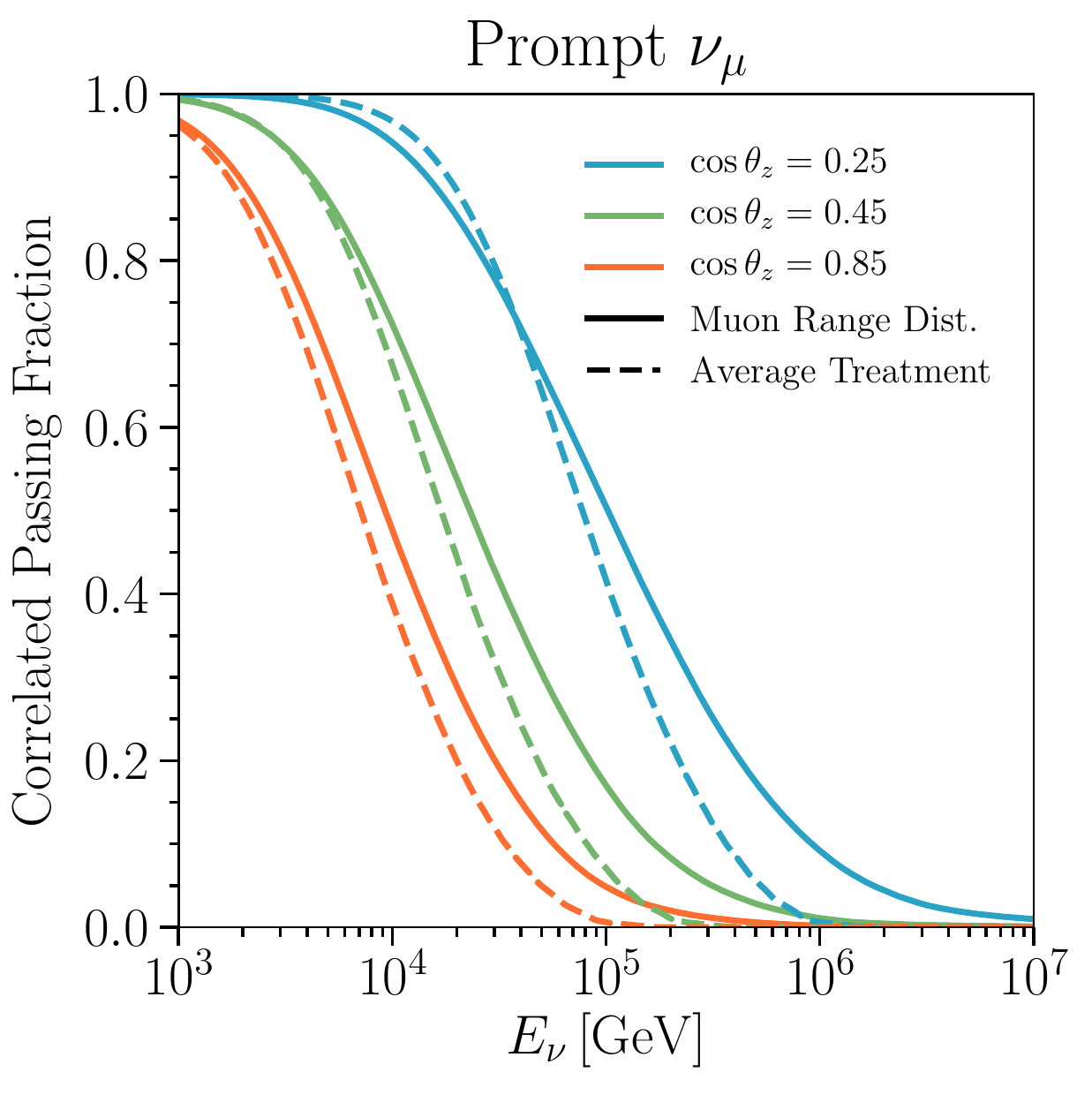}
    }
\caption{\textbf{\textit{Correlated passing fractions: effect of the treatment of muon losses in ice.}} Same as Fig.~\ref{fig:nue_passing-preach-effect} but only for the correlated part of the passing fraction for $\nu_\mu$.}
\label{fig:nu-mu-correlated-preach-effect}
\end{figure}

Atmospheric muon neutrinos (and antineutrinos) from hadron decays are always produced alongside a muon. Therefore, the passing fraction of atmospheric muon neutrinos has a correlated suppression factor in addition to the uncorrelated one, described in the previous section. This extra suppression depends entirely on the probability of the sibling muon to trigger the veto, which in turn depends on the energy and incoming direction of the muon~\cite{Schonert:2008is}. For two-body decays, as in the case of pions and most kaons, the muon energy is $\Emi = E_p - E_\nu$, conditional on the muon neutrino energy $E_\nu$ from the decay of a parent particle $p$ with energy $E_p$. Thus, the energy spectrum of these muons is $dN_{p, \mu}^{2 \text{-body}}/d\Emi(E_p, E_\nu, \Emi) = \delta(E_p-E_\nu+\Emi)$. For more generic $n$-body decays, as in the case of charmed hadrons, the probability of non-detection can be written as
\begin{equation}
\label{eq:pnomusib}
\Pzmsib \left(\theta_z|E_p, E_\nu \right) = 1 - \int d\Emi \, \Prob_{\rm det}\left(\Emi, \theta_z\right) \frac{dN_{p, \mu}}{d\Emi}(E_p, E_\nu, \Emi) ~.
\end{equation}

\begin{figure}
\centering
    \subfloat{
    \includegraphics[width=0.5\linewidth]{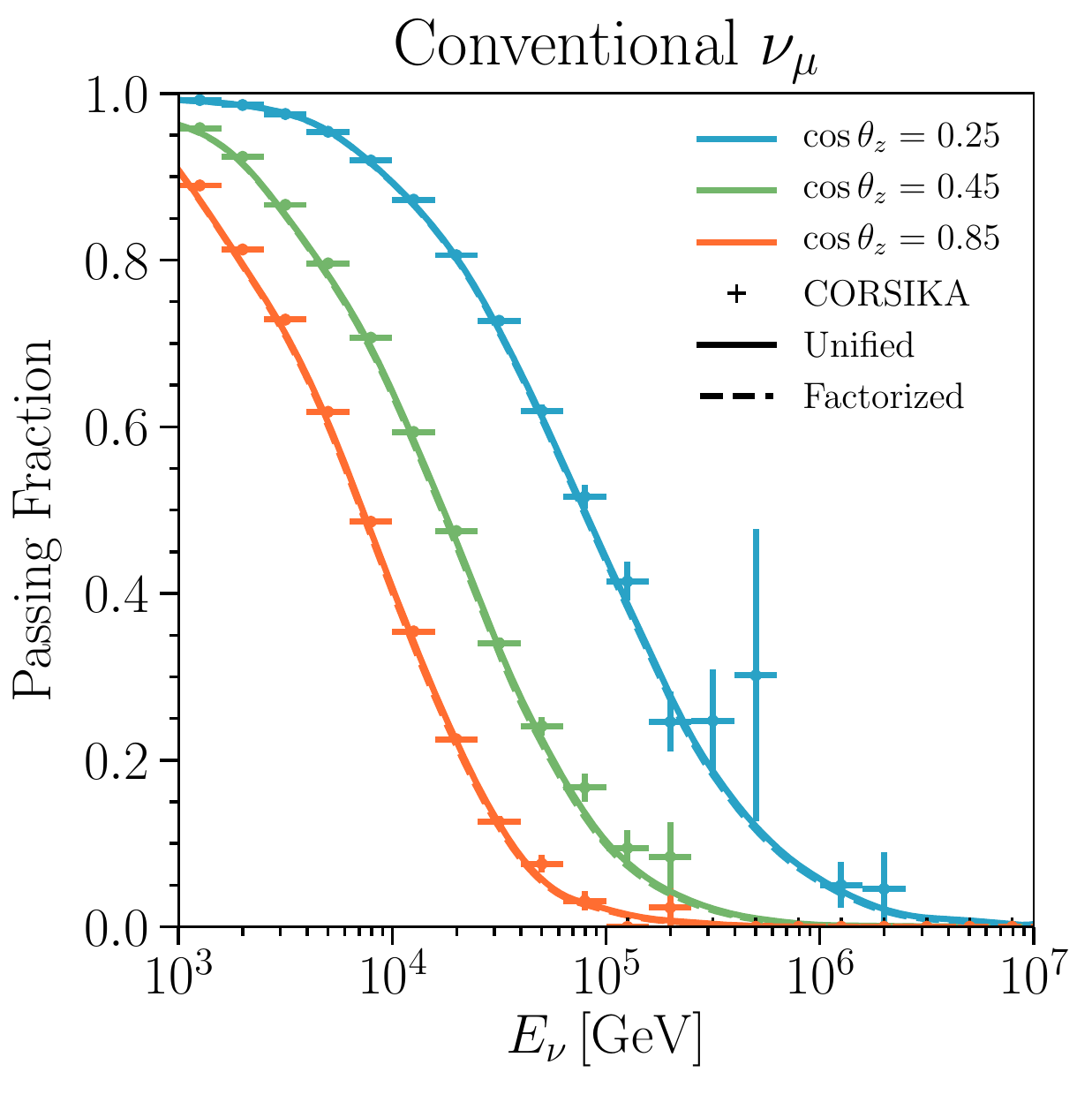}
    }
    \subfloat{
    \includegraphics[width=0.5\linewidth]{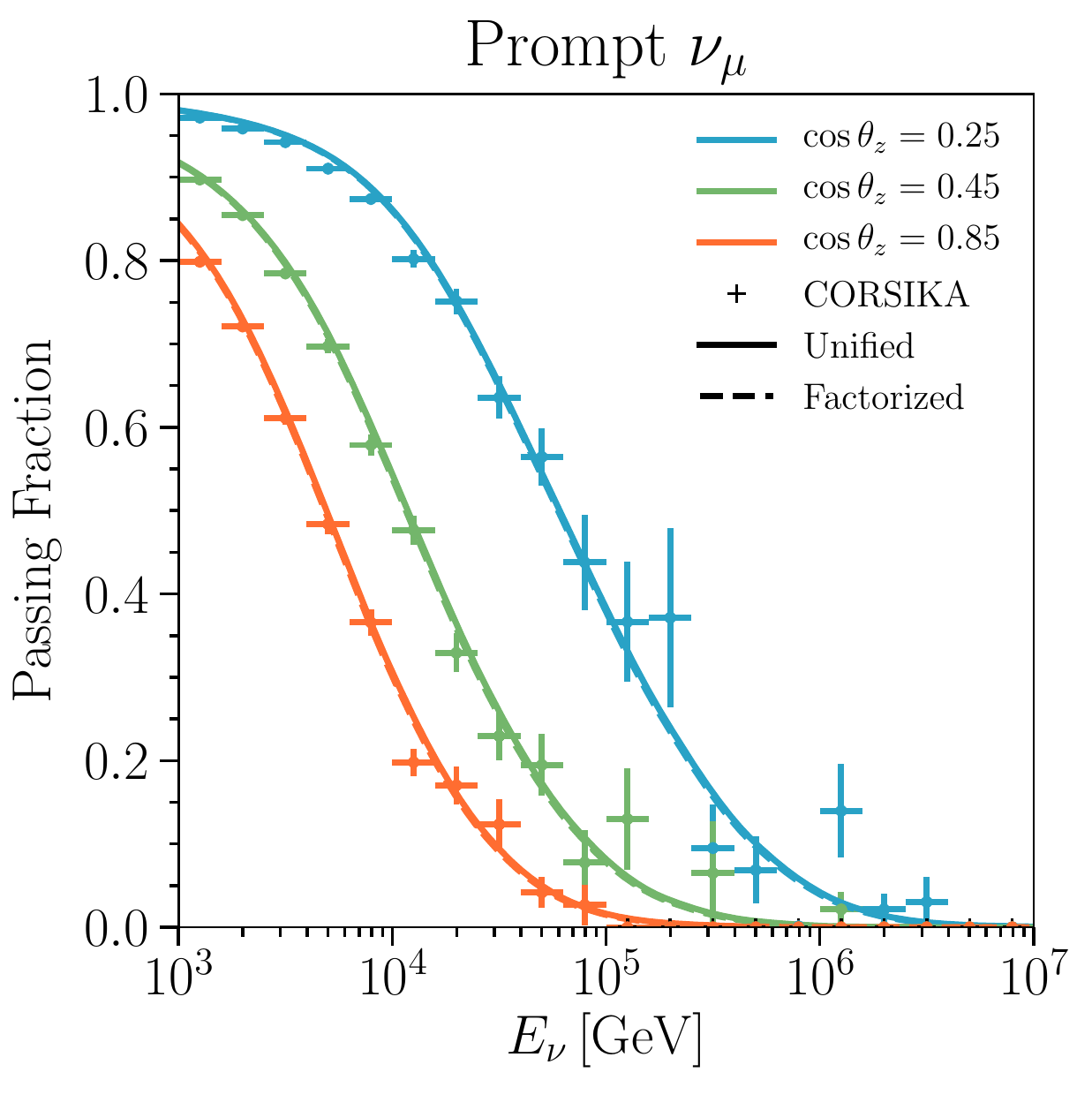}
    }
\caption{\textbf{\textit{Passing fractions: differences between the unified and the factorized treatments for $\boldsymbol{\nu_\mu}$.}} Same as Fig.~\ref{fig:nue_passing-double-counting} regarding the approximations on the energy of the shower which gives rise to uncorrelated muons. Comparison of the unified treatment (solid), Eq.~(\ref{eq:GUE}), and the approximate treatment (dashed), Eq.~(\ref{eq:PGJKvS}), which factorizes the correlated and uncorrelated passing fractions. The result is driven by the correlated part.}
\label{fig:nu-mu-unified-effect}
\end{figure}

Now, it is necessary to express the neutrino flux in terms of the parent fluxes and their decays. As described in Sec.~\ref{sec:pfnue}, the decay probability of a parent $p$ at slant depth $X$ is $dX/\lambda_p(E_p, X)$. Then, the differential flux of parent particles at slant depth $X$, $\phi_{A, p}(E_p, X)$, can be defined in terms of the spectrum of parent particles with energy $E_p$ from a prototypical cosmic-ray shower, initiated from nucleus $A$, with energy $\ECR$ and differential flux $\phi_{A}(\ECR)$, as
\begin{equation}
\label{eq:phip}
\phi_{A, p}(E_p, X) = \int d\ECR \, \frac{dN_{A, p}}{dE_p}(\ECR, E_p, X) \, \phi_{\rm CR} (\ECR) ~.
\end{equation}
If we now express the energy spectrum of neutrinos with energy $E_\nu$ from the decay of a parent particle $p$ as $dN_{p, \nu}/dE_\nu(E_p,E_\nu)$, then the differential flux of atmospheric neutrinos at the surface of the Earth, $\phi_\nu(E_\nu, \theta_z)$, is given by
\begin{equation}
\label{eq:nufluxcor}
\phi_\nu(E_\nu, \theta_z) = \sum_A \sum_{p} \int dE_p  \int \frac{dX}{\lambda_p(E_p, X)} \, \frac{dN_{p, \nu}}{dE_\nu}(E_p, E_\nu) \, \phi_{A, p}(E_p, X),
\end{equation}
and therefore, we can define the correlated passing fraction as
\begin{equation}
\label{eq:Pcor}
\Ppcor(E_\nu, \theta_z) = \frac{1}{\phi_\nu(E_\nu, \theta_z)} \, \sum_A \sum_{p} \int dE_p \int \frac{dX}{\lambda_p(E_p, X)} \, \frac{dN_{p, \nu}}{dE_\nu}(E_p, E_\nu) \, \phi_{A, p}(E_p, X) \, \Pzmsib \left(\theta_z|E_p, E_\nu \right).
\end{equation}

In analogy with Fig.~\ref{fig:nue_passing-preach-effect}, Fig.~\ref{fig:nu-mu-correlated-preach-effect} shows the effect of the correct treatment of the muon range distribution on the correlated passing fraction defined in Eq.~(\ref{eq:Pcor}). As described in section~\ref{sec:preach_plight}, accounting for the stochasticity of muon energy losses smears the $\Prob_{\rm det}$ distribution from a Heaviside function, such that muons with lower $\Emi$ may be vetoed and muons with higher $\Emi$ may not be so. This effect is seen in the passing fraction, which drops at lower energies due to the increased vetoing probability of lower $\Emi$ muons and increases at higher energies due to the decreased vetoing probability of higher $\Emi$ muons. The transition energy is higher for more horizontal trajectories and in the relevant energy range, for more vertical trajectories, the effect is a general suppression of the passing fraction. Note also that in the case of a Heaviside $P_{\rm det}$, there is a shoulder due to the different contribution of pions and kaons to the neutrino flux. This gets smeared out when taking the muon range distribution into account.

\begin{figure}
\centering
    \subfloat{
    \includegraphics[width=0.5\linewidth]{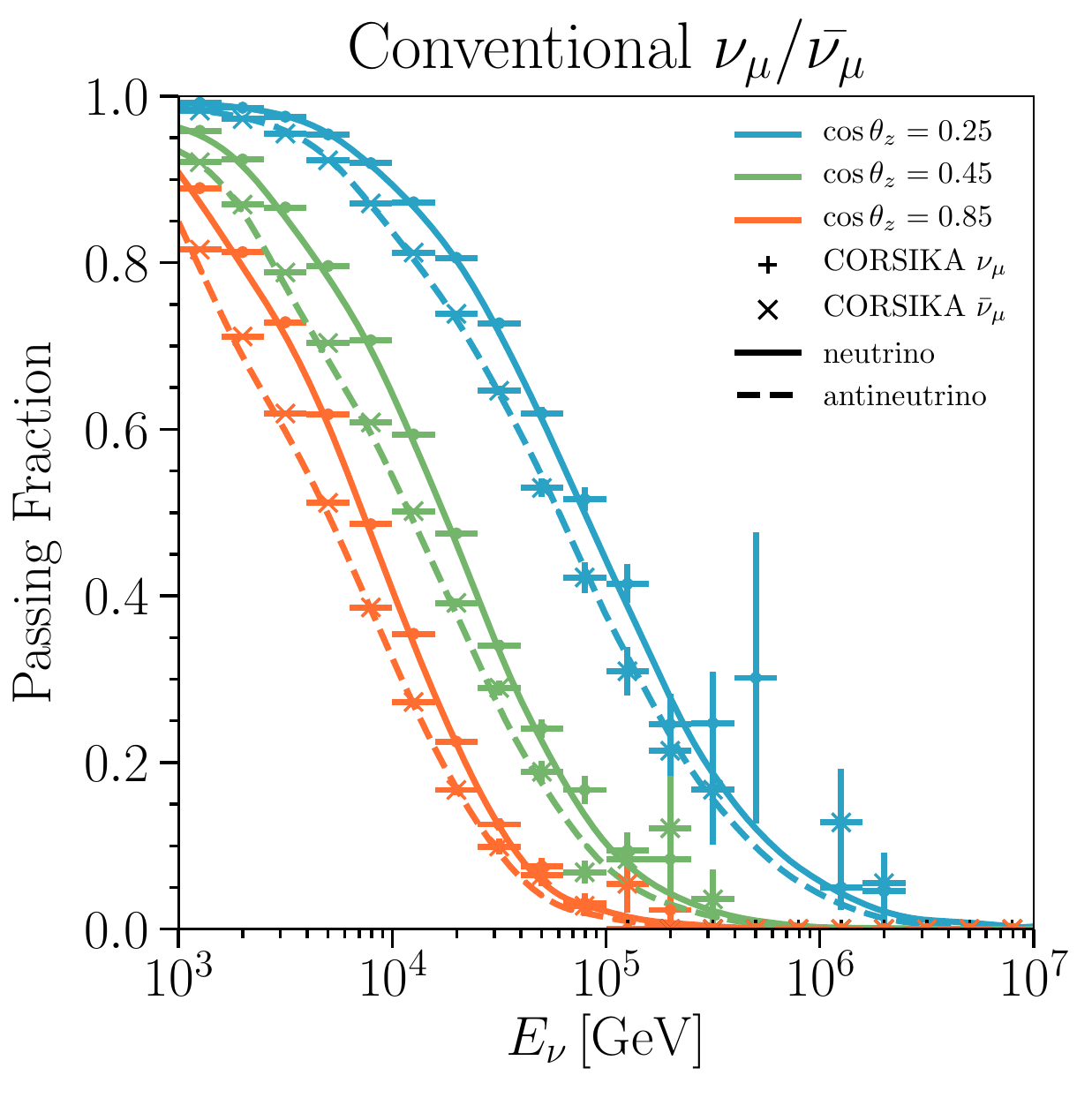}
    }
    \subfloat{
    \includegraphics[width=0.5\linewidth]{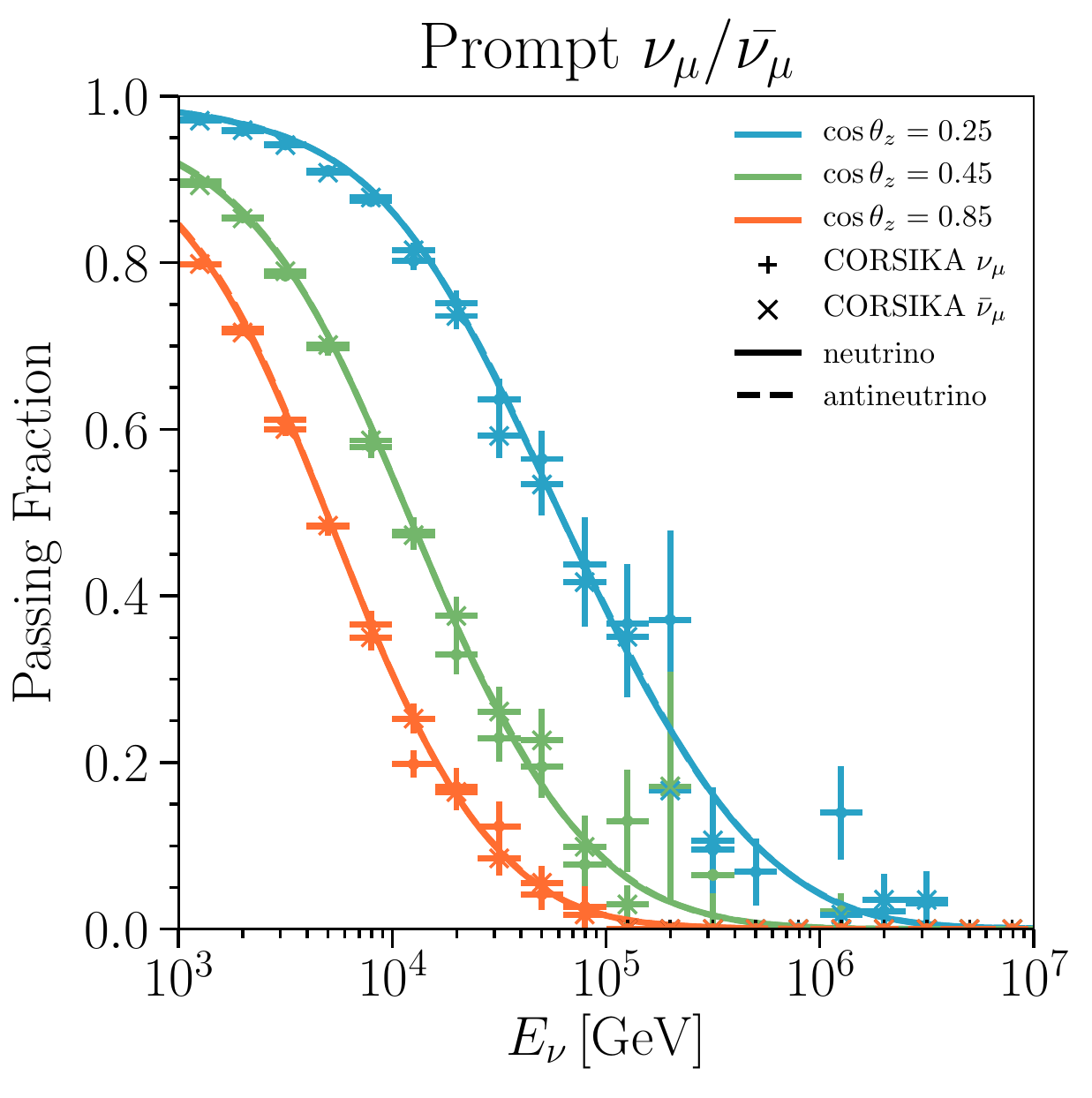}
    }
\caption{\textbf{\textit{Passing fractions: neutrinos versus antineutrinos.}} Same as Fig.~\ref{fig:nu-e-neutrino-vs-antineutrino} but for $\nu_\mu$ and $\bar\nu_\mu$. Note that for prompt neutrinos there are not differences between $\nu_\mu$ and $\bar\nu_\mu$.}
\label{fig:nu-mu-neutrino-vs-antineutrino}
\end{figure}

\pagebreak

As discussed above, for two-body decays $\Emi = E_p - E_\nu$ and assuming Eq.~(\ref{eq:PrPlSGRS}) for the muon detection probability, Eq.~(\ref{eq:Pcor}) simplifies to
\begin{equation}
\label{eq:PcorSGRS}
\Prob_{\rm pass}^{\rm cor, SGRS}(E_\nu, \theta_z) = \frac{1}{\phi_\nu(E_\nu, \theta_z)} \, \sum_A \sum_{p} \int dE_p  \int \frac{dX}{\lambda_p(E_p, X)} \, \frac{dN_{p, \nu}}{dE_\nu}(E_p, E_\nu) \, \phi_{A, p}(E_p, X) \,  
\left[1 - \Prob_{\rm det}^{\rm SGRS}\left(E_p-E_\nu , \theta_z\right)\right]~.
\end{equation}
In this way, an approximate passing probability for muon neutrinos was defined as~\cite{Gaisser:2014bja}
\begin{equation}
\label{eq:PGJKvS}
\Prob_{\rm pass}^{\rm GJKvS} (E_\nu, \theta_z) \equiv \Prob_{\rm pass}^{\rm cor, SGRS}(E_\nu, \theta_z) \, \Prob_{\rm pass}^{\rm uncor, GJKvS}(E_\nu, \theta_z) ~,
\end{equation}
where we have used Eqs.~(\ref{eq:PuncorGJKvS}) and~(\ref{eq:PcorSGRS}).

However, the two-body approximation is not appropriate for the calculation of the passing fractions for the prompt fluxes, as neutrinos (and antineutrinos) come mainly from $D^{\pm}$, $D^0$, $\bar{D}^0$, $\Lambda_c^+$, and $D_s^{\pm}$ decays~\cite{Fedynitch:2015zma}. A correct treatment of $n$-body decays requires evaluating the muon distributions from the decays of these particles and using Eq.~(\ref{eq:pnomusib}). In this work, $dN_{p, \mu}/d\Emi$ was generated for $K^0_L$, $D^+$, $D^0$, and $D^+_s$.\footnote{The decay distributions for their antiparticles are identical.}

The ``factorized'' approach in Eq.~(\ref{eq:PGJKvS}) can be further improved by combining Eqs.~(\ref{eq:PuncorGUE}) and~(\ref{eq:Pcor}), which accounts for the correction in the energy of the prototypical shower, $\ECR - E_p$. The new passing fraction can be defined as
\begin{align}
\label{eq:GUE}
\Prob_{\rm pass}(E_\nu, \theta_z) = \frac{1}{\phi_\nu(E_\nu, \theta_z)} \, \sum_A \sum_{p} & \int dE_p  \int \frac{dX}{\lambda_p(E_p, X)} \int d\ECR \, \frac{dN_{p, \nu}}{dE_\nu}(E_p, E_\nu) \, \frac{dN_{A, p}}{dE_p}(\ECR, E_p, X) \, \phi_A(\ECR) \nonumber \\
& \times \Pzmsib \left(\theta_z|E_p, E_\nu \right) \, \Pzmproto\left(N_\mu=0;  \bar N_{\mu, A}(\ECR - E_p, \theta_z) \right),
\end{align}
where we have substituted Eq.~(\ref{eq:phip}) into Eq.~(\ref{eq:Pcor}). The numerator in this equation defines $\phi_\nu^{\rm pass}(E_\nu, \theta_z)$. In this way, Eq.~(\ref{eq:GUE}) represents our master equation to obtain the passing fractions for atmospheric neutrinos. It can be applied directly to the case of electron neutrinos where no sibling muons are produced and thus, $\Pzmsib \left(\theta_z | E_p, E_\nu \right) = 1$. In that case, Eq.~(\ref{eq:GUE}) reduces to Eq.~(\ref{eq:PuncorGUE}). More explicitly, we can substitute Eqs.~(\ref{eq:PrPl}),~(\ref{eq:pnomuproto}), and~(\ref{eq:pnomusib}) to get
\begin{align}
\label{eq:GUEexplicit}
\Prob_{\rm pass} (E_\nu, \theta_z) = & \frac{1}{\phi_\nu(E_\nu, \theta_z)} \, \sum_A \sum_{p} \int dE_p \int \frac{dX}{\lambda_p(E_p, X)} \int d\ECR \, \frac{dN_{p, \nu}}{dE_\nu}(E_p, E_\nu) \, \frac{dN_{A, p}}{dE_p}(\ECR, E_p, X) \, \phi_A(\ECR) \nonumber \\
& \times \left[ 1 - \int \, d\Emi \, \frac{dN_{p, \mu}}{d\Emi}(E_p, E_\nu, \Emi) \, \int d\Emf \, \Prob_{\rm light}(\Emf)  \, \Prob_{\rm reach}\left(\Emf | \Emi, \theta_z\right) \right] \, e^{-N_{A, \mu}(\ECR - E_p, \theta_z)}  ~,
\end{align}

A comparison between the results for muon neutrinos obtained using Eq.~(\ref{eq:PGJKvS}) (factorizing the correlated and uncorrelated passing fractions) and those obtained with the more accurate Eq.~(\ref{eq:GUE}), or equivalently Eq.~(\ref{eq:GUEexplicit}), is shown in Fig.~\ref{fig:nu-mu-unified-effect}. The effect of the subtraction of $E_p$ from $\ECR$ is smaller than in the case of electron neutrinos, shown in Fig.~\ref{fig:nue_passing-double-counting}, due to $\Pzmsib$ being a much more dominant factor than $\Pzmproto$. Note that, for the uncorrelated part of the passing fraction, this subtraction is more important at higher energies, for which the correlated part is very small. Thus, the relative effect on the muon neutrino passing fraction is much smaller. Finally, comparisons of the passing fractions calculated for muon neutrinos and muon antineutrinos are shown in Fig.~\ref{fig:nu-mu-neutrino-vs-antineutrino} and exhibit very good agreement with the Monte Carlo results, as in the case of electron neutrinos (Fig.~\ref{fig:nu-e-neutrino-vs-antineutrino}). 

\subsection{Passing fraction for tau neutrinos}
\label{sec:pfnutau}

For down-going neutrinos at energies above $\sim$~TeV, the major atmospheric $\nu_\tau$ component is the prompt flux from $D_s$ decays into $\nu_\tau$ and $\tau$ and subsequent $\tau$ decays~\cite{Pasquali:1998xf}. The additional bottom-quark ($B^+$ and $B^0$) contribution represents a $\lesssim 10\%$ correction~\cite{Pasquali:1998xf, Bhattacharya:2016jce}. The calculation of the uncorrelated passing fraction is still given by Eq.~(\ref{eq:PuncorGUE}). However, the calculation of the correlated passing fraction is different from the conventional $\nu_\mu$ calculation. Conventional $\nu_\mu$ are produced mainly via two-body decays where $\Emi = E_p - E_\nu$. Prompt $\nu_\tau$, on the other hand, are more similar to to prompt $\nu_\mu$ in which correlated muons are produced from $n$-body decay. These muon decay distributions must be evaluated in order to calculate the $\nu_\tau$ passing fraction. At $\sim$~TeV energies, $\tau$ leptons are produced high in the atmosphere and can be assumed to decay without energy losses. Although Eq.~(\ref{eq:Pcor}) applies, $\Pzmsib$ has to be specifically computed. In addition, there is a subleading contribution from the oscillated atmospheric $\nu_\mu$ flux~\cite{Martin:2003us, Bulmahn:2010pg}. The passing fraction for oscillated $\nu_\tau$ is identical to that for $\nu_\mu$, though the event signature would be different in the detector. There is another, $4-6$ orders of magnitude more subleading contribution due to tau pair production from muon losses in the atmosphere~\cite{Bulmahn:2010pg}. In this case, there is a correlated muon accompanying the $\nu_\tau$ from tau decays and the always present uncorrelated part.

At the relevant energies in this work, the atmospheric $\nu_\tau$ flux is about an order of magnitude smaller than the prompt $\nu_e$ and $\nu_\mu$ fluxes. Given our current knowledge of the atmospheric prompt flux, for which only upper bounds exist, and given the subleading nature of the $\nu_\tau$ component, we do not compute $\nu_\tau$ passing fractions and do not discuss this flux any further. 

\subsection{Improvements with respect to previous calculations}
\label{sec:improvements}

\begin{figure}
\centering
    \subfloat{
        \includegraphics[width=0.5\linewidth]{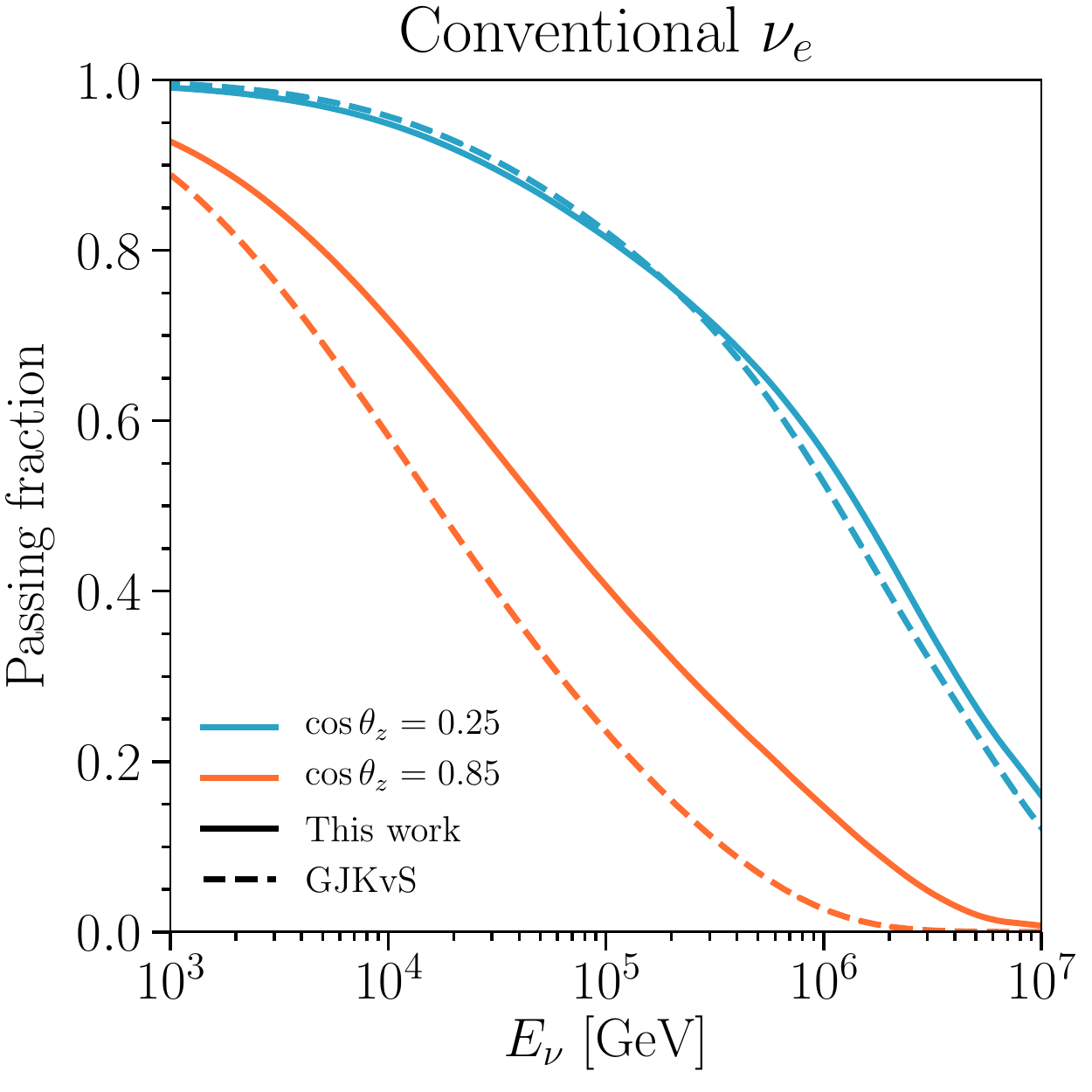}
    }
    \subfloat{
        \includegraphics[width=0.5\linewidth]{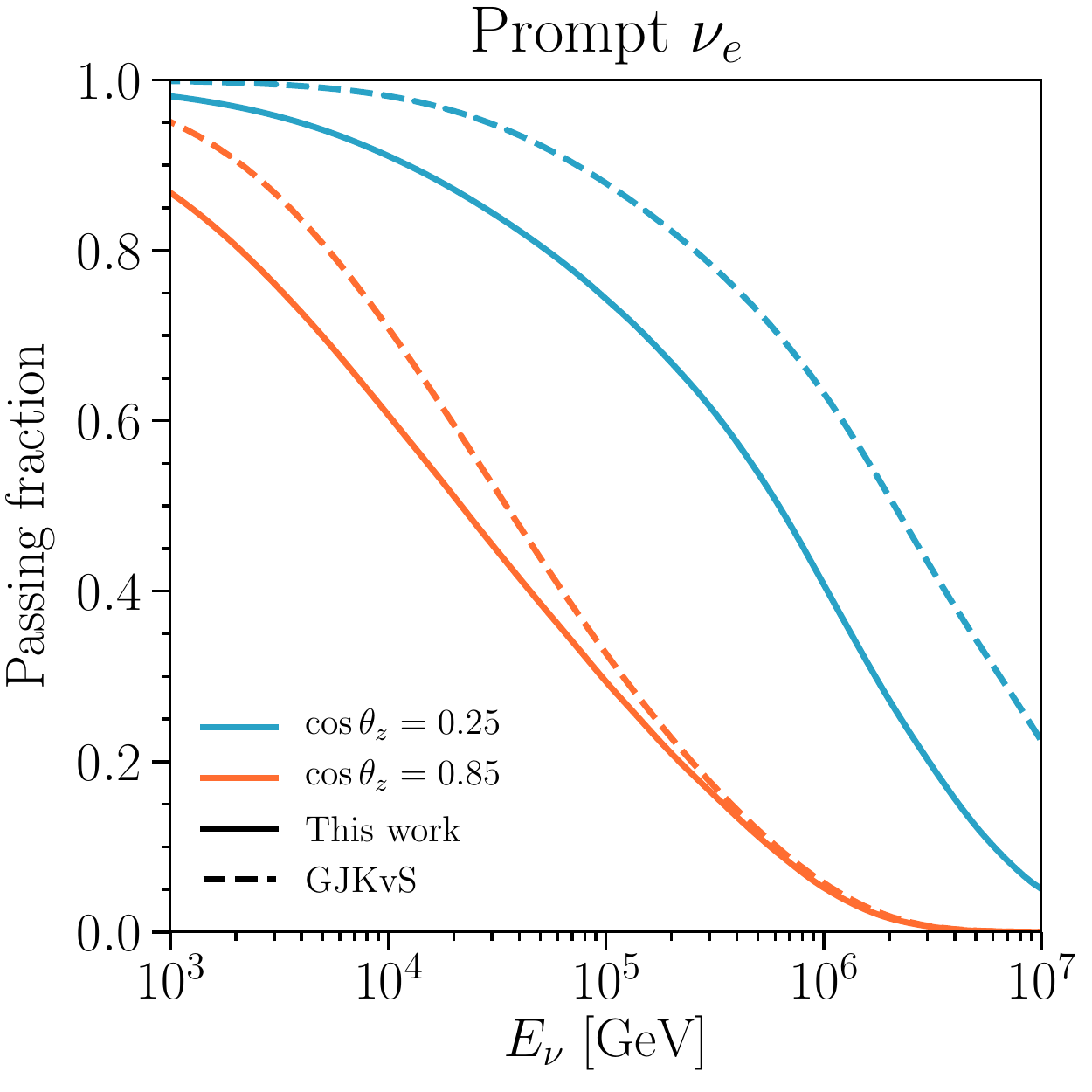}
    } \\[2ex]
    \subfloat{
        \includegraphics[width=0.5\linewidth]{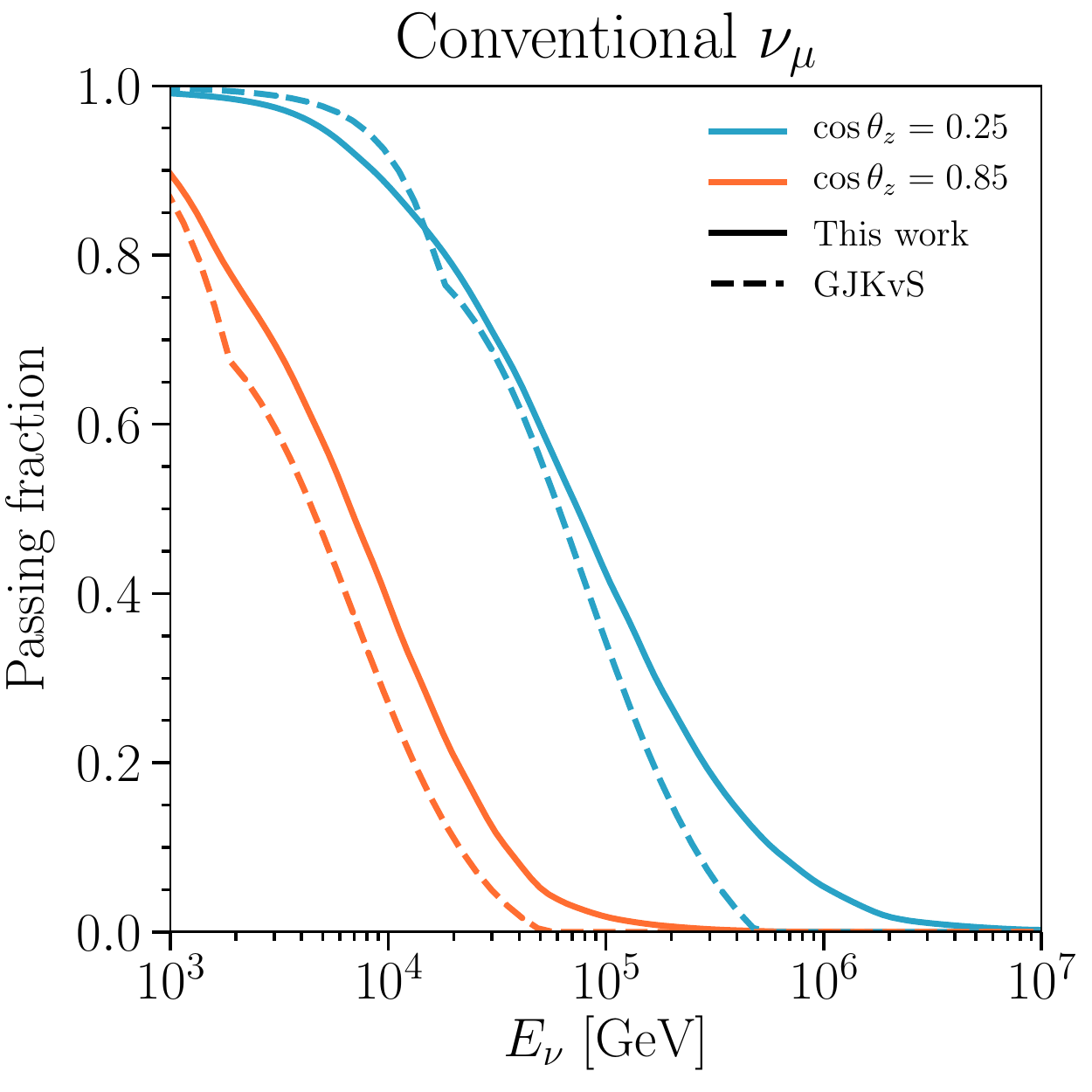}
    }
    \subfloat{
        \includegraphics[width=0.5\linewidth]{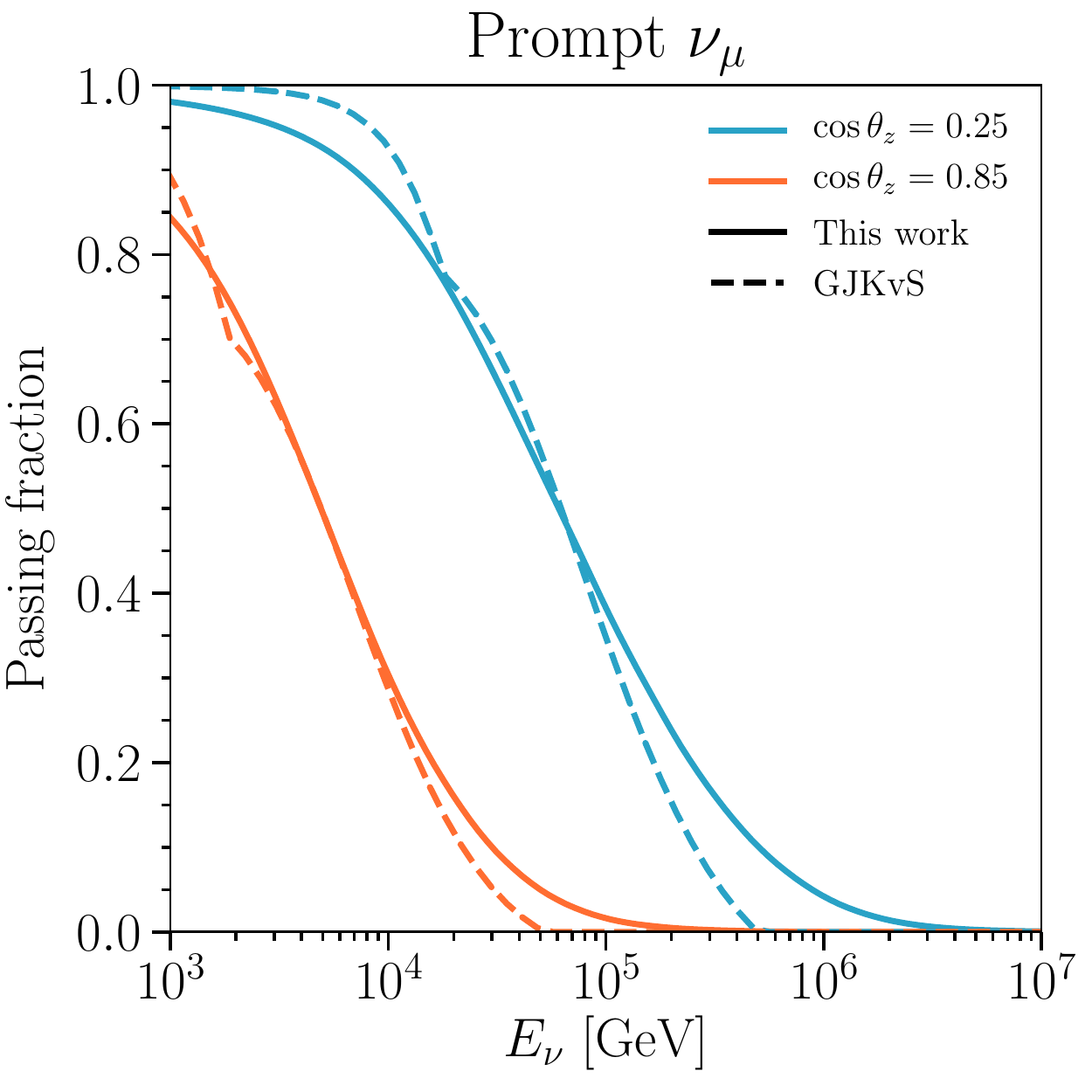}
    }    
\caption{\textbf{\textit{Passing fractions: comparison with previous work}}. Results are shown for two values of $\cos\theta_z =$ (from top to bottom): 0.25 (blue) and 0.85 (orange); with the calculation in this work (solid) and with that in Ref.~\cite{Gaisser:2014bja} (dashed). Our results are obtained with the H3a primary cosmic-ray spectrum~\cite{Gaisser:2011cc}, the SIBYLL~2.3c hadronic-interaction model~\cite{Riehn:2017mfm}, the MSIS-90-E atmosphere-density model at the South Pole on July 1, 1997~\cite{Labitzke:1985, Hedin:1991}, and assuming $d_{\rm det} = 1.95$~km in ice and $\Prob_{\rm light}(\Emf) = \Theta(\Emf - 1\,{\rm TeV})$. They include all of the effects discussed in previous sections, Eq.~(\ref{eq:GUE}). \textit{Top-left panel:} Conventional $\nu_e$ passing fraction. \textit{Top-right panel:} Prompt $\nu_e$ passing fraction. \textit{Bottom-left panel:} Conventional $\nu_\mu$ passing fraction. \textit{Bottom-right panel:} Prompt $\nu_\mu$ passing fraction.} 
\label{fig:nue-passing-comparison-old}
\end{figure}

The first proposal to calculate the suppression of the down-going atmospheric neutrino flux for high-energy neutrino telescopes only accounted for accompanying muons produced from the same decaying parent meson~\cite{Schonert:2008is}. Therefore, it was only applied to atmospheric muon neutrinos. The calculation was based on a several analytical approximations, applicable to neutrinos from pion and kaon decays and assuming a power-law primary cosmic-ray spectrum. As described above, the treatment of energy losses did not account for muon range distributions, and the criteria for muons to trigger the veto was a Heaviside function at some threshold energy. This allowed the final passing fractions to be expressed analytically.

The analytic treatment was generalized by including muons from other branches of the shower in which the neutrino was produced~\cite{Gaisser:2014bja}, as explained in section~\ref{sec:pfnue}. This allowed part of the flux of down-going, atmospheric electron neutrinos to be excluded from the sample. This method used analytic approximations to describe the average properties of muons generated by cosmic rays for both conventional and prompt neutrino fluxes. These approximations were shown to correctly describe the results of the \CORSIKA{} simulation. Indeed, $\bar{N}_{A, \mu}(\ECR, \theta_z)$ was calculated by fitting a parameterized yield function to the \CORSIKA{} simulation, which used the SIBYLL~2.1~\cite{Ahn:2009wx} hadronic-interaction model for conventional atmospheric neutrinos and DPMJET-2.55~\cite{Ranft:1999fy, Ranft:1999qe, Berghaus:2007hp} for prompt atmospheric neutrinos. Along with the results from Ref.~\cite{Schonert:2008is}, this extra contribution led to Eq.~(\ref{eq:PGJKvS}). By not subtracting $E_p$ from $\ECR$, the passing fraction for muon neutrinos accounted for the correlated muon more than once. In the case of the passing fraction for electron neutrinos, it led to the overestimation of the energy of the shower producing detectable muons. This allowed the correlated and uncorrelated rates to be factorized.\footnote{Moreover, note that in Ref.~\cite{Gaisser:2014bja}, the correlated part as computed for conventional neutrinos was also used for prompt neutrinos.}

Our approach includes the parent spectra from cosmic-ray interactions in the atmosphere and led to the more accurate Eq.~(\ref{eq:GUE}). It is a fully consistent definition of the passing fraction for both electron and muon neutrinos, as described in section~\ref{sec:pf}. In our treatment we also account, in full detail, for the stochasticity of muon energy losses by computing the muon range distributions. We stress that our approach also allows considering different detector configurations (depth, type of medium, muon vetoing efficiency), cosmic-ray primary spectra, hadronic-interaction models and atmosphere-density models, so that a fully consistent treatment of systematic uncertainties on the passing fractions and final fluxes can be performed.

In Fig.~\ref{fig:nue-passing-comparison-old} we compare results using the code provided by Ref.~\cite{Gaisser:2014bja} with those obtained by the default setup used in this work for conventional (left) and prompt (right) fluxes of electron (top) and muon (bottom) atmospheric neutrinos. The comparison is shown for two zenith angles: $\cos\theta_z = $ 0.25 (upper and blue curves) and 0.85 (lower and orange curves). The differences for both conventional $\nu_e$ and $\nu_\mu$ fluxes are larger for more vertical directions, for which we obtain larger passing fractions. This is due to two approximations adopted in Ref.~\cite{Gaisser:2014bja}, as discussed in section~\ref{sec:pf}. On the one hand, the approximate treatment of the muon range distribution, taken as the median range, tends to overestimate the passing fraction in horizontal directions and underestimate it in more vertical ones, as explained in section~\ref{sec:pfnue}. On the other hand, the overestimation of the energy of the cosmic-ray shower producing the uncorrelated muons always produces a larger suppression of the passing fractions. These two factors, which partially cancel for more horizontal directions, explain the results for the conventional atmospheric neutrino fluxes. Moreover, accounting for the stochastic behavior of muon losses with full muon range distributions explains the disappearance of the shoulder, which is present in the dashed curves for the conventional $\nu_\mu$ flux and marks the transition from pion to kaon dominated atmospheric $\nu_\mu$. In the case of prompt $\nu_e$, however, the result of the comparison is less straightforward to interpret. In principle, from Figs.~\ref{fig:nue_passing-double-counting} and~\ref{fig:nue_passing-preach-effect}, we would expect our prompt $\nu_e$ passing fractions to be slightly larger, but this is not the case. This could be explained by two facts: the passing fractions obtained with DPMJET-2.55 are larger than with SIBYLL~2.3c (see Fig.~\ref{fig:nue-hadronic-model-effect}, although not exactly for the same comparison), and the passing fractions as obtained in Ref.~\cite{Gaisser:2014bja} are larger than the results from the \CORSIKA{} simulation (see Fig.~4 of Ref.~\cite{Gaisser:2014bja}). Overall, this results in smaller prompt $\nu_e$ passing fractions using our approach and inputs. Regardless, the results for the prompt $\nu_\mu$ passing fractions are very similar to those for the conventional flux. This is expected as Ref.~\cite{Gaisser:2014bja} applied the correlated passing fraction calculated for conventional neutrinos in Ref.~\cite{Schonert:2008is} for prompt neutrinos as well. This also explains the shoulder that appears in their prompt $\nu_\mu$ passing fractions (lower-right, dashed), which should not be present even with the Heaviside approximation for $\Prob_{\rm det}$. 

\section{Systematic uncertainties in the atmospheric neutrino passing fractions}
\label{sec:systematics}

In this section we study different sources of uncertainties in the calculations of the atmospheric neutrino passing fractions presented above. We first study the uncertainties in the propagation of muons on their way to the detector, which affect the determination of $\Prob_{\rm det}$. We next compute the passing fractions for different models of cosmic-ray spectra as well as for different hadronic-interaction models, to bracket the allowed variations from these inputs. Finally, we have also evaluated the impact on our results of different density models of the Earth's atmosphere, as well as of variations in the density profile for different locations and epochs of the year. These uncertainties on the passing fractions translate into uncertainties in the neutrino fluxes.

\subsection{Muon energy losses}
\label{sec:muonlosses}

The energy losses when muons traverse a medium are determined by ionization losses, $e^+ e^-$ pair production, bremsstrahlung, and photonuclear scattering. At energies $\lesssim$~TeV, losses are dominated by ionization. At higher energies, radiative processes become more important. In order of importance, pair production, bremsstrahlung, and photonuclear losses are dominant, with this hierarchy extending up to at least $\sim 10^7$~GeV. Above that energy, photonuclear interactions may become as important as bremsstrahlung~\cite{Chirkin:2004hz}. Muon energy losses are well understood up to energies of about 10~TeV, with uncertainties below the percent level. At higher energies, the extrapolation of the photonuclear cross section has the largest uncertainties.

The default photonuclear cross section used in \MMC{} is based on deep-inelastic scattering formalism where soft (non-perturbative) and hard (perturbative) physics are included~\cite{Dutta:2000hh}. This approach makes use of a data-driven parameterization of the proton structure function~\cite{Abramowicz:1991xz, Abramowicz:1997ms} and also includes nuclear shadowing.\footnote{Shadowing takes into account the attenuation of quark density in a nucleus due to the coherent scattering of the virtual photon off the nucleons in the small Bjorken-$x$ region.} The photonuclear cross section has also been described using a generalized vector dominance model for the soft component~\cite{Bezrukov:1981ci} and a framework of the color dipole model for the hard component~\cite{Bugaev:2002gy, Bugaev:2003sw}.

To evaluate the impact of uncertainties in the photonuclear cross section on the calculation of the atmospheric neutrino passing fractions, we have compared the outcomes obtained using these two different approaches. The default case in \MMC{} results in smaller photonuclear losses at the relevant energies for this work. Therefore, the passing fractions are expected to be smaller, as muons would have a higher probability to reach the detector with slightly higher energies. Although not completely negligible, the resulting uncertainties have a maximum absolute difference of $0.01$ for the conventional and prompt $\nu_e$ and $\nu_\mu$ fluxes.

\begin{figure}[h!]
\centering
    \subfloat{
        \includegraphics[width=0.5\linewidth]{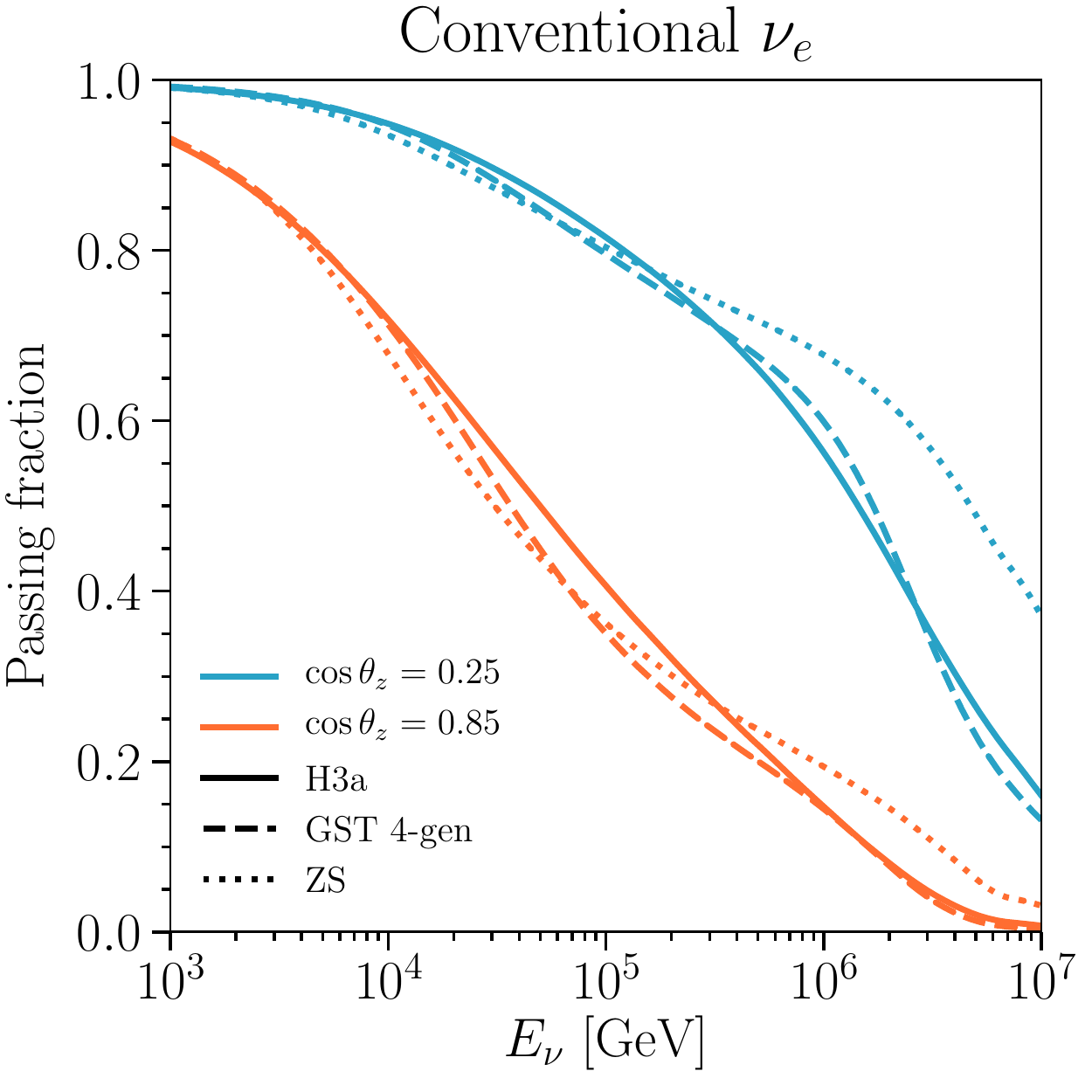}
    }
    \subfloat{
        \includegraphics[width=0.5\linewidth]{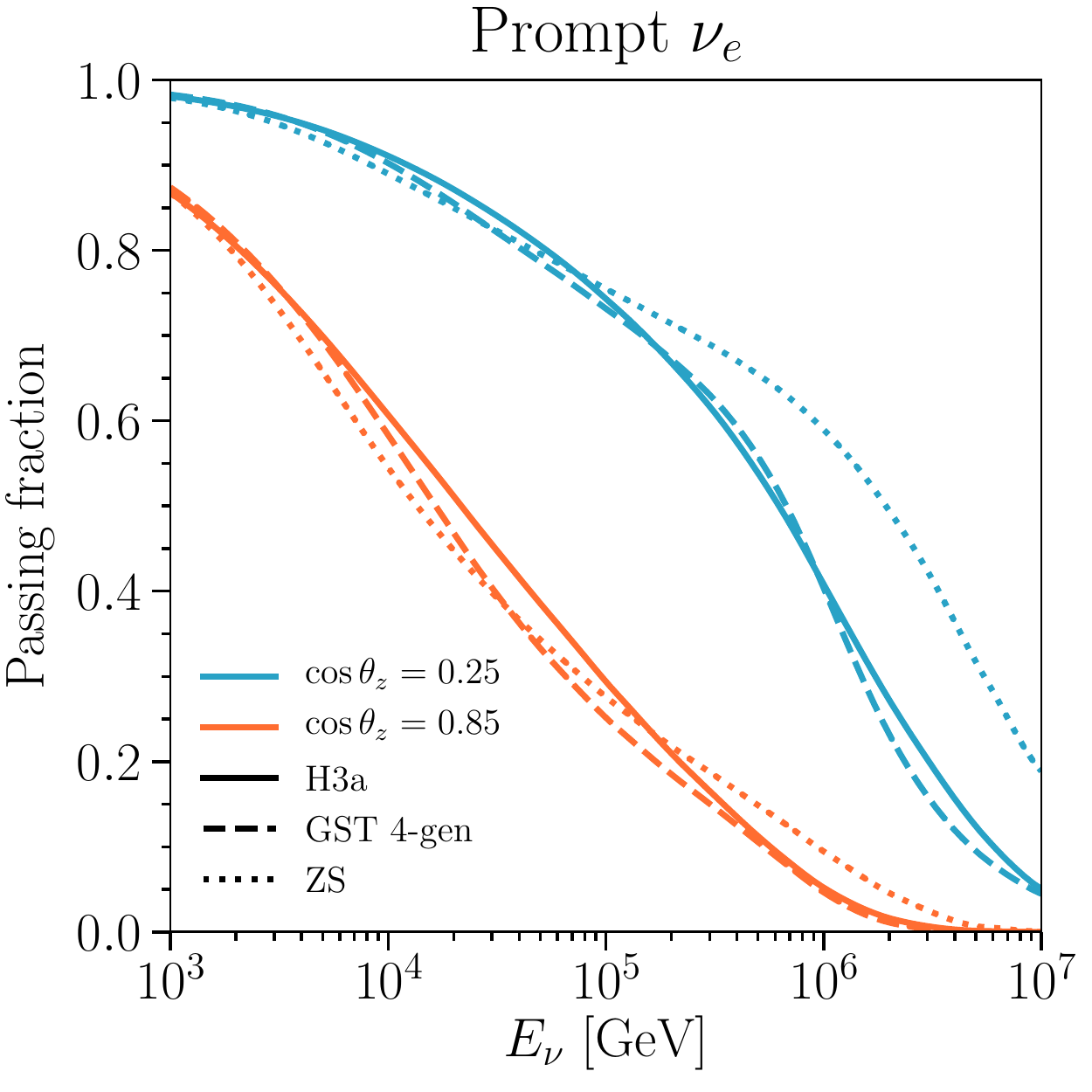}
    }\\[2ex]
\subfloat{
        \includegraphics[width=0.5\linewidth]{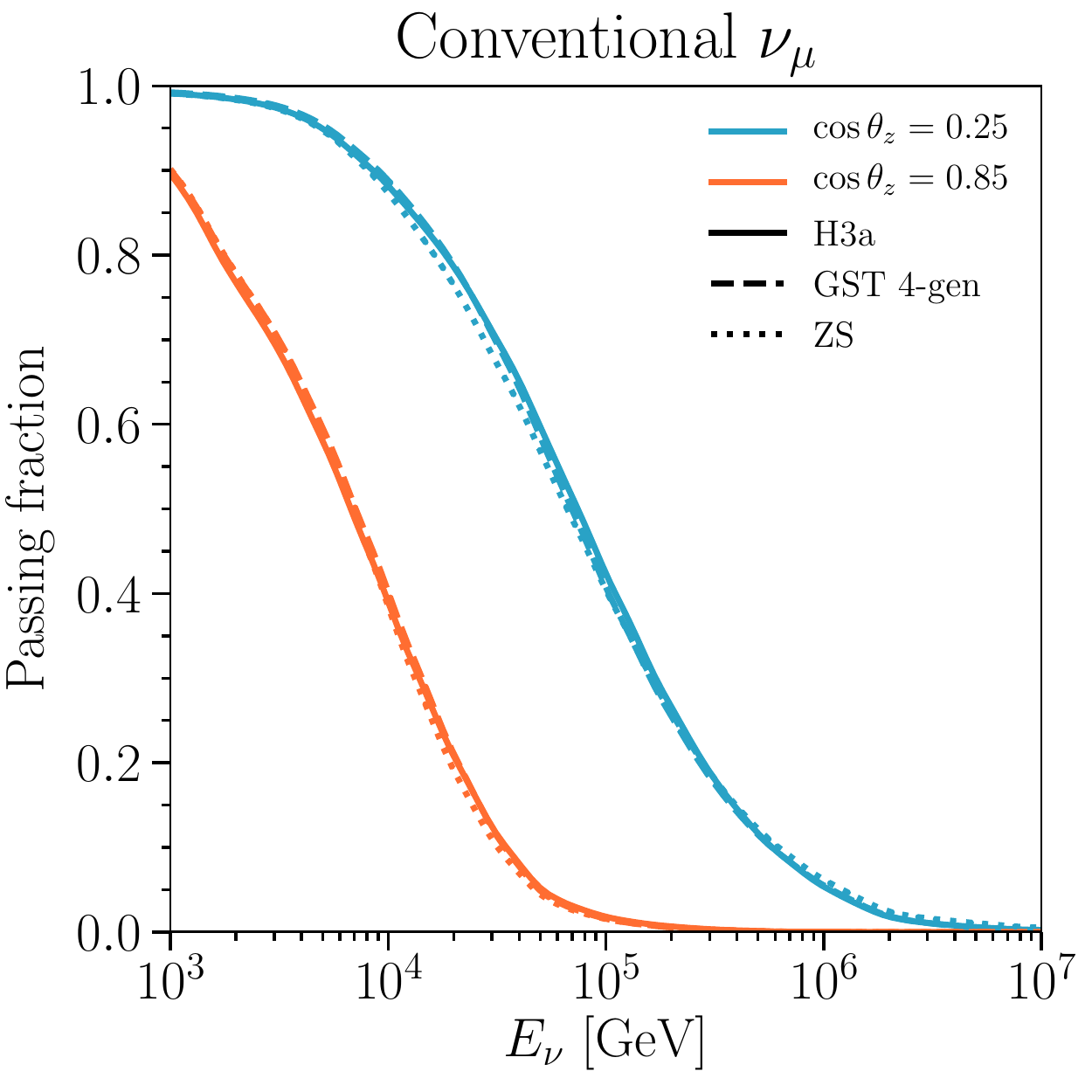}
    }
    \subfloat{
        \includegraphics[width=0.5\linewidth]{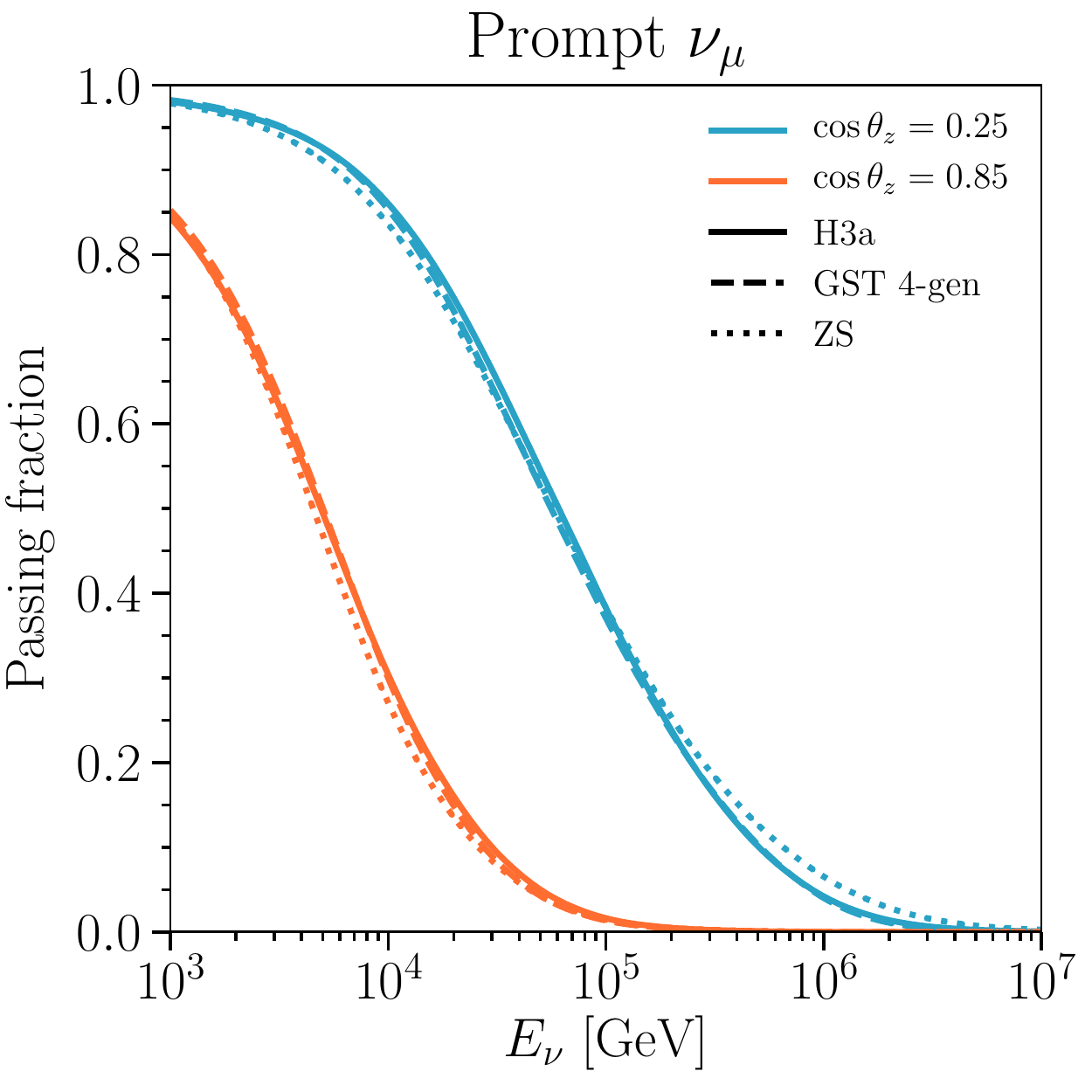}
    }    
\caption{\textbf{\textit{Passing fractions: effect of primary cosmic-ray spectrum}}. Results are shown for two values of $\cos\theta_z$ (from top to bottom): 0.25 (blue) and 0.85 (orange); for four CR models: Hillas-Gaisser H3a~\cite{Gaisser:2011cc} (solid), Gaisser, Stanev and Tilav (GST 4-gen)~\cite{Gaisser:2013bla} (dashed) and Zatsepin-Solkolskaya (ZS)~\cite{Zatsepin:2006ci} (dotted).
\textit{Top-left panel:} Conventional $\nu_e$ passing fraction. \textit{Top-right panel:} Prompt $\nu_e$ passing fraction. \textit{Bottom-left panel:} Conventional $\nu_\mu$ passing fraction. \textit{Bottom-right panel:} Prompt $\nu_\mu$ passing fraction.}
\label{fig:nue-cr-model-effect} \vspace{1.5cm}
\end{figure}

\subsection{Primary cosmic-ray spectra}
\label{sec:CR}

Throughout this work, we have considered the Hillas-Gaisser three-population model~\cite{Gaisser:2011cc}, H3a, as a the default spectrum to show our results. In this model, each population contains five groups of nuclei which cut off exponentially at a characteristic rigidity.\footnote{For a discussion on data of primary cosmic-ray spectra and on different models, we refer the reader to Refs.~\cite{Fedynitch:2012fs, Dembinski:2017zsh}.}

An alternative model was proposed by Gaisser, Stanev, and Tilav (GST)~\cite{Gaisser:2013bla} with a much lower rigidity cutoff for the first and second populations, which have a harder spectra, and a third population with only iron above the ankle. In this model (GST 4-gen), a fourth, pure-proton population was added to obtain a better agreement with the measurements of the depth of shower maximum above $10^9$~GeV, which disagrees with pure iron composition. A purely proton-iron mixture, however, is also in tension with Auger data~\cite{Abraham:2010yv, Aab:2014aea}.

Another three-population model to describe the (per nucleon) energy range $(10 - 10^{8})$~GeV was proposed by Zatsepin and Sokolskaya (ZS)~\cite{Zatsepin:2006ci}. The first (lowest energy) component comes from expanding shells of nova explosions, the second one from explosions of isolated stars into the interstellar medium and the third (highest energy) one from explosions of massive stars into their own stellar wind, which produces a superbubble of hot and low-density gas surrounded by a shell of neutral hydrogen. Note that this model also implies a heavy composition at the highest energies.

A new data-driven phenomenological model, Global Spline Fit (GSF), has recently been proposed to describe the energy range $(10^{10} - 10^{11})$~GeV~\cite{Dembinski:2017zsh}. It combines direct and air-shower measurements and parameterizes the cosmic-ray energy spectrum and its composition with smooth curves for all elements from proton to nickel. It is provided with a band that represents the uncertainties of the input data. As of this writing, its full implementation is not yet available in \MCEq, but once released can easily be incorporated in this framework.

In this section, we compare the atmospheric neutrino passing fractions obtained with three models of the primary cosmic-ray spectrum. The passing fractions for the conventional (left) and prompt (right) fluxes of electron (top) and muon (bottom) neutrinos are depicted in Fig.~\ref{fig:nue-cr-model-effect}. As in previous figures, the comparison is shown for two zenith angles: $\cos\theta_z = $ 0.25 (upper and blue curves) and 0.85 (lower and orange curves). The largest variations occur for the electron neutrino flux, both conventional and prompt, although the passing fractions never have an absolute difference larger than $0.05$, except for the most horizontal trajectories when using the ZS primary cosmic-ray spectrum. In this latter case, the passing fractions are significantly larger than with the other spectra above $\sim 10^6$~GeV. Note, however, that the ZS spectrum does not explain the cosmic-ray data above $10^{8}$~GeV and thus, the neutrino flux is not expected to be correctly predicted above a few $10^6$~GeV or even lower. In the case of the conventional and prompt muon neutrino fluxes, the differences among the passing fractions using the various primary cosmic-ray spectra are much smaller. This can be understood from the importance of the correlated muon, which depends less on variations in the shower development. 

Overall, systematic uncertainties from the choice of the primary cosmic-ray spectrum are non-negligible for electron neutrino passing fractions, but are small in the case of muon neutrinos.

\pagebreak 

\subsection{Hadronic-interaction models}
\label{sec:hadronic}

Given the very high energies of extensive air showers produced in the atmosphere by interactions of cosmic rays, the description of the cascade development crucially relies on the theoretical modeling of hadronic interactions at these high energies which are orders of magnitude higher than those reached at colliders. Several hadronic-interaction models exist, in which contributions of soft and hard parton dynamics to inelastic hadron interactions are described within Regge field theory~\cite{Gribov:1968fc}. All these models are tuned to agree with collider data, although the predictions for the characterization of extensive air showers depend on extrapolations. Therefore, the properties of these cascades depend on the different assumptions and approaches of the models. It is important then to estimate the variations in the resulting atmospheric neutrino passing fractions depending on the hadronic-interaction model.

The default model used throughout this work for the description of multiparticle production and for the development of extensive air showers is the recently released SIBYLL~2.3c~\cite{Riehn:2017mfm}. This model, based on the dual parton model~\cite{Capella:1977me, Capella:1981xr, Capella:1992yb}, represents an update from the earlier SIBYLL~2.3~\cite{Engel:2015dxa, Riehn:2015oba} in order to obtain a better agreement with NA49 data~\cite{Anticic:2010yg}. SIBYLL~2.3c provides a better description of kaon production spectra than previous versions~\cite{Ahn:2009wx, Engel:2015dxa, Riehn:2015oba}, but its predictions for the production of charmed particles and the characterization of atmospheric cascades are similar.

An alternative phenomenological model, also based on Regge field theory, is the updated quark-gluon string model with jets (QGSJET-II-04)~\cite{Ostapchenko:2010vb}, which has also been calibrated with LHC data and builds on the original QGSJET model~\cite{Kalmykov:1997te}.

A third phenomenological approach to describe cosmic-ray air shower development and heavy-ion interactions is the Monte Carlo event generator for minimum bias hadronic interactions EPOS-LHC~\cite{Werner:2005jf, Pierog:2013ria},\footnote{EPOS stands for Energy conserving mechanical multiple scattering approach, based on Partons (parton ladders), Off-shell remnants and Splitting of parton ladders.} which is also calibrated with LHC data. The model is also based on the parton-based Regge field theory developed for NEXUS~\cite{Drescher:2000ha}, which described soft interactions with the VENUS model~\cite{Werner:1993uh} and semi-hard scattering with the QGSJET model~\cite{Kalmykov:1997te}.

Another Monte Carlo model, also based on the two-component, dual-parton model~\cite{Capella:1977me, Capella:1981xr, Capella:1992yb}, which describes hadron-hadron, hadron-nucleus, nucleus-nucleus, and neutrino-nucleus interactions, is DPMJET-III~\cite{Roesler:2000he}. This model unifies all features of the previous event generators DPMJET-II~\cite{Ranft:1994fd, Ranft:1999fy, Ranft:1999qe}, DTNUC-2~\cite{Engel:1996yb, Roesler:1998wy}, and PHOJET 1.12~\cite{Engel:1994vs, Engel:1995yda}. Within this model particle production takes place via the fragmentation of colorless parton-parton chains, which are constructed from the quark content of the interacting hadrons and nuclei. In addition to describe cosmic-ray cascade simulations, the model was also tuned to work for central collisions and to explain several features of the CERN-SPS heavy-ion data. We consider this model to calculate passing fractions for prompt atmospheric neutrinos. For conventional atmospherics, it gives similar passing fractions as SIBYLL~2.3c. Although we do not discuss it in this section, the \CORSIKA{} simulation in Ref.~\cite{Gaisser:2014bja} used DPMJET-2.55~\cite{Berghaus:2007hp} for prompt atmospheric neutrinos (in Fig.~\ref{fig:nue-passing-comparison-old}). It features, in comparison to DPMJET-II, an improved description of collision scaling in charm production and is compared with HERA-B data~\cite{Abt:2007zg}. 

QGSJET-II-04 is the model that predicts the smallest value for the position of the maximum in extended air showers,\footnote{Although it is larger than previous QGSJET versions.} $X_{\rm max}$, while SIBYLL~2.3c predicts the largest $X_{\rm max}$.\footnote{Although it is smaller than SIBYLL~2.3.} This has to do with the energy dependence of the inelasticity in each model~\cite{Ostapchenko:2016wtv}. On another hand, SIBYLL~2.3c predicts the largest number of muons of all post-LHC models at energies above $10^9$~GeV, closer to the observed excess than other models~\cite{Aab:2014pza, Apel:2017thr, Abbasi:2018fkz}. Note, however, that at energies below $\sim 10^8$~GeV, no excess of muons has been observed~\cite{Gonzalez:2016ntp, Fomin:2016kul}, although the lateral distributions of muons are not well reproduced by most hadronic-interaction models~\cite{Abbasi:2012kza}. We refer the reader to Refs.~\cite{Fedynitch:2012fs, Aab:2016hkv, Ostapchenko:2016wtv, Riehn:2017mfm, Dedenko:2017lag, Dedenko:2017hbu, Pierog:2018nkf} for discussions on the impact of different hadronic-interaction models on cosmic-ray data.

\begin{figure}[h!]
\centering
    \subfloat{
        \includegraphics[width=0.5\linewidth]{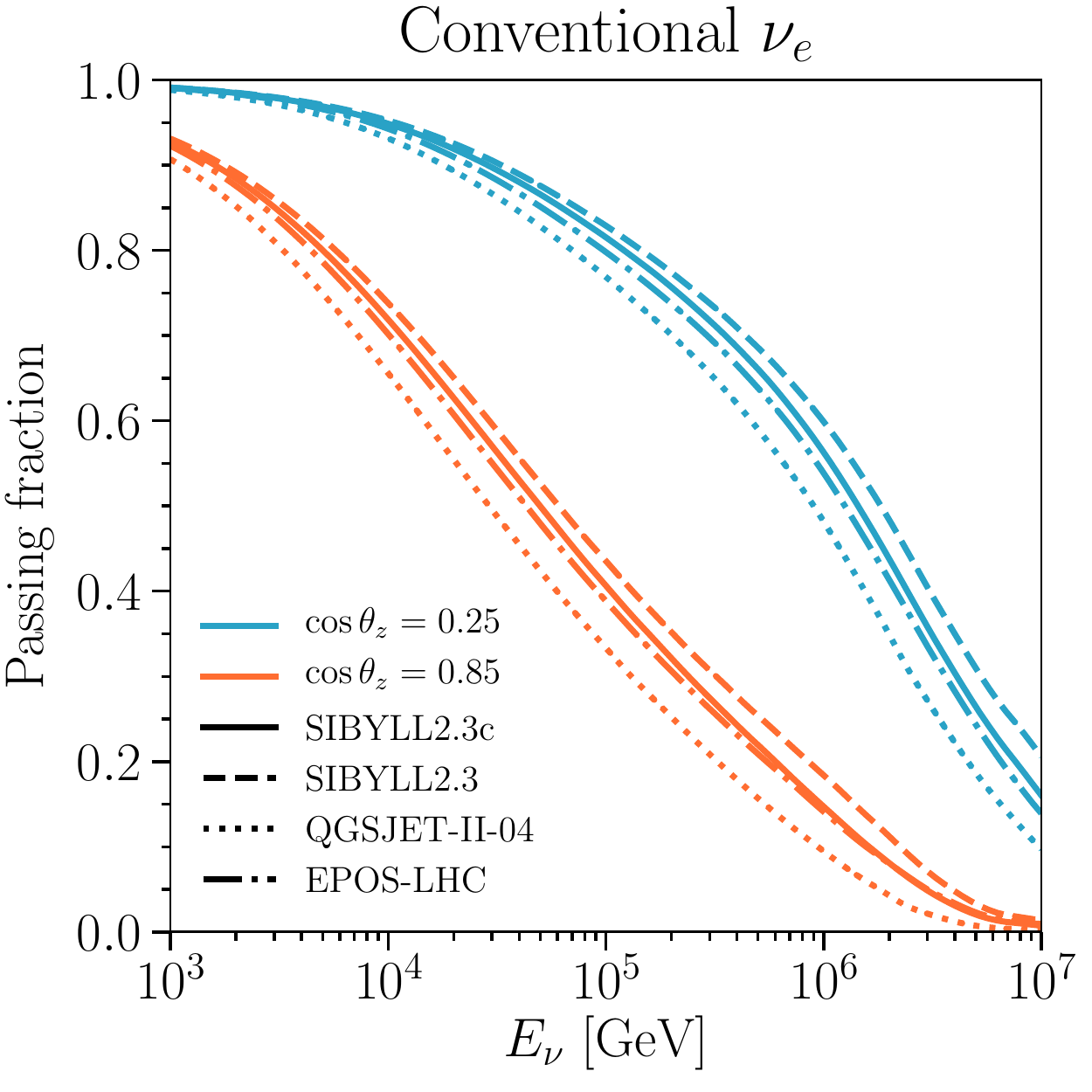}
    }
    \subfloat{
        \includegraphics[width=0.5\linewidth]{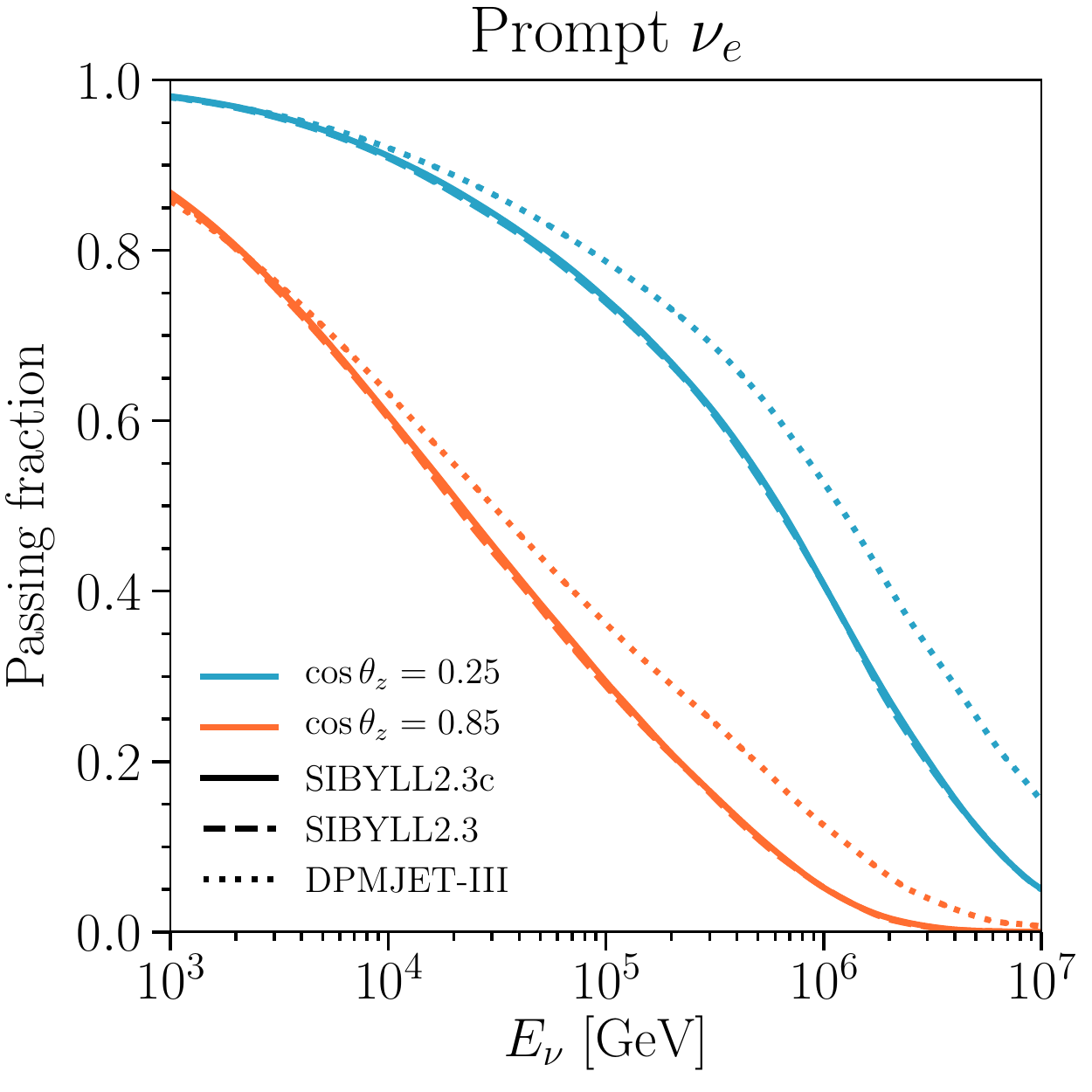}
    }\\[2ex]
    \subfloat{
        \includegraphics[width=0.5\linewidth]{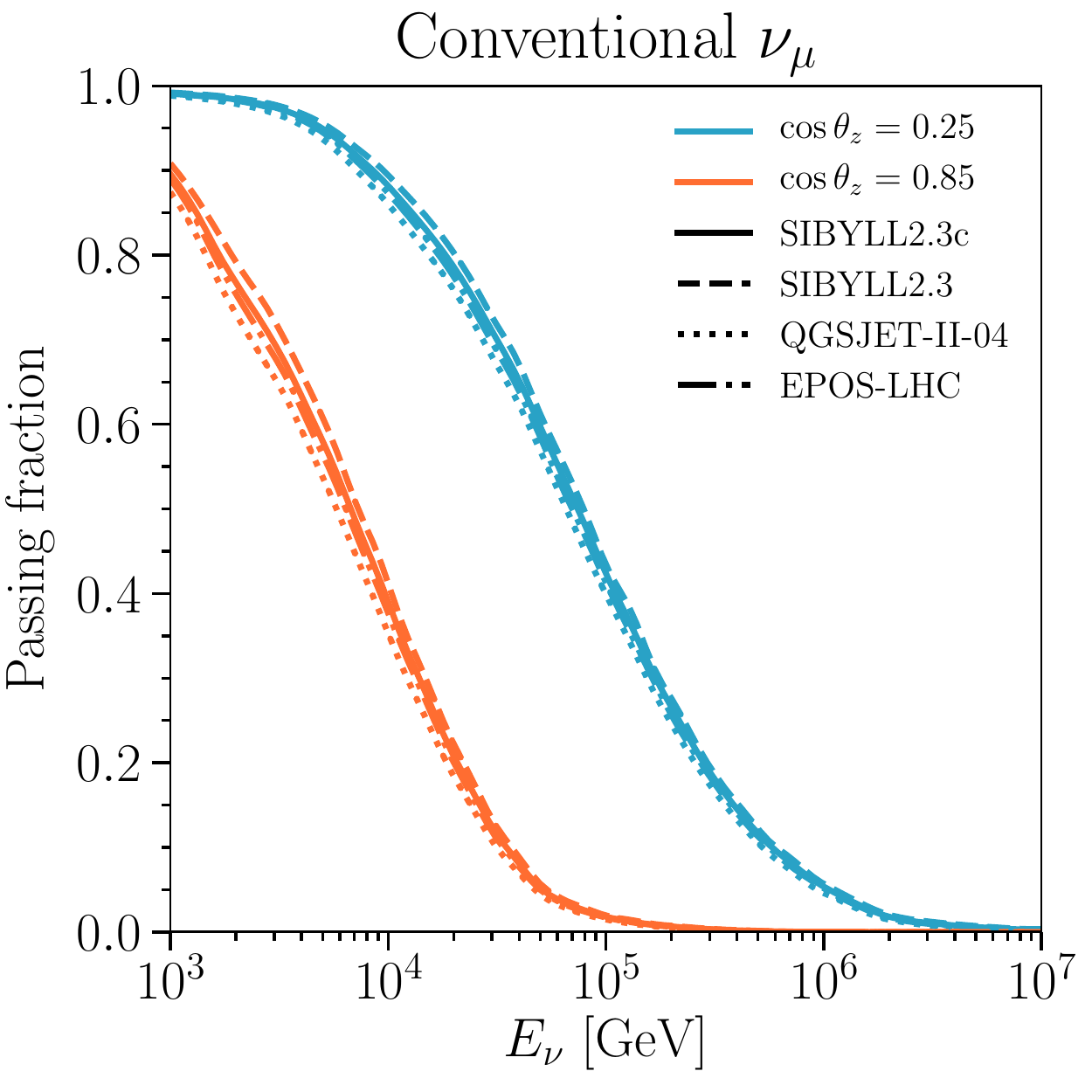}
    }
    \subfloat{
        \includegraphics[width=0.5\linewidth]{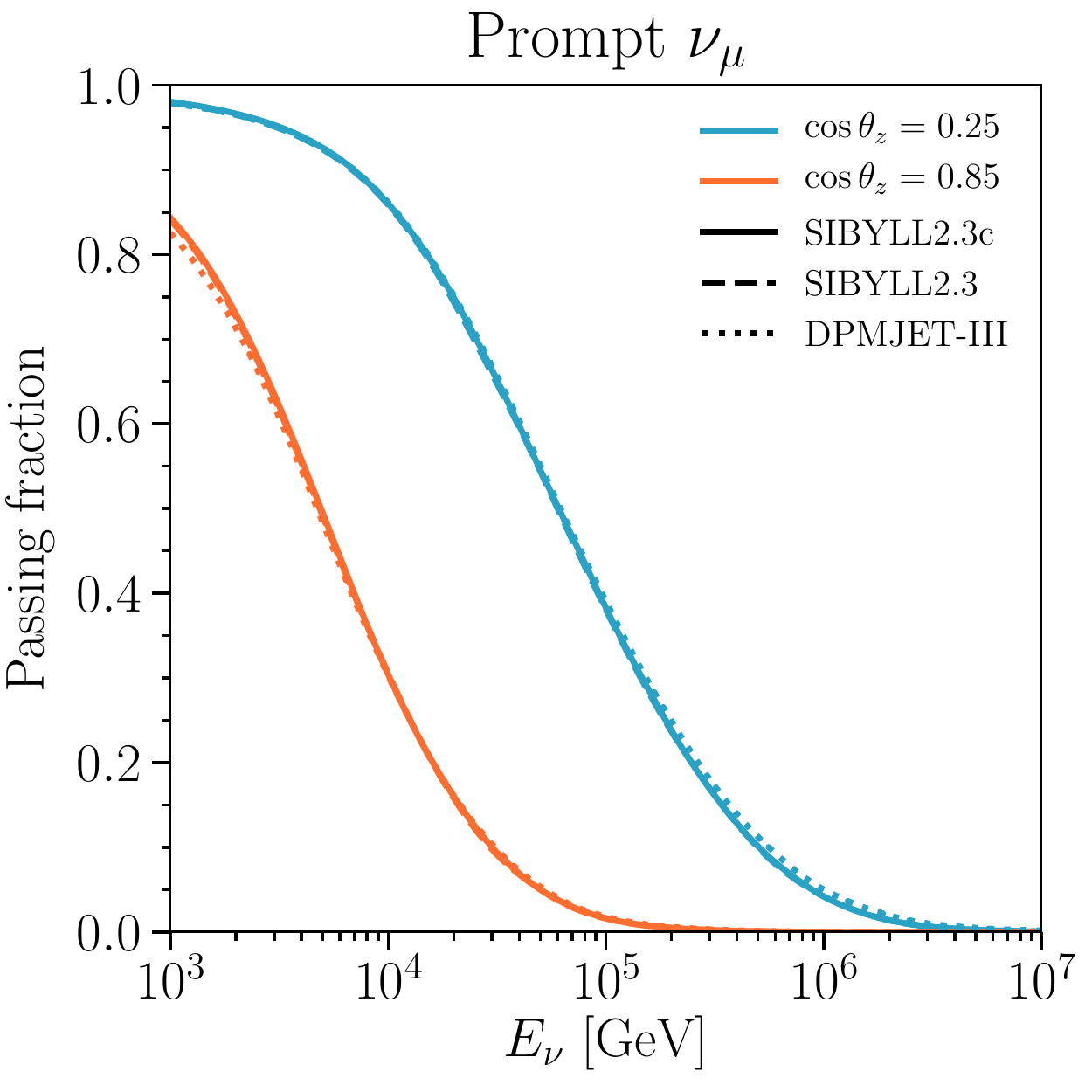}
    }    
\caption{\textbf{\textit{Passing fractions: effect of hadronic-interaction model}}. Results are shown for two values of $\cos\theta_z$ (from top to bottom): 0.25 (blue) and 0.85 (orange); for different hadronic-interaction models: SIBYLL~2.3c~\cite{Riehn:2017mfm} (solid), SIBYLL~2.3~\cite{Engel:2015dxa, Riehn:2015oba} (dashed), QGSJET-II-04~\cite{Ostapchenko:2010vb} (dotted in left panel), EPOS-LHC~\cite{Pierog:2013ria} (dash-dotted in left panel) and DPMJET-III\cite{Roesler:2000he} (dotted in right panel).
\textit{Top-left panel:} Conventional $\nu_e$ passing fraction. \textit{Top-right panel:} Prompt $\nu_e$ passing fraction. \textit{Bottom-left panel:} Conventional $\nu_\mu$ passing fraction. \textit{Bottom-right panel:} Prompt $\nu_\mu$ passing fraction.} \vspace{1cm}
\label{fig:nue-hadronic-model-effect}
\end{figure}

In Fig.~\ref{fig:nue-hadronic-model-effect} we compare the results for the atmospheric neutrino passing fractions using different hadronic-interaction models. We show the passing fractions for the conventional (left) and prompt (right) fluxes of electron (top) and muon (bottom) neutrinos. As in previous plots, the comparison is shown for two zenith angles: $\cos\theta_z = $ 0.25 (upper and blue curves) and 0.85 (lower and orange curves). For conventional neutrinos, the differences are larger for electron neutrinos. This is expected due to the more well-understood, correlated muon. All models predict similar conventional passing fractions, except QGSJET-II-04 which predicts the smallest and has an absolute difference of $\sim 0.1$. This can be understood from its harder muon spectrum~\cite{Riehn:2017mfm}. In the case of prompt $\nu_e$, the passing fractions obtained when using DPMJET-III are larger than those with SIBYLL~2.3 or SIBYLL~2.3c. For prompt $\nu_\mu$, the differences from model to model are negligible.

Overall, we conclude that the choice of the hadronic-interaction model introduces negligible systematic uncertainties in the passing fractions for atmospheric muon neutrinos, but they can be important for electron neutrinos.

\begin{figure}[h!]
\centering
    \subfloat{
        \includegraphics[width=0.5\linewidth]{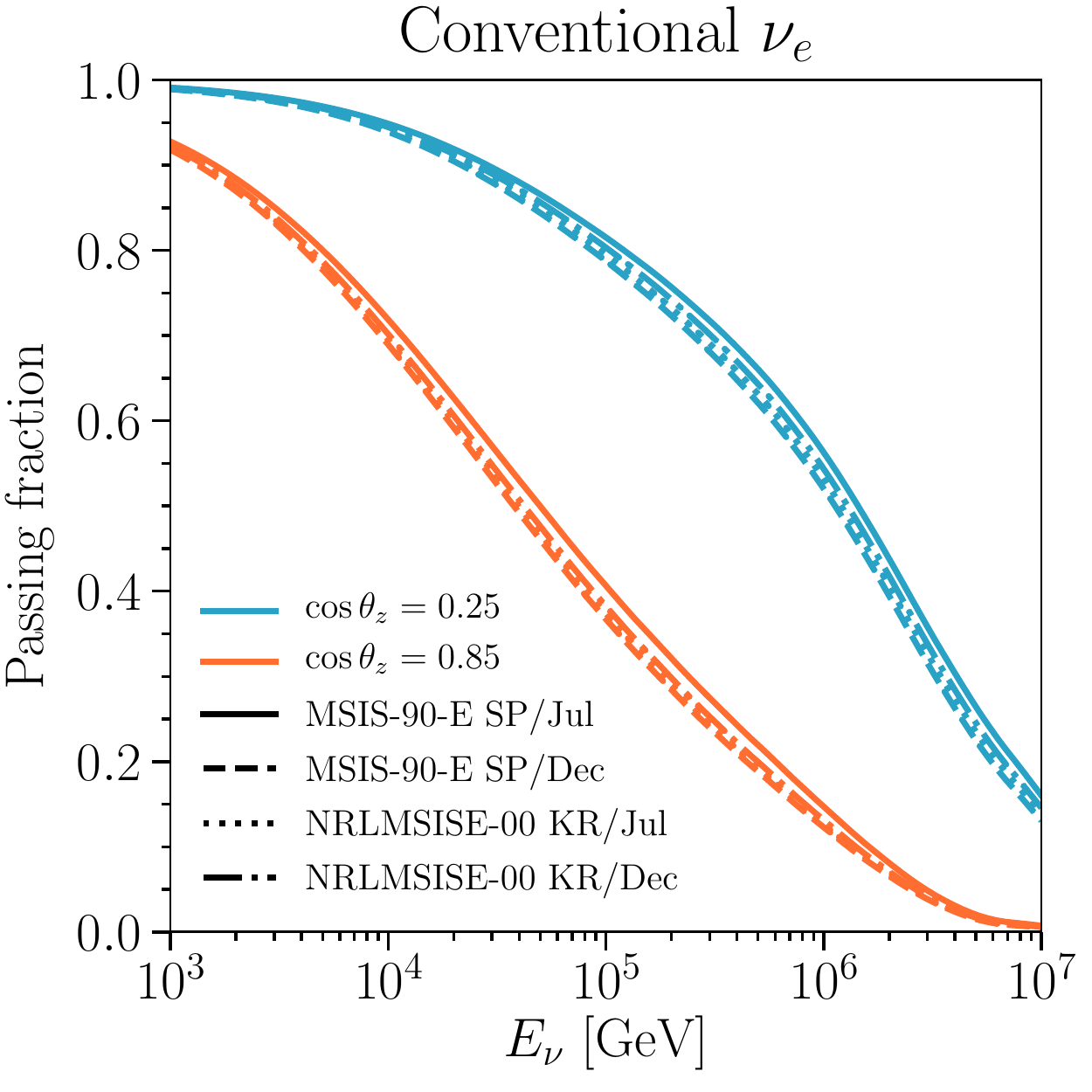}
    }
    \subfloat{
        \includegraphics[width=0.5\linewidth]{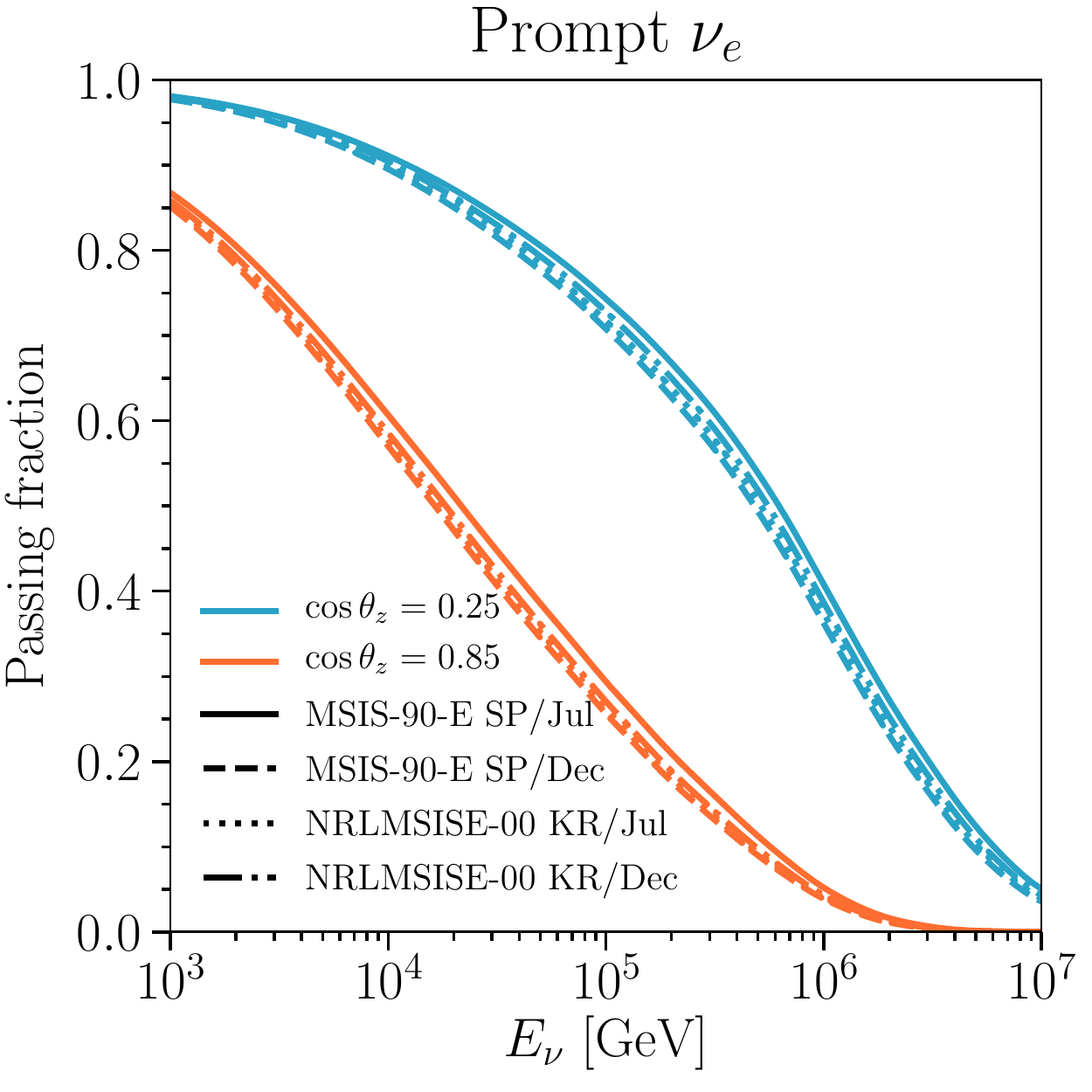}
    }\\[2ex]
    \subfloat{
        \includegraphics[width=0.5\linewidth]{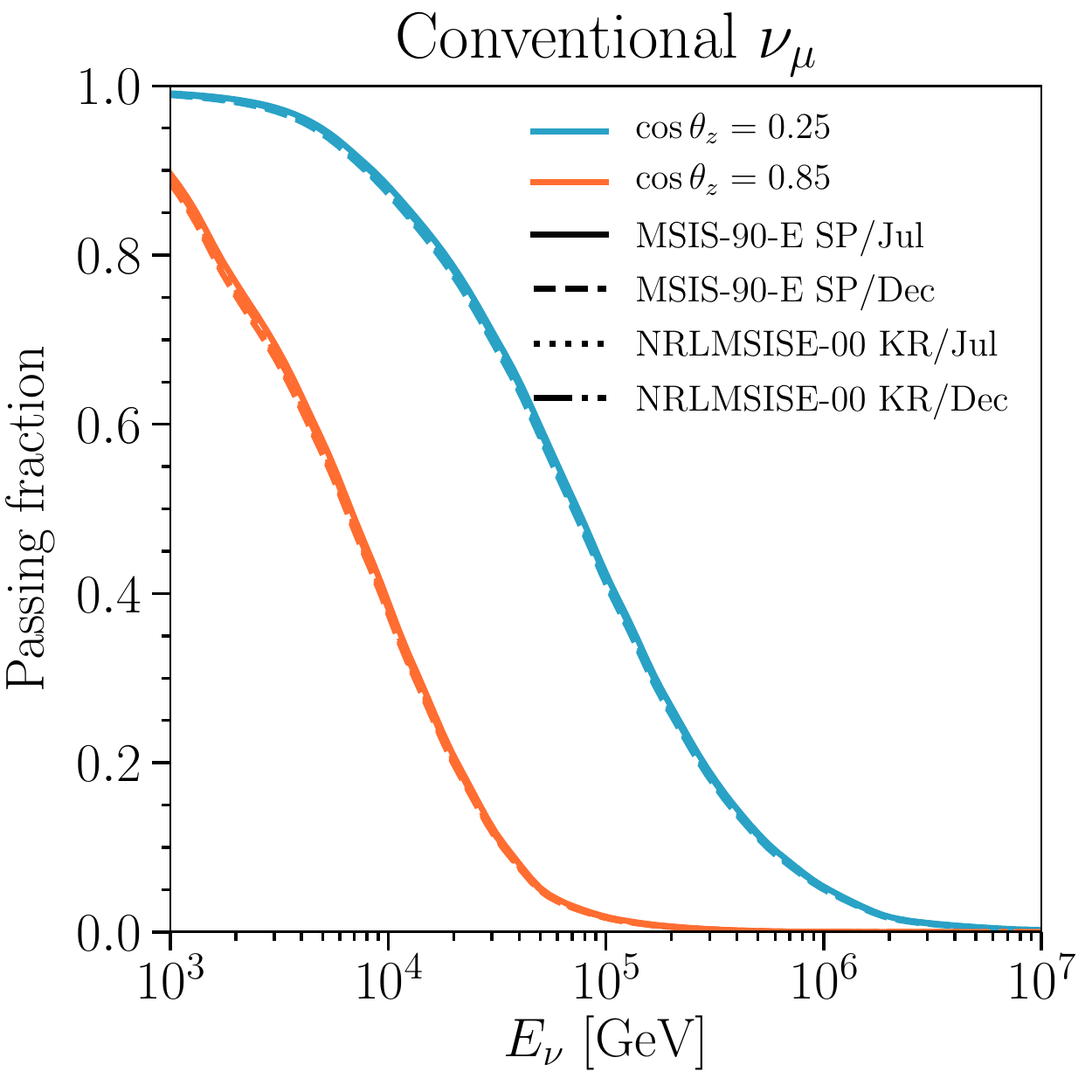}
    }
    \subfloat{
        \includegraphics[width=0.5\linewidth]{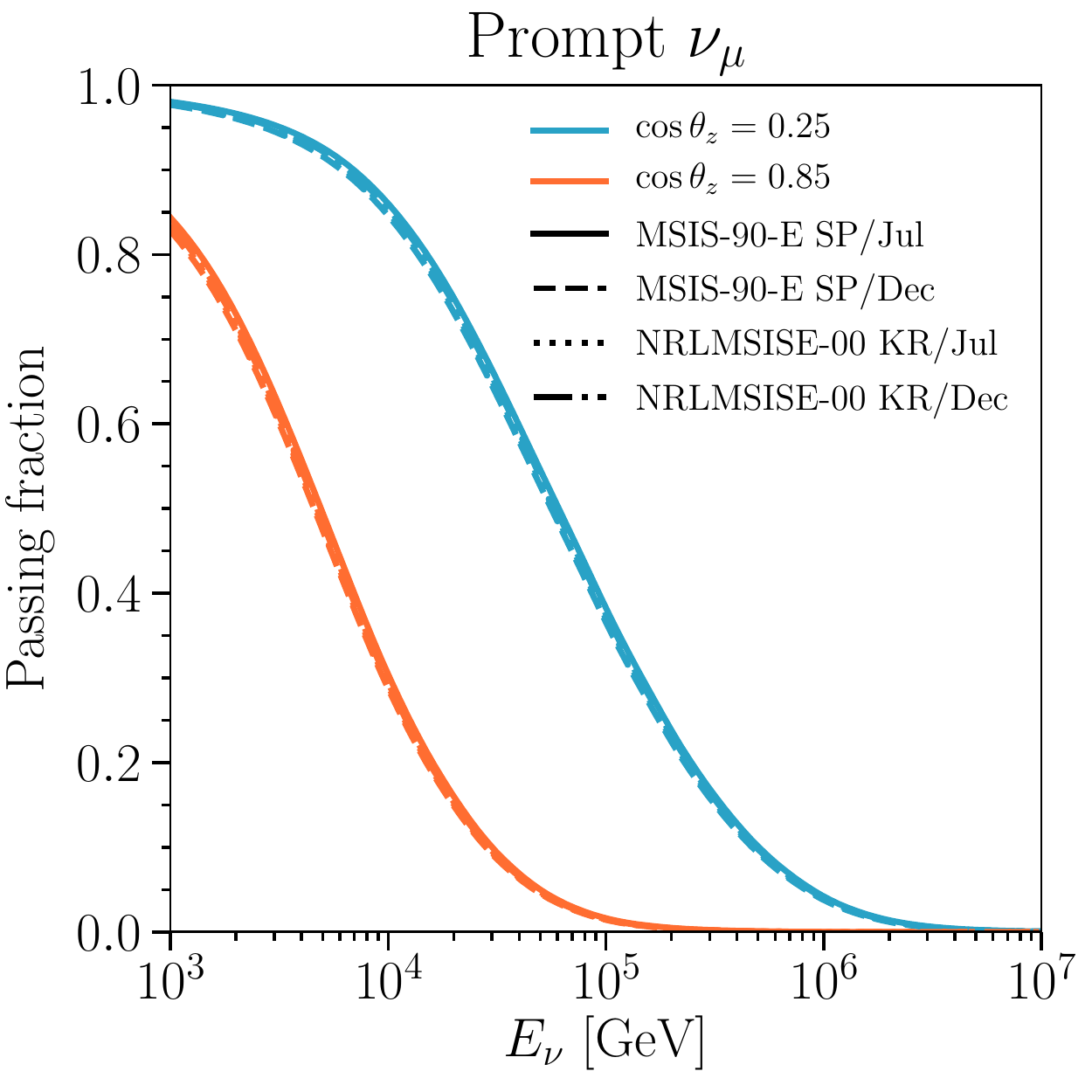}
    }    
\caption{\textbf{\textit{Passing fractions: effect of the atmosphere-density model}}. Results are shown for two values of $\cos\theta_z$ (from top to bottom): 0.25 (blue) and 0.85 (orange); for two locations: South Pole (solid and dashed) and Karlsruhe (dotted and dash-dotted); and for two different months, July (solid and dotted) and December (dashed and dash-dotted). For the South Pole calculations we use the MSIS-90-E model~\cite{Labitzke:1985, Hedin:1991}, whereas for the Karlsruhe case we use NRLMSIS-00~\cite{Picone:2002}. \textit{Top-left panel:} Conventional $\nu_e$ passing fraction. \textit{Top-right panel:} Prompt $\nu_e$ passing fraction. \textit{Bottom-left panel:} Conventional $\nu_\mu$ passing fraction. \textit{Bottom-right panel:} Prompt $\nu_\mu$ passing fraction.} \vspace{1cm}
\label{fig:atmosphere-effect}
\end{figure}

\subsection{Atmosphere-density models}
\label{sec:atmosphere}

The two atmosphere-density models we consider are included in \MCEq, one of which is also included in \CORSIKA~\cite{Heck:2018}. They are five-layer profiles with a volume composition of 78.1\% $N_2$, 21\% $O_2$ and 0.9\% $Ar$. The mass overburden of the lower four layers follows an exponential profile dependent on the altitude, while the upper layer decreases linearly with altitude up to a maximum of 112.8~km~\cite{Heck:1998vt}. 

All the results presented in this work have been obtained using the MSIS-90-E model~\cite{Labitzke:1985, Hedin:1991} for the South Pole atmosphere on July 01, 1997~\cite{Heck:2018}.\footnote{Note that in \MCEq{} the season tag is ``June".} This model was also used to perform the \CORSIKA{} simulation shown in some figures in section~\ref{sec:pf}. 

In this section we evaluate the effects of seasonal and location variations in the atmosphere on the passing fractions. To check the seasonal dependence, we consider the MSIS-90-E model for the South Pole atmosphere on December 31, 1997~\cite{Heck:2018}. To understand the dependence of our results on the atmosphere-density model, we also consider the NRLMSIS-00 empirical model~\cite{Picone:2002} for the South Pole and for Karlsruhe, during summer and winter. This is a more recent model based on MSIS-90-E. In \MCEq{} different locations can be chosen when using NRLMSIS-00, but not MSIS-90-E.

Differences in the passing fractions from model to model are below the percent level, so we do not explicitly show them. We next study seasonal and location differences. We consider the MSIS-90-E model for the South Pole and the NRLMSIS-00 model for Karlsruhe, at two different times of the year. In Fig.~\ref{fig:atmosphere-effect} we illustrate how the seasonal and location variations in the atmosphere's density affect the passing fractions. As done for other comparisons, we show the passing fractions for the conventional (left) and prompt (right) atmospheric fluxes of electron (top) and muon (bottom) neutrinos; and for two zenith angles: $\cos\theta_z = $ 0.25 (upper and blue curves) and 0.85 (lower and orange curves). Seasonal effects at the South Pole are not negligible for electron neutrino passing fractions, with absolute differences up to $\sim 0.05$. Seasonal variations are much smaller at Karlsruhe, which results in the electron neutrino passing fractions lying within the seasonal variations for the South Pole. Analogous to previous section discussions about the other sources of systematic uncertainties, location and seasonal differences have very little impact on the muon neutrino passing fractions.

\section{Detector configuration}
\label{sec:detector}

In this work we have considered a particular, simplified, detector configuration. In the previous section we highlighted and studied the main sources of systematic uncertainties entering into the calculations. In this section we describe the different characteristics of the detector which also affects the final efficiency.

\subsection{Depth}
\label{sec:depth}

\begin{figure}[h!]
\centering
    \subfloat{
        \includegraphics[width=0.5\linewidth]{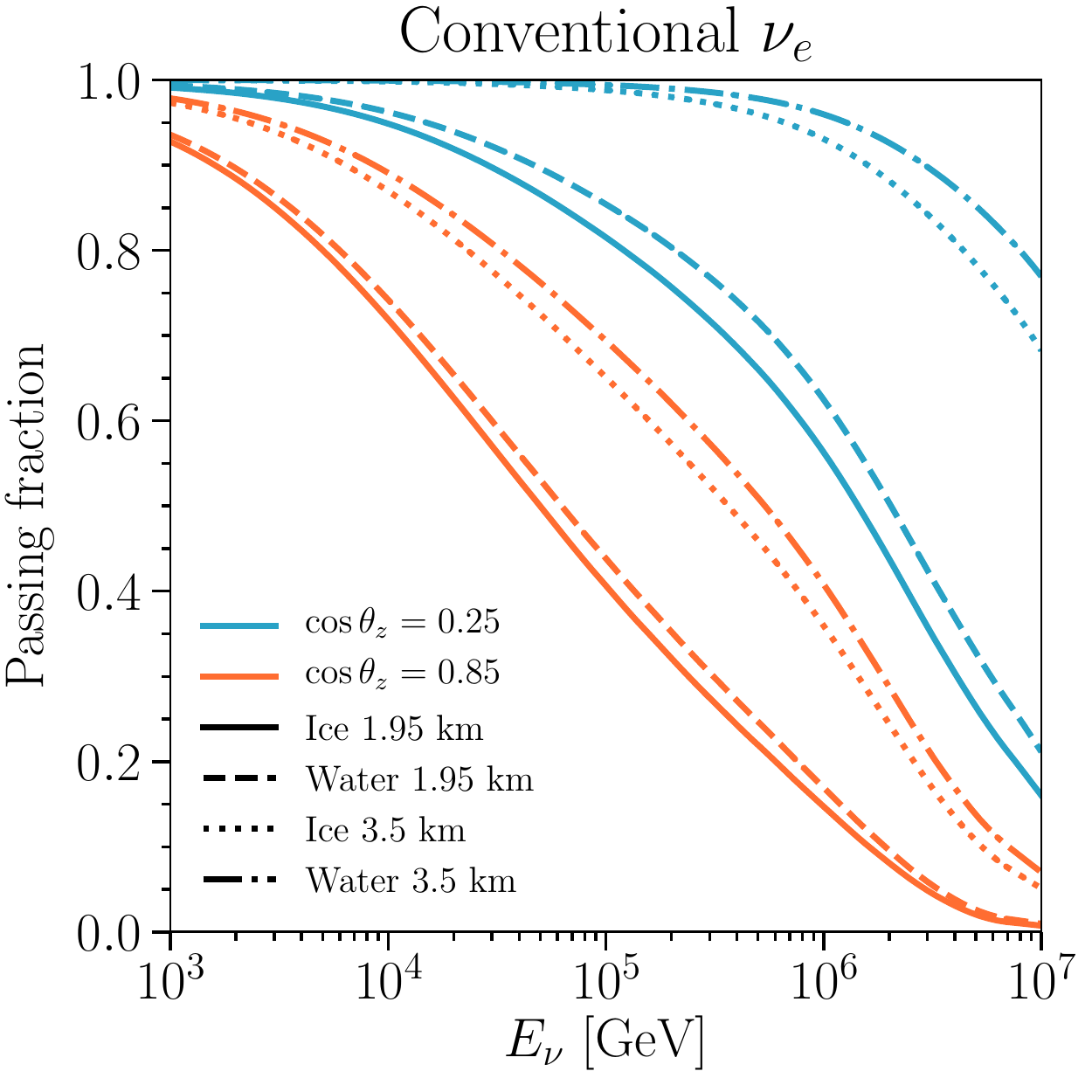}
    }
    \subfloat{
        \includegraphics[width=0.5\linewidth]{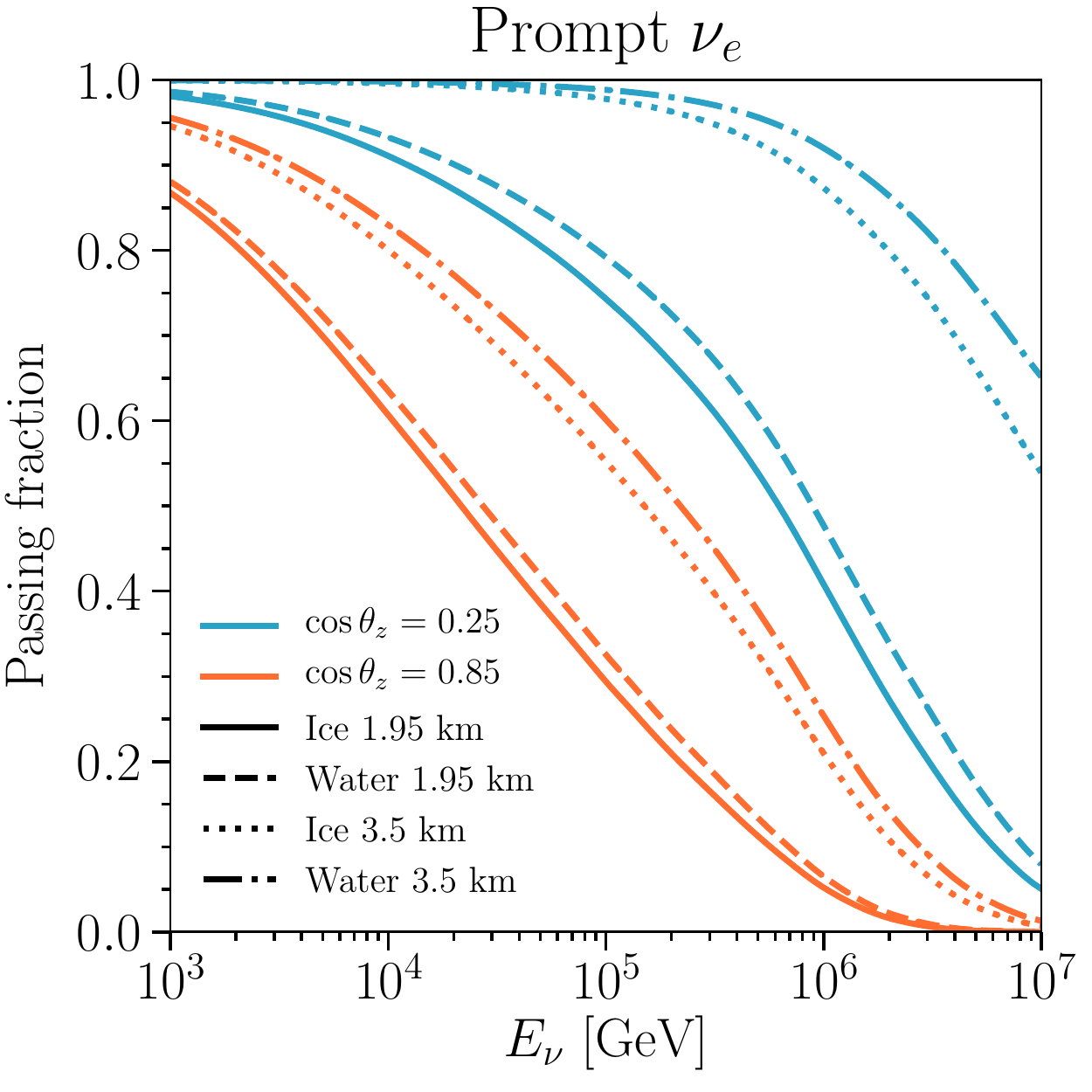}
    }\\[2ex]
    \subfloat{
        \includegraphics[width=0.5\linewidth]{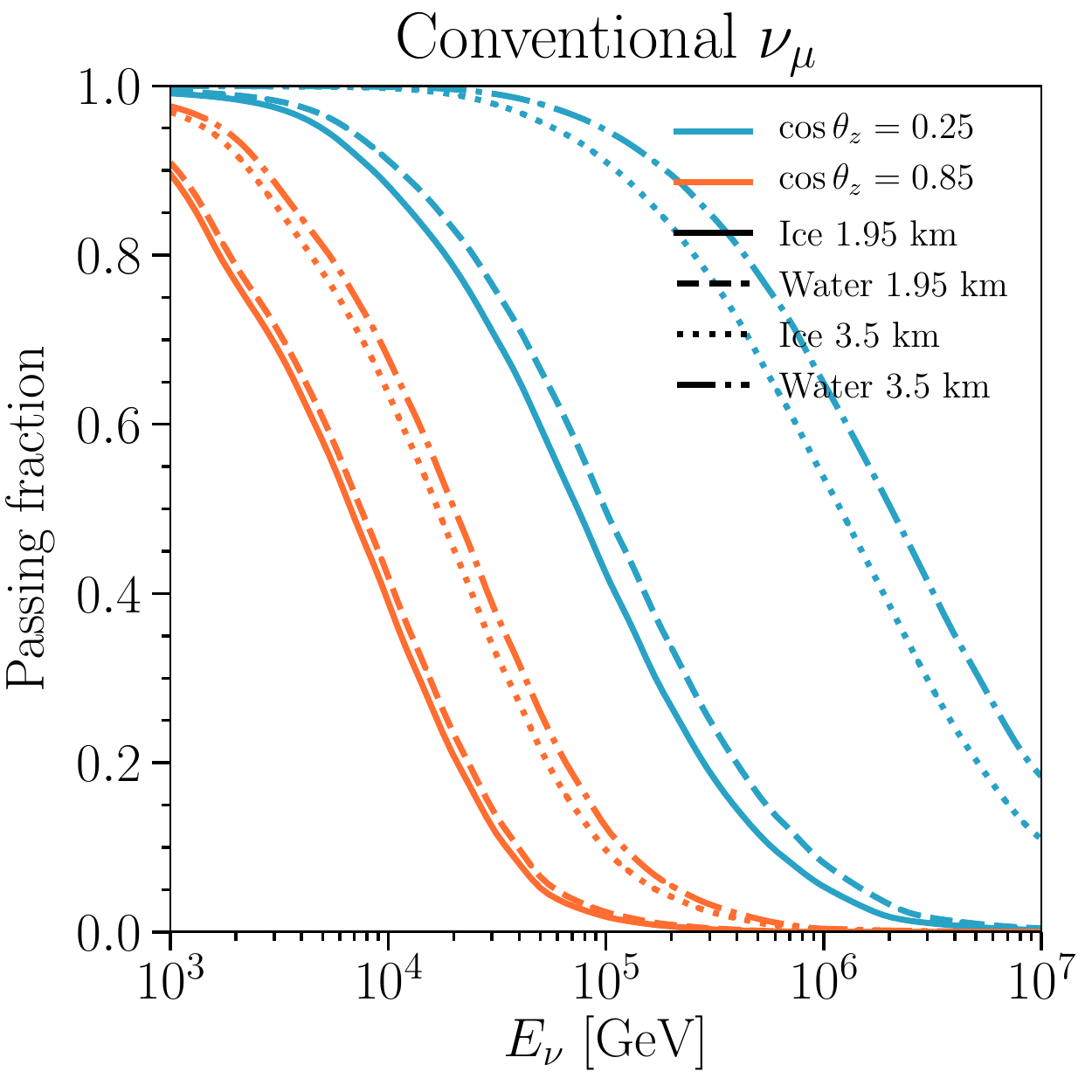}
    }
    \subfloat{
        \includegraphics[width=0.5\linewidth]{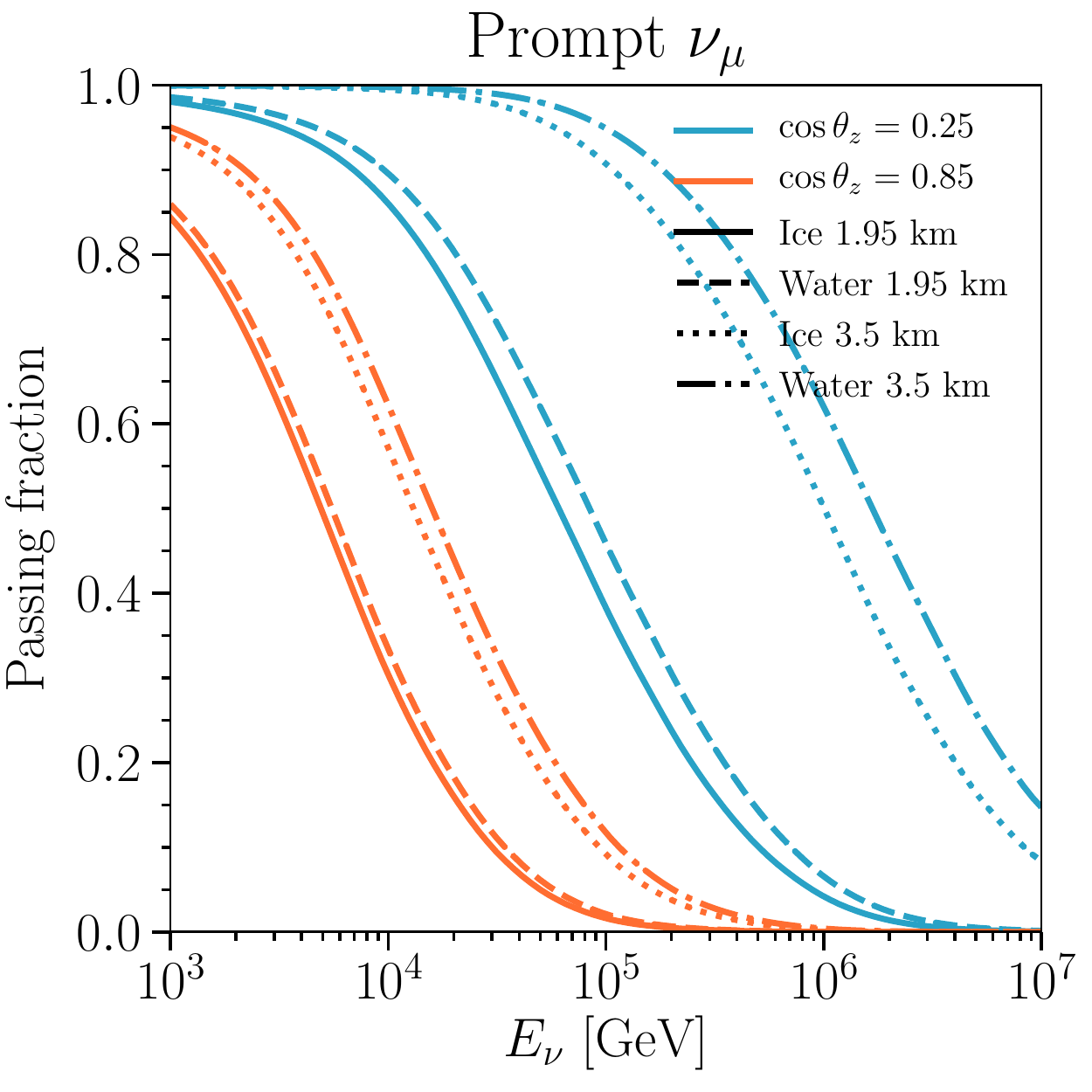}
    }    
\caption{\textbf{\textit{Passing fractions: effect of depth/medium}}. Results are shown for two values of $\cos\theta_z$ (from top to bottom): 0.25 (blue) and 0.85 (orange); for two depths: $d_{\rm det} = 1.95$~km (solid and dashed) and $d_{\rm det} = 3.5$~km (dotted and dash-dotted); and for two different media: ice (solid and dotted) and water (dashed and dot-dashed). \textit{Top-left panel:} Conventional $\nu_e$ passing fraction. \textit{Top-right panel:} Prompt $\nu_e$ passing fraction. \textit{Bottom-left panel:} Conventional $\nu_\mu$ passing fraction. \textit{Bottom-right panel:} Prompt $\nu_\mu$ passing fraction.} \vspace{1.5cm}
\label{fig:medium-effect}
\end{figure}

Throughout this work, we have assumed the detector's veto to be located at a depth of 1.95~km in ice, corresponding to the center of the IceCube detector~\cite{Achterberg:2006md}. Nevertheless, the exact modeling of the veto geometry has an important impact on the calculation of the atmospheric neutrino passing fractions. To illustrate the effect of the depth, in Fig~\ref{fig:medium-effect} we depict the passing fractions for two different values: $d_{\rm det} = 1.95$~km (solid and dashed) and $d_{\rm det} = 3.5$~km (dotted and dash-dotted). As in previous plots, we show the passing fractions for the conventional (left) and prompt (right) fluxes of electron (top) and muon (bottom) neutrinos; and for two zenith angles: $\cos\theta_z = $ 0.25 (upper and blue curves) and 0.85 (lower and orange curves). As expected, the deeper the detector is located, the more energy the muon loses, increasing the probability to not trigger the veto. Therefore, the vetoing power of this technique is reduced significantly and the atmospheric neutrino passing fractions are larger for deeper detectors. This highlights the importance of taking into account the geometry of the veto properly when applying this technique and becomes relevant for other neutrino telescopes.

\pagebreak

\subsection{Surrounding medium}
\label{sec:medium}

In section~\ref{sec:preach_plight}, we have described the probability for muons with a given energy at the Earth's surface to reach and trigger the detector's veto, $\Prob_{\rm det}$. This is dictated in-part by how much energy the muons lose when traversing the surrounding medium of the detector. The nature of muon energy losses depend on the medium of propagation. Although we have studied energy losses in ice in detail, which applies to the IceCube detector~\cite{Achterberg:2006md}, another type of neutrino detectors which uses water is represented by ANTARES~\cite{Collaboration:2011nsa} and KM3NeT~\cite{Adrian-Martinez:2016fdl} in the Mediterranean sea, and by Baikal-GVD~\cite{Baikal2017} in Lake Baikal.

The KM3NeT-Italy infrastructure, at the former NEMO site, is located at a depth of 3.5~km, whereas Baikal-GVD centers at a depth of 1~km. Therefore, for an appropriate evaluation of the impact of the medium on our results, in Fig~\ref{fig:medium-effect} we also compare the passing fractions in ice (solid and dotted) and water (dashed and dash-dotted) for the detector's veto at depth $d_{\rm det} = 1.95$~km (solid and dashed) and $d_{\rm det} = 3.5$~km (dotted and dash-dotted), with the rest of the parameters and inputs fixed. We note that the differences caused by the propagation of muons in solid or liquid water are not dramatic. These differences are more pronounced for more horizontal trajectories. Conventional and prompt fluxes exhibit similar differences. Finally, even though the changes due to the propagation medium are not large, the difference in depths between IceCube and KM3NeT results in a very important reduction of rejection power for the latter. Similarly, and although not explicitly shown, the shallower location of the Baikal-GVD implies that the atmospheric neutrino passing fractions for this configuration are the smallest ones of the three cases and thus, applying this technique might be especially relevant for Baikal-GVD. 

\subsection{Muon veto trigger probability}
\label{sec:response}

\begin{figure}[h!]
\centering
    \subfloat{
        \includegraphics[width=0.5\linewidth]{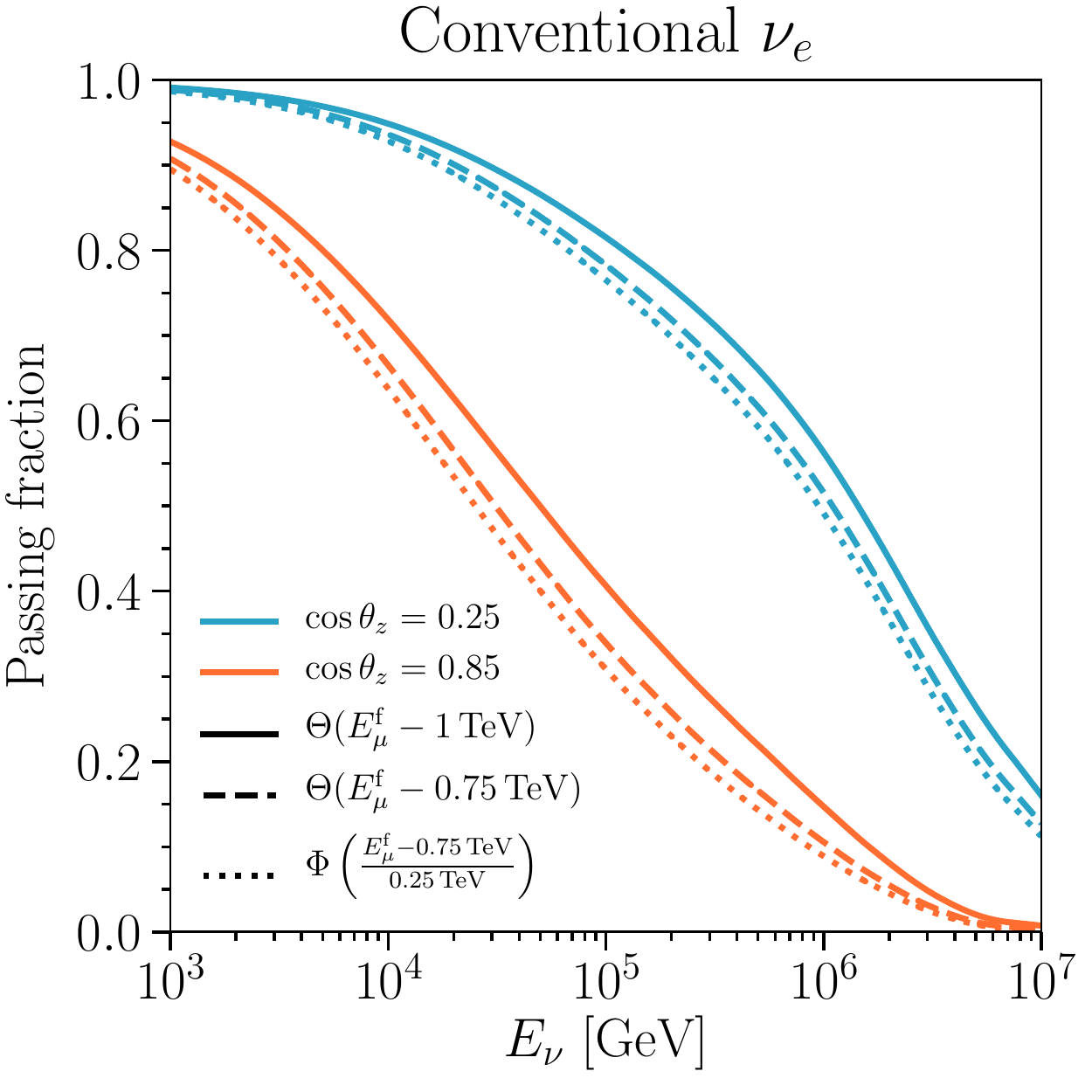}
    }
    \subfloat{
        \includegraphics[width=0.5\linewidth]{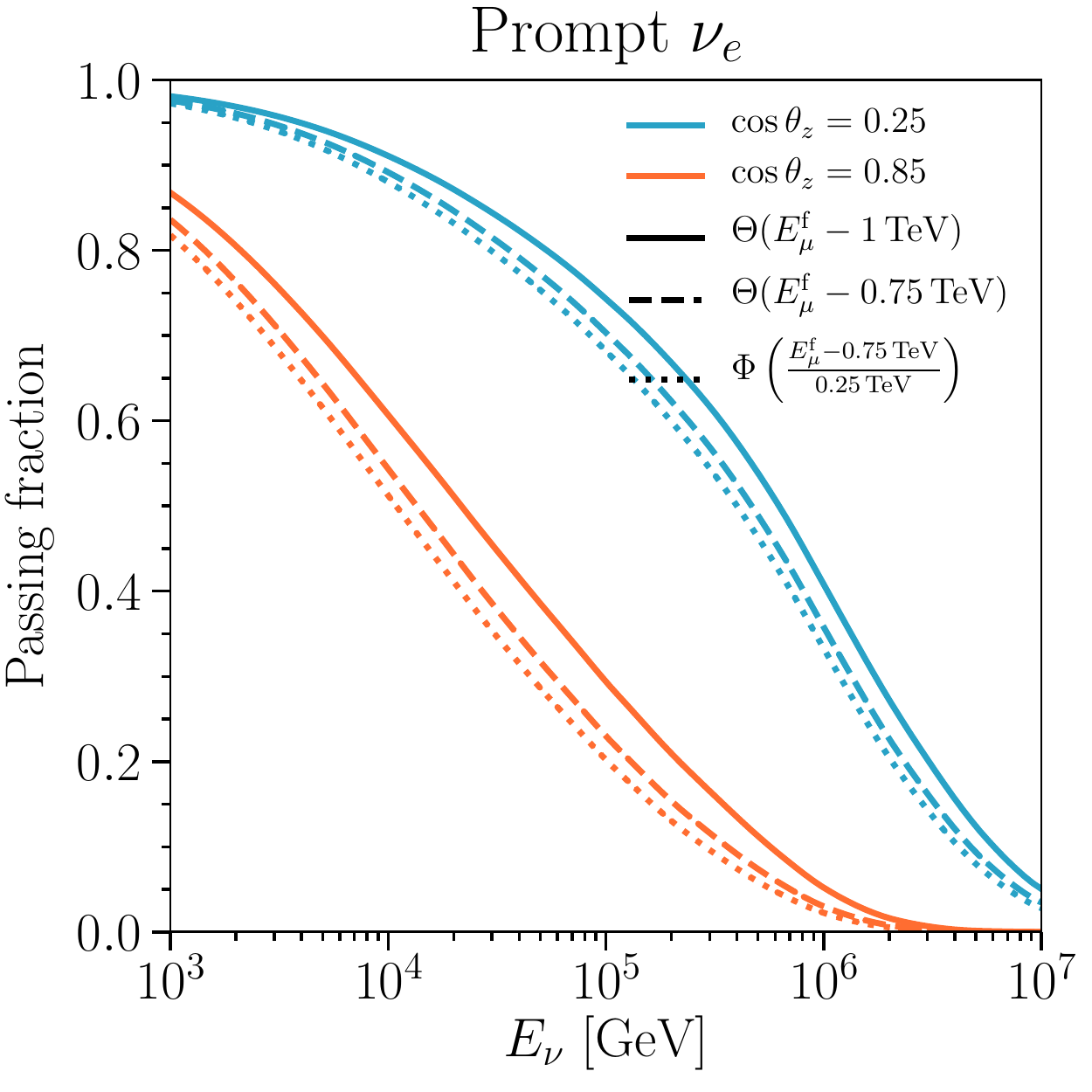}
    }\\[2ex]
    \subfloat{
        \includegraphics[width=0.5\linewidth]{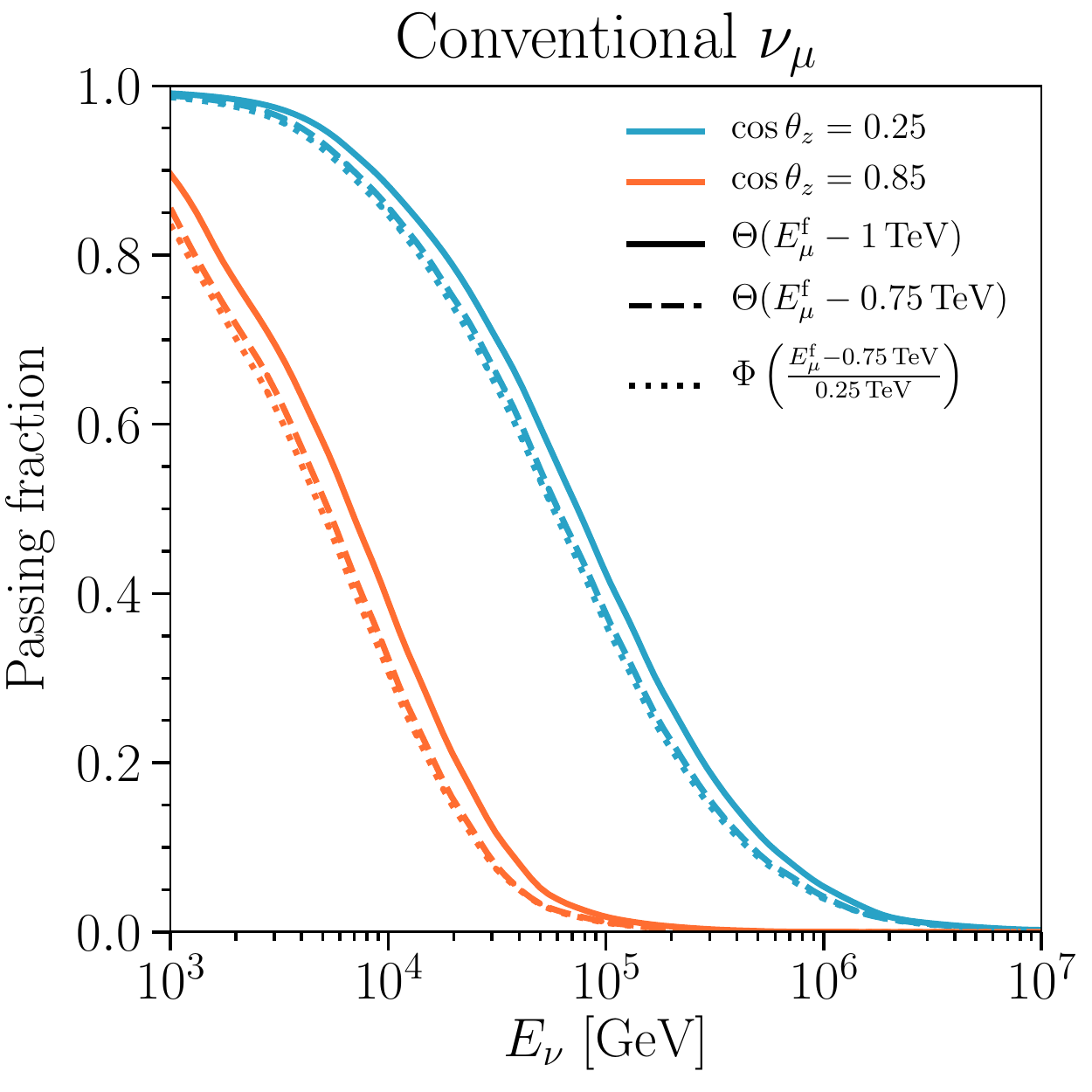}
    }
    \subfloat{
        \includegraphics[width=0.5\linewidth]{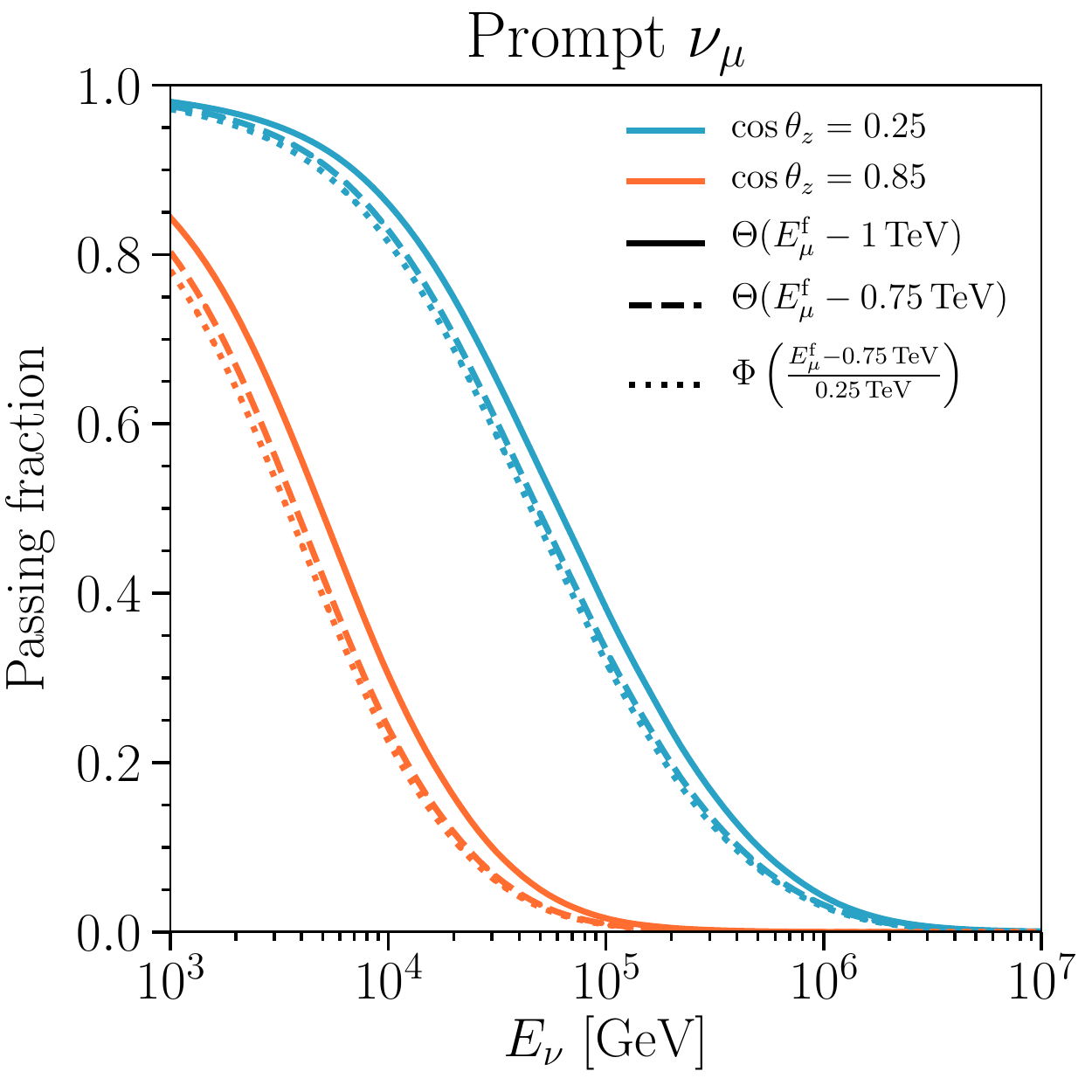}
    }    
\caption{\textbf{\textit{Passing fractions: effect of $\boldsymbol{\Prob_{\rm light}}$}}. Results are shown for two values of $\cos\theta_z$ (from top to bottom): 0.25 (blue) and 0.85 (orange); for three different $\Prob_{\rm light}(\Emf)$: a Heaviside with a muon threshold at 1~TeV, $\Theta\left(\Emf - 1 \, {\rm TeV}\right)$ (solid), a Heaviside with a muon threshold at 0.75~TeV, $\Theta\left(\Emf - 0.75 \, {\rm TeV}\right)$ (dashed) and a sigmoid $\Phi\left(\left(\Emf - 0.75 \, {\rm TeV}\right)/0.25 \, {\rm TeV}\right)$ (dotted). \textit{Top-left panel:} Conventional $\nu_e$ passing fraction. \textit{Top-right panel:} Prompt $\nu_e$ passing fraction. \textit{Bottom-left panel:} Conventional $\nu_\mu$ passing fraction. \textit{Bottom-right panel:} Prompt $\nu_\mu$ passing fraction.} \vspace{1.5cm}
\label{fig:nue-plight-effect}
\end{figure}

Once muons reach the detector, the key quantity to determine whether they can be correlated with a neutrino event is the probability to trigger the veto, $\Prob_{\rm light}$. This is described in section~\ref{sec:preach_plight}. Although we have shown $\Prob_{\rm det}$ for two different $\Prob_{\rm light}$ settings (Fig.~\ref{fig:pdet}), thus far we have only shown passing fractions assuming a Heaviside function. This efficiency function depends on the veto's configuration, so it is important to study the results using different functional forms.

We illustrate the effects of the modeling of the veto's response in Fig.~\ref{fig:nue-plight-effect}, where we depict the passing fractions for the conventional (left) and prompt (right) fluxes of atmospheric electron (top) and muon (bottom) neutrinos, for three different forms of $\Prob_{\rm light}$: Heaviside with a muon threshold at 1~TeV, $\Theta\left(\Emf - 1 \, {\rm TeV}\right)$ (solid), Heaviside with a muon threshold at 0.75~TeV, $\Theta\left(\Emf - 0.75 \, {\rm TeV}\right)$ (dashed), and sigmoid $\Phi\left(\left(\Emf - 0.75 \, {\rm TeV}\right)/0.25 \, {\rm TeV}\right)$ (dotted). In each plot, we show the passing fractions for two zenith angles: $\cos\theta_z = $ 0.25 (upper and blue curves) and 0.85 (lower and orange curves). As expected, reducing the energy threshold reduces the passing fractions, as more muons can be correlated with the neutrino event. The difference in absolute passing fractions are slightly larger for electron neutrinos, as the less energetic uncorrelated muons are more likely to trigger the veto. This effect is the largest change in passing fractions for muon neutrinos for a given detector. Combining this information with the fact that the muon neutrinos dominate the atmospheric flux implies that this small change in the passing fractions have a larger impact on the final passing rate. On the other hand, a smoother (more realistic) efficiency function around the nominal threshold energy, as described by the sigmoid, slightly reduces the passing fraction with respect to the abrupt form represented by the Heaviside function. For the examples shown, this effect is small. As discussed in section~\ref{sec:preach_plight}, this should be revisited in the extended formalism that considers the total energy of muon bundles and their properties.

Overall, absolute changes in the passing fractions are not very large when varying the threshold energy from 1~TeV to 0.75~TeV and the effect of a smooth $\Prob_{\rm light}$ is also mild; as reflected by our toy $\Prob_{\rm light}$ studies. Note that this later statement does not imply that the change in passing flux is negligible. Finally, each experiment that uses this calculation ought to carefully estimate their $\Prob_{\rm light}$, and significant changes of the passing fractions presented in this work can arise due to this.

\newpage

\section{Discussion and conclusions}
\label{sec:conclusions}

After the discovery of the first high-energy neutrinos of extraterrestrial origin by the IceCube neutrino observatory~\cite{Aartsen:2013bka, Aartsen:2013jdh, Aartsen:2014gkd, Aartsen:2015zva, Aartsen:2017mau, Aartsen:2015rwa, Aartsen:2016xlq, Aartsen:2014muf}, the next step is to accurately measure this flux and determine its cosmic origin. Accessing its contribution at lower energies would greatly help in this task and, in order to do so, a precise modeling of the background is required. In particular, a technique that substantially reduces the background atmospheric neutrino flux was proposed a few years ago~\cite{Schonert:2008is, Gaisser:2014bja}. The idea relies on the fact that some of the muons produced in the same air shower as atmospheric neutrinos might reach the detector and trigger the veto in coincidence with the neutrino event. In this way, the atmospheric neutrino background can be significantly reduced. Building upon this idea, we have presented a new calculation of the atmospheric neutrino passing fraction, that is, the fraction of the neutrino flux that contributes to the background in searches for astrophysical neutrinos. This calculation unifies the calculation for electron and muon neutrinos and is suited for including systematic uncertainties and different detector characteristics. As a summary of our results, in Fig.~\ref{fig:fluxes} we have depicted the atmospheric neutrino passing fluxes for 10~TeV and 100~TeV, for conventional and prompt electron and muon neutrinos. An astrophysical $\nu_\mu$ flux is included for comparison~\cite{Aartsen:2017mau}.

In section~\ref{sec:shower_physics}, we indicated our default settings and the numerical tool (\MCEq) used to describe atmospheric lepton fluxes. One of the important ingredients in our calculations is the probability for a muon to reach the detector and trigger the veto, which is defined in section~\ref{sec:preach_plight} and shown in Fig.~\ref{fig:pdet}. After the preliminary descriptions, we introduced a new definition of the passing fractions for electron and muon neutrinos, and discussed the implications for tau neutrinos in section~\ref{sec:pf}. This is the core of this work. Our definition allows for a unified treatment of the correlated~\cite{Schonert:2008is} and uncorrelated~\cite{Gaisser:2014bja} parts of the atmospheric neutrino passing fractions. In that section, we also studied the effect of some previous approximations and have compared our results with those from a \CORSIKA{} simulation, finding excellent agreement. All these comparisons are depicted in Figs.~\ref{fig:nue_passing-double-counting}--~\ref{fig:nu-e-neutrino-vs-antineutrino} for electron neutrinos and in Figs.~\ref{fig:nu-mu-correlated-preach-effect}--~\ref{fig:nu-mu-neutrino-vs-antineutrino} for muon neutrinos. In general, we find non-negligible differences between our approach with and without a number of approximations. Moreover, these differences have non-trivial energy and zenith angle dependence. In section~\ref{sec:improvements}, we presented the overall comparison between our results and those from the original proposal~\cite{Schonert:2008is, Gaisser:2014bja} (Fig.~\ref{fig:nue-passing-comparison-old}), with a critical discussion on the modifications and improvements. 

After presenting our approach to compute the passing fractions, the second goal of this work was to evaluate how systematic uncertainties propagate into our calculations and how our results depend on the veto configuration. The former is described in section~\ref{sec:systematics}, where we studied the impact of different sources of systematic uncertainties: from the treatment of muon-energy losses in section~\ref{sec:muonlosses}, from the choice of primary cosmic-ray spectrum in section~\ref{sec:CR} (Fig.~\ref{fig:nue-cr-model-effect}), from the choice of the hadronic-interaction model in section~\ref{sec:hadronic} (Fig.~\ref{fig:nue-hadronic-model-effect}), and from the choice of the atmosphere-density model in section~\ref{sec:atmosphere} (Fig.~\ref{fig:atmosphere-effect}). We conclude that systematic uncertainties, for both the conventional and prompt passing fractions, are non-negligible in the case of electron neutrinos, but they are small for muon neutrinos. Notice, however, that the atmospheric muon neutrino flux is an order of magnitude larger than the electron neutrino flux at these energies. Therefore, small changes in the muon passing fractions can have more important effects on the passing fluxes than larger changes in the electron neutrino passing fraction. The detector-dependent effects on the passing fractions are discussed in section~\ref{sec:detector}, namely: depth of the detector in section~\ref{sec:depth}, surrounding medium in section~\ref{sec:medium} (Fig.~\ref{fig:medium-effect}), and muon veto trigger probability of muons in section~\ref{sec:response} (Fig.~\ref{fig:nue-plight-effect}). The depth of the detector has a crucial impact on our results. Therefore, the detector geometry is important. Consequently, the atmospheric neutrino passing fractions computed for different present and planned large-scale neutrino telescopes, such as IceCube~\cite{Achterberg:2006md}, ANTARES~\cite{Collaboration:2011nsa}, KM3NeT~\cite{Adrian-Martinez:2016fdl}, and Baikal-GVD~\cite{Baikal2017}, are significantly different. Moreover, although the medium (water or ice) is not a critical factor, the differences in the passing fractions are not negligible. Finally, we have also shown that considering different veto efficiencies for muons can have a non-negligible impact on the results; this needs to be carefully modeled by a detailed detector simulation. 

The variation of the different inputs in our calculation mainly affect the prediction of the passing fractions for electrons neutrinos, while for muon neutrinos the effect tends to be smaller. Nevertheless, the final uncertainties on the passing fluxes may not be proportional to that on the passing fractions. A consistent treatment of uncertainties involves the convolution of the passing fraction and neutrino flux uncertainties, which can be easily implemented within our approach.

In this work, a new framework for calculating the atmospheric neutrino passing fractions has been proposed. Equation~(\ref{eq:GUE}) unifies the formalism introduced in previous works into a fully consistent treatment for electron, muon, and tau neutrinos. Our calculation allows for easily testing the effect of different primary cosmic-ray spectra, hadronic-interaction model, atmosphere-density model, muon-energy loss, and detector configuration. To facilitate this, a new \Python{} package is available and described in appendix~\ref{app:code}, and the results with our default settings are provided in tabulated form in appendix~\ref{app:tables}. It should be a widely-applicable tool for current and future large-scale neutrino telescopes that could help to pave the road into the era of precision sub-PeV neutrino astrophysics.

\section*{Acknowledgments}

We would like to thank Anatoli Fedynitch for his valuable help with \MCEq{}. We would also like to thank Kyle Jero and Jakob van Santen for generating the \CORSIKA{} dataset, Tyce DeYoung, Tom Gaisser, Maria Vittoria Garzelli, and Hallsie Reno for useful discussions, and Jean DeMerit for carefully proofreading our work. 
CAA is supported by U.S. National Science Foundation (NSF) grant No. PHY-1505858.
SPR is supported by a Ram\'on y Cajal contract, by the Spanish MINECO under grants FPA2017-84543-P, FPA2014-54459-P,  SEV-2014-0398, by the Generalitat Valenciana under grant PROMETEOII/2014/049 and by the European Union's Horizon 2020 research and innovation program under the Marie Sk\l odowska-Curie grant agreements No. 690575 and 674896. SPR is also partially supported by the Portuguese FCT through the CFTP-FCT Unit 777 (PEst-OE/FIS/UI0777/2013).  
AS, LW, and TY are supported in part by NSF under grant PHY-1607644 and by the University of Wisconsin Research Committee with funds granted by the Wisconsin Alumni Research Foundation. 
SPR would like to thank WIPAC and CAA, AS, and TY would also like to thank IFIC for hospitality. 
CAA, SPR, and TY would like to further thank the VietNus 2017 participants and organizers for hosting a wonderful and intellectually stimulating workshop, where this project was initiated. 

\bibliography{nuveto}

\newpage

\appendix

\section{Atmospheric neutrino passing fraction code}
\label{app:code}

The atmospheric neutrino passing fraction code, \nuveto, is a \Python{} package that can be found at \\

\noindent \url{https://github.com/tianluyuan/nuVeto/} \\

\noindent and implements our master equations, Eqs.~(\ref{eq:Pcor}) and~(\ref{eq:GUE}). It calculates the passing fraction or passing flux of atmospheric neutrinos and antineutrinos as a function of energy, flavor, zenith angle, and detector depth for any given primary flux, hadronic-interaction model, and atmosphere-density profile implemented in \MCEq. It is also possible to generate any detection probability, given by Eq.~(\ref{eq:PrPl}), for different detector configurations in water and ice using the $\Prob_{\rm reach}$ distributions that are provided for those two media.

The package requires \MCEq; follow the directions for installing \MCEq{} before trying to use the code. Another dependency that is only needed for generating Eq.~(\ref{eq:PrPl}) is \href{https://pandas.pydata.org/}{\texttt{pandas}}. Once installed, a simple example to get started is the following:
\begin{lstlisting}[language=Python]
from nuVeto.nuveto import passing
enu = 1e5*Units.GeV
cos_theta = 0.5
pf = passing(enu, cos_theta, kind=`conv_numu',
            pmodel=(pm.HillasGaisser2012, `H3a'),
            hadr=`SIBYLL2.3c', depth=1950*Units.m,
            density=(`CORSIKA', (`SouthPole',`June')))
\end{lstlisting}
Here, \lstinline|passing| is a wrapper function that builds and calls the \lstinline|nuVeto| calculation object. All parameters that affect the total flux as calculated in \MCEq{} are passed to the constructor of the \lstinline|nuVeto| class, and other parameters such as the neutrino energy, zenith angle and medium, are passed directly to the calculation function \lstinline|nuVeto.get_fluxes|. This reduces the computing time by calling \MCEq{} only when necessary.

Equation~(\ref{eq:PrPl}) is constructed from a convolution of $\Prob_{\rm reach}$ and $\Prob_{\rm light}$. $\Prob_{\rm reach}$ is provided for both water and ice in \lstinline|resources/mu/mmc/*.pklz|. The detector response probability must be defined in \lstinline|resources/mu/pl.py| as a function of the muon energy at the detector. The function must follow the proper format as the other functions in that file and begin with \lstinline|pl_|, as in \lstinline|def pl_myplight(emu)|. Since this is a detector-dependent quantity, it must be computed separately from the \nuveto{} package. However, several default assumptions of $\Prob_{\rm light}$ are included as described in Sec.~\ref{sec:response}. Once $\Prob_{\rm light}$ is defined, execute the following commands to construct and place Eq.~(\ref{eq:PrPl}) in the proper location,
\begin{lstlisting}[language=Bash]
cd nuVeto/resources/mu
./mu.py -o ../../prpl/mypdet.pkl --plight pl_myplight mmc/ice_allm97.pklz
\end{lstlisting} 
To use the newly generated file, it must be passed as a string to the prpl argument,
\begin{lstlisting}[language=Python]
passing(enu, cos_theta, prpl=`mypdet')
\end{lstlisting}\lstinline||

\section{Tabulated atmospheric neutrino passing fractions}
\label{app:tables}

In this section we present tabulated passing fractions for our default settings in the text, namely: the H3a cosmic-ray spectrum~\cite{Gaisser:2011cc}, the SIBYLL~2.3c~\cite{Riehn:2017mfm} hadronic-interaction model, and the MSIS-90-E atmosphere-density model at the South Pole on July 1, 1997~\cite{Labitzke:1985, Hedin:1991}. We consider a detector located in ice at a depth of $d_{\rm det} = 1.95$~km. For the propagation of muons we use the default settings in \MMC~\cite{Chirkin:2004hz}. For $\Prob_{\rm light}$ we use a Heaviside function with a muon threshold at $E_\mu^{\rm th} = 1$~TeV.

\newpage 

\begin{table}[h!]
\centering
\begin{tabular*}{\textwidth}{l @{\extracolsep{\fill}} r r r r r r r r r r}
\toprule
& \multicolumn{10}{c}{$\cos \theta_z$}\\
\cmidrule{2-11}
$E_\nu$ [GeV] & 0.1 & 0.2 & 0.3 & 0.4 & 0.5 & 0.6 & 0.7 & 0.8 & 0.9 & 1.0 \\ 
\midrule
$1.0\cdot10^{3}$ & 1.00 & 1.00 & 0.98 & 0.97 & 0.96 & 0.94 & 0.94 & 0.93 & 0.92 & 0.92 \\ \hline
$1.6\cdot10^{3}$ & 1.00 & 0.99 & 0.98 & 0.96 & 0.94 & 0.92 & 0.91 & 0.90 & 0.89 & 0.89 \\ \hline
$2.6\cdot10^{3}$ & 1.00 & 0.99 & 0.97 & 0.94 & 0.91 & 0.89 & 0.88 & 0.86 & 0.86 & 0.85 \\ \hline
$4.3\cdot10^{3}$ & 1.00 & 0.99 & 0.95 & 0.91 & 0.88 & 0.85 & 0.84 & 0.82 & 0.81 & 0.80 \\ \hline
$7.0\cdot10^{3}$ & 1.00 & 0.98 & 0.93 & 0.88 & 0.84 & 0.81 & 0.79 & 0.77 & 0.76 & 0.75 \\ \hline
$1.1\cdot10^{4}$ & 1.00 & 0.98 & 0.91 & 0.85 & 0.80 & 0.75 & 0.73 & 0.71 & 0.69 & 0.68 \\ \hline
$1.8\cdot10^{4}$ & 1.00 & 0.96 & 0.88 & 0.80 & 0.74 & 0.69 & 0.67 & 0.64 & 0.63 & 0.62 \\ \hline
$3.0\cdot10^{4}$ & 1.00 & 0.95 & 0.84 & 0.75 & 0.69 & 0.63 & 0.61 & 0.57 & 0.56 & 0.55 \\ \hline
$4.8\cdot10^{4}$ & 1.00 & 0.93 & 0.80 & 0.70 & 0.62 & 0.56 & 0.54 & 0.51 & 0.49 & 0.48 \\ \hline
$7.8\cdot10^{4}$ & 1.00 & 0.91 & 0.76 & 0.64 & 0.56 & 0.50 & 0.47 & 0.44 & 0.43 & 0.42 \\ \hline
$1.3\cdot10^{5}$ & 1.00 & 0.89 & 0.71 & 0.59 & 0.50 & 0.44 & 0.41 & 0.38 & 0.37 & 0.36 \\ \hline
$2.1\cdot10^{5}$ & 1.00 & 0.86 & 0.66 & 0.53 & 0.44 & 0.38 & 0.35 & 0.32 & 0.31 & 0.30 \\ \hline
$3.4\cdot10^{5}$ & 1.00 & 0.82 & 0.60 & 0.47 & 0.38 & 0.32 & 0.29 & 0.26 & 0.25 & 0.25 \\ \hline
$5.5\cdot10^{5}$ & 0.99 & 0.78 & 0.54 & 0.40 & 0.32 & 0.26 & 0.24 & 0.21 & 0.20 & 0.20 \\ \hline
$8.9\cdot10^{5}$ & 0.99 & 0.72 & 0.47 & 0.34 & 0.26 & 0.20 & 0.19 & 0.16 & 0.15 & 0.15 \\ \hline
$1.4\cdot10^{6}$ & 0.98 & 0.65 & 0.39 & 0.26 & 0.19 & 0.15 & 0.13 & 0.11 & 0.11 & 0.10 \\ \hline
$2.3\cdot10^{6}$ & 0.97 & 0.57 & 0.30 & 0.18 & 0.13 & $9.3\cdot10^{-2}$ & $8.2\cdot10^{-2}$ & $6.8\cdot10^{-2}$ & $6.4\cdot10^{-2}$ & $6.1\cdot10^{-2}$ \\ \hline
$3.8\cdot10^{6}$ & 0.96 & 0.48 & 0.21 & 0.11 & $7.3\cdot10^{-2}$ & $4.9\cdot10^{-2}$ & $4.2\cdot10^{-2}$ & $3.4\cdot10^{-2}$ & $3.2\cdot10^{-2}$ & $3.0\cdot10^{-2}$ \\ \hline
$6.2\cdot10^{6}$ & 0.94 & 0.39 & 0.14 & $6.6\cdot10^{-2}$ & $3.7\cdot10^{-2}$ & $2.3\cdot10^{-2}$ & $1.9\cdot10^{-2}$ & $1.4\cdot10^{-2}$ & $1.3\cdot10^{-2}$ & $1.2\cdot10^{-2}$ \\ \hline
$1.0\cdot10^{7}$ & 0.91 & 0.30 & $8.9\cdot10^{-2}$ & $3.8\cdot10^{-2}$ & $2.0\cdot10^{-2}$ & $1.2\cdot10^{-2}$ & $9.8\cdot10^{-3}$ & $7.4\cdot10^{-3}$ & $6.8\cdot10^{-3}$ & $6.4\cdot10^{-3}$ \\ \hline
\bottomrule
\end{tabular*}
\caption{\textbf{\textit{Conventional electron neutrino}} passing fraction for  $E_\nu = [10^3, \, 10^7]$~GeV and $\cos\theta_z = [0.1, \, 1.0]$.} \vspace{-6mm}
\label{tbl:conventional_nue}
\end{table}

\begin{table}[h!]
\centering
\begin{tabular*}{\textwidth}{l @{\extracolsep{\fill}} r r r r r r r r r r}
\toprule
& \multicolumn{10}{c}{$\cos \theta_z$}\\
\cmidrule{2-11}
$E_\nu$ [GeV] & 0.1 & 0.2 & 0.3 & 0.4 & 0.5 & 0.6 & 0.7 & 0.8 & 0.9 & 1.0 \\ 
\midrule
$1.0\cdot10^{3}$ & 1.00 & 1.00 & 0.98 & 0.96 & 0.95 & 0.93 & 0.93 & 0.92 & 0.91 & 0.91 \\ \hline
$1.6\cdot10^{3}$ & 1.00 & 0.99 & 0.97 & 0.95 & 0.93 & 0.91 & 0.90 & 0.88 & 0.88 & 0.87 \\ \hline
$2.6\cdot10^{3}$ & 1.00 & 0.99 & 0.96 & 0.93 & 0.90 & 0.87 & 0.86 & 0.84 & 0.83 & 0.83 \\ \hline
$4.3\cdot10^{3}$ & 1.00 & 0.99 & 0.94 & 0.90 & 0.86 & 0.83 & 0.81 & 0.79 & 0.78 & 0.77 \\ \hline
$7.0\cdot10^{3}$ & 1.00 & 0.98 & 0.92 & 0.87 & 0.82 & 0.78 & 0.76 & 0.74 & 0.73 & 0.72 \\ \hline
$1.1\cdot10^{4}$ & 1.00 & 0.97 & 0.89 & 0.82 & 0.77 & 0.72 & 0.70 & 0.67 & 0.66 & 0.65 \\ \hline
$1.8\cdot10^{4}$ & 1.00 & 0.96 & 0.86 & 0.78 & 0.71 & 0.66 & 0.64 & 0.61 & 0.59 & 0.58 \\ \hline
$3.0\cdot10^{4}$ & 1.00 & 0.94 & 0.82 & 0.73 & 0.65 & 0.60 & 0.57 & 0.54 & 0.53 & 0.52 \\ \hline
$4.8\cdot10^{4}$ & 1.00 & 0.92 & 0.78 & 0.67 & 0.59 & 0.53 & 0.50 & 0.47 & 0.46 & 0.45 \\ \hline
$7.8\cdot10^{4}$ & 1.00 & 0.90 & 0.73 & 0.61 & 0.53 & 0.46 & 0.44 & 0.41 & 0.39 & 0.38 \\ \hline
$1.3\cdot10^{5}$ & 1.00 & 0.88 & 0.68 & 0.55 & 0.47 & 0.40 & 0.38 & 0.35 & 0.33 & 0.32 \\ \hline
$2.1\cdot10^{5}$ & 1.00 & 0.84 & 0.63 & 0.50 & 0.41 & 0.34 & 0.32 & 0.29 & 0.28 & 0.27 \\ \hline
$3.4\cdot10^{5}$ & 1.00 & 0.80 & 0.57 & 0.43 & 0.35 & 0.29 & 0.26 & 0.23 & 0.22 & 0.22 \\ \hline
$5.5\cdot10^{5}$ & 0.99 & 0.76 & 0.51 & 0.37 & 0.29 & 0.23 & 0.21 & 0.18 & 0.17 & 0.17 \\ \hline
$8.9\cdot10^{5}$ & 0.99 & 0.70 & 0.43 & 0.30 & 0.22 & 0.17 & 0.16 & 0.13 & 0.13 & 0.12 \\ \hline
$1.4\cdot10^{6}$ & 0.98 & 0.62 & 0.35 & 0.22 & 0.16 & 0.12 & 0.11 & $8.8\cdot10^{-2}$ & $8.3\cdot10^{-2}$ & $7.9\cdot10^{-2}$ \\ \hline
$2.3\cdot10^{6}$ & 0.97 & 0.53 & 0.26 & 0.15 & 0.10 & $7.1\cdot10^{-2}$ & $6.2\cdot10^{-2}$ & $5.1\cdot10^{-2}$ & $4.7\cdot10^{-2}$ & $4.5\cdot10^{-2}$ \\ \hline
$3.8\cdot10^{6}$ & 0.95 & 0.44 & 0.18 & $9.1\cdot10^{-2}$ & $5.5\cdot10^{-2}$ & $3.6\cdot10^{-2}$ & $3.0\cdot10^{-2}$ & $2.4\cdot10^{-2}$ & $2.2\cdot10^{-2}$ & $2.1\cdot10^{-2}$ \\ \hline
$6.2\cdot10^{6}$ & 0.93 & 0.35 & 0.11 & $5.1\cdot10^{-2}$ & $2.7\cdot10^{-2}$ & $1.6\cdot10^{-2}$ & $1.3\cdot10^{-2}$ & $9.4\cdot10^{-3}$ & $8.5\cdot10^{-3}$ & $7.9\cdot10^{-3}$ \\ \hline
$1.0\cdot10^{7}$ & 0.89 & 0.26 & $6.9\cdot10^{-2}$ & $2.7\cdot10^{-2}$ & $1.3\cdot10^{-2}$ & $7.5\cdot10^{-3}$ & $6.0\cdot10^{-3}$ & $4.4\cdot10^{-3}$ & $4.0\cdot10^{-3}$ & $3.7\cdot10^{-3}$ \\ \hline
\bottomrule
\end{tabular*}
\caption{\textbf{\textit{Conventional electron antineutrino}} passing fraction for $E_\nu = [10^3, \, 10^7]$~GeV and $\cos\theta_z = [0.1, \, 1.0]$.}
\label{tbl:conventional_antinue}
\end{table}

\begin{table}[h!]
\centering
\begin{tabular*}{\textwidth}{l @{\extracolsep{\fill}} r r r r r r r r r r}
\toprule
& \multicolumn{10}{c}{$\cos \theta_z$}\\
\cmidrule{2-11}
$E_\nu$ [GeV] & 0.1 & 0.2 & 0.3 & 0.4 & 0.5 & 0.6 & 0.7 & 0.8 & 0.9 & 1.0 \\ 
\midrule
$1.0\cdot10^{3}$ & 1.00 & 1.00 & 0.98 & 0.97 & 0.95 & 0.93 & 0.92 & 0.90 & 0.89 & 0.87 \\ \hline
$1.6\cdot10^{3}$ & 1.00 & 0.99 & 0.98 & 0.95 & 0.91 & 0.88 & 0.85 & 0.82 & 0.79 & 0.78 \\ \hline
$2.6\cdot10^{3}$ & 1.00 & 0.99 & 0.96 & 0.90 & 0.84 & 0.79 & 0.76 & 0.73 & 0.71 & 0.69 \\ \hline
$4.3\cdot10^{3}$ & 1.00 & 0.99 & 0.92 & 0.83 & 0.76 & 0.70 & 0.67 & 0.63 & 0.60 & 0.58 \\ \hline
$7.0\cdot10^{3}$ & 1.00 & 0.97 & 0.86 & 0.75 & 0.67 & 0.59 & 0.55 & 0.50 & 0.47 & 0.44 \\ \hline
$1.1\cdot10^{4}$ & 1.00 & 0.94 & 0.79 & 0.65 & 0.55 & 0.47 & 0.42 & 0.37 & 0.34 & 0.31 \\ \hline
$1.8\cdot10^{4}$ & 1.00 & 0.90 & 0.70 & 0.54 & 0.42 & 0.33 & 0.28 & 0.24 & 0.21 & 0.19 \\ \hline
$3.0\cdot10^{4}$ & 1.00 & 0.85 & 0.59 & 0.40 & 0.29 & 0.21 & 0.17 & 0.14 & 0.11 & $9.5\cdot10^{-2}$ \\ \hline
$4.8\cdot10^{4}$ & 1.00 & 0.77 & 0.46 & 0.28 & 0.18 & 0.12 & $8.7\cdot10^{-2}$ & $6.2\cdot10^{-2}$ & $4.9\cdot10^{-2}$ & $3.9\cdot10^{-2}$ \\ \hline
$7.8\cdot10^{4}$ & 1.00 & 0.69 & 0.35 & 0.18 & 0.10 & $5.8\cdot10^{-2}$ & $4.1\cdot10^{-2}$ & $3.0\cdot10^{-2}$ & $2.4\cdot10^{-2}$ & $1.9\cdot10^{-2}$ \\ \hline
$1.3\cdot10^{5}$ & 0.99 & 0.58 & 0.24 & 0.10 & $5.5\cdot10^{-2}$ & $2.8\cdot10^{-2}$ & $2.1\cdot10^{-2}$ & $1.5\cdot10^{-2}$ & $1.2\cdot10^{-2}$ & $9.3\cdot10^{-3}$ \\ \hline
$2.1\cdot10^{5}$ & 0.98 & 0.46 & 0.15 & $5.7\cdot10^{-2}$ & $3.0\cdot10^{-2}$ & $1.4\cdot10^{-2}$ & $1.0\cdot10^{-2}$ & $7.3\cdot10^{-3}$ & $6.0\cdot10^{-3}$ & $4.3\cdot10^{-3}$ \\ \hline
$3.4\cdot10^{5}$ & 0.96 & 0.35 & $9.0\cdot10^{-2}$ & $3.0\cdot10^{-2}$ & $1.5\cdot10^{-2}$ & $6.4\cdot10^{-3}$ & $4.5\cdot10^{-3}$ & $3.3\cdot10^{-3}$ & $2.7\cdot10^{-3}$ & $1.9\cdot10^{-3}$ \\ \hline
$5.5\cdot10^{5}$ & 0.93 & 0.25 & $5.1\cdot10^{-2}$ & $1.4\cdot10^{-2}$ & $6.5\cdot10^{-3}$ & $1.8\cdot10^{-3}$ & $1.2\cdot10^{-3}$ & $1.1\cdot10^{-3}$ & $7.8\cdot10^{-4}$ & $5.4\cdot10^{-4}$ \\ \hline
$8.9\cdot10^{5}$ & 0.90 & 0.17 & $2.7\cdot10^{-2}$ & $5.0\cdot10^{-3}$ & $1.8\cdot10^{-3}$ & $8.2\cdot10^{-4}$ & $4.7\cdot10^{-4}$ & $2.8\cdot10^{-4}$ & $2.3\cdot10^{-4}$ & $1.4\cdot10^{-4}$ \\ \hline
$1.4\cdot10^{6}$ & 0.84 & 0.11 & $1.0\cdot10^{-2}$ & $2.7\cdot10^{-3}$ & $9.8\cdot10^{-4}$ & $4.5\cdot10^{-4}$ & $2.3\cdot10^{-4}$ & $1.4\cdot10^{-4}$ & $1.1\cdot10^{-4}$ & $9.0\cdot10^{-5}$ \\ \hline
$2.3\cdot10^{6}$ & 0.77 & $6.2\cdot10^{-2}$ & $6.0\cdot10^{-3}$ & $1.4\cdot10^{-3}$ & $5.1\cdot10^{-4}$ & $2.2\cdot10^{-4}$ & $1.2\cdot10^{-4}$ & $6.1\cdot10^{-5}$ & $3.8\cdot10^{-5}$ & $4.3\cdot10^{-5}$ \\ \hline
$3.8\cdot10^{6}$ & 0.69 & $3.0\cdot10^{-2}$ & $3.1\cdot10^{-3}$ & $6.2\cdot10^{-4}$ & $2.1\cdot10^{-4}$ & $8.6\cdot10^{-5}$ & $5.7\cdot10^{-5}$ & $2.3\cdot10^{-5}$ & $1.1\cdot10^{-5}$ & $1.9\cdot10^{-5}$ \\ \hline
$6.2\cdot10^{6}$ & 0.59 & $1.8\cdot10^{-2}$ & $1.5\cdot10^{-3}$ & $2.5\cdot10^{-4}$ & $7.6\cdot10^{-5}$ & $2.6\cdot10^{-5}$ & $2.0\cdot10^{-5}$ & $7.5\cdot10^{-6}$ & $2.1\cdot10^{-6}$ & $6.2\cdot10^{-6}$ \\ \hline
$1.0\cdot10^{7}$ & 0.50 & $1.1\cdot10^{-2}$ & $6.7\cdot10^{-4}$ & $1.1\cdot10^{-4}$ & $2.4\cdot10^{-5}$ & $8.9\cdot10^{-6}$ & $5.7\cdot10^{-6}$ & $3.3\cdot10^{-6}$ & $7.6\cdot10^{-7}$ & $1.7\cdot10^{-6}$ \\ \hline
\bottomrule
\end{tabular*}
\caption{\textbf{\textit{Conventional muon neutrino}} passing fraction for $E_\nu = [10^3, \, 10^7]$~GeV and $\cos\theta_z = [0.1, \, 1.0]$.} \vspace{-6mm}
\label{tbl:conventional_numu}
\end{table}

\begin{table}[h!]
\centering
\begin{tabular*}{\textwidth}{l @{\extracolsep{\fill}} r r r r r r r r r r}
\toprule
& \multicolumn{10}{c}{$\cos \theta_z$}\\
\cmidrule{2-11}
$E_\nu$ [GeV] & 0.1 & 0.2 & 0.3 & 0.4 & 0.5 & 0.6 & 0.7 & 0.8 & 0.9 & 1.0 \\ 
\midrule
$1.0\cdot10^{3}$ & 1.00 & 0.99 & 0.98 & 0.95 & 0.93 & 0.91 & 0.89 & 0.87 & 0.85 & 0.83 \\ \hline
$1.6\cdot10^{3}$ & 1.00 & 0.99 & 0.96 & 0.93 & 0.89 & 0.84 & 0.80 & 0.76 & 0.74 & 0.71 \\ \hline
$2.6\cdot10^{3}$ & 1.00 & 0.99 & 0.94 & 0.87 & 0.80 & 0.73 & 0.70 & 0.66 & 0.64 & 0.62 \\ \hline
$4.3\cdot10^{3}$ & 1.00 & 0.98 & 0.89 & 0.78 & 0.70 & 0.64 & 0.60 & 0.57 & 0.54 & 0.52 \\ \hline
$7.0\cdot10^{3}$ & 1.00 & 0.96 & 0.82 & 0.70 & 0.61 & 0.54 & 0.49 & 0.45 & 0.42 & 0.39 \\ \hline
$1.1\cdot10^{4}$ & 1.00 & 0.93 & 0.74 & 0.60 & 0.50 & 0.41 & 0.37 & 0.32 & 0.29 & 0.27 \\ \hline
$1.8\cdot10^{4}$ & 1.00 & 0.88 & 0.65 & 0.49 & 0.38 & 0.29 & 0.25 & 0.21 & 0.18 & 0.16 \\ \hline
$3.0\cdot10^{4}$ & 1.00 & 0.82 & 0.55 & 0.36 & 0.26 & 0.18 & 0.15 & 0.12 & $9.7\cdot10^{-2}$ & $8.0\cdot10^{-2}$ \\ \hline
$4.8\cdot10^{4}$ & 1.00 & 0.74 & 0.42 & 0.25 & 0.16 & 0.10 & $7.3\cdot10^{-2}$ & $5.2\cdot10^{-2}$ & $4.1\cdot10^{-2}$ & $3.2\cdot10^{-2}$ \\ \hline
$7.8\cdot10^{4}$ & 0.99 & 0.66 & 0.32 & 0.16 & $8.8\cdot10^{-2}$ & $4.8\cdot10^{-2}$ & $3.4\cdot10^{-2}$ & $2.4\cdot10^{-2}$ & $1.9\cdot10^{-2}$ & $1.5\cdot10^{-2}$ \\ \hline
$1.3\cdot10^{5}$ & 0.99 & 0.55 & 0.21 & $8.9\cdot10^{-2}$ & $4.6\cdot10^{-2}$ & $2.3\cdot10^{-2}$ & $1.7\cdot10^{-2}$ & $1.2\cdot10^{-2}$ & $9.5\cdot10^{-3}$ & $7.2\cdot10^{-3}$ \\ \hline
$2.1\cdot10^{5}$ & 0.97 & 0.44 & 0.13 & $4.9\cdot10^{-2}$ & $2.4\cdot10^{-2}$ & $1.1\cdot10^{-2}$ & $7.8\cdot10^{-3}$ & $5.5\cdot10^{-3}$ & $4.6\cdot10^{-3}$ & $3.2\cdot10^{-3}$ \\ \hline
$3.4\cdot10^{5}$ & 0.96 & 0.33 & $7.9\cdot10^{-2}$ & $2.5\cdot10^{-2}$ & $1.2\cdot10^{-2}$ & $4.9\cdot10^{-3}$ & $3.4\cdot10^{-3}$ & $2.4\cdot10^{-3}$ & $1.9\cdot10^{-3}$ & $1.4\cdot10^{-3}$ \\ \hline
$5.5\cdot10^{5}$ & 0.93 & 0.23 & $4.3\cdot10^{-2}$ & $1.1\cdot10^{-2}$ & $4.9\cdot10^{-3}$ & $1.3\cdot10^{-3}$ & $8.6\cdot10^{-4}$ & $7.7\cdot10^{-4}$ & $5.3\cdot10^{-4}$ & $3.6\cdot10^{-4}$ \\ \hline
$8.9\cdot10^{5}$ & 0.88 & 0.15 & $2.2\cdot10^{-2}$ & $3.7\cdot10^{-3}$ & $1.3\cdot10^{-3}$ & $5.2\cdot10^{-4}$ & $2.9\cdot10^{-4}$ & $1.7\cdot10^{-4}$ & $1.4\cdot10^{-4}$ & $8.1\cdot10^{-5}$ \\ \hline
$1.4\cdot10^{6}$ & 0.83 & $9.3\cdot10^{-2}$ & $7.6\cdot10^{-3}$ & $1.8\cdot10^{-3}$ & $5.8\cdot10^{-4}$ & $2.5\cdot10^{-4}$ & $1.2\cdot10^{-4}$ & $7.1\cdot10^{-5}$ & $5.3\cdot10^{-5}$ & $4.2\cdot10^{-5}$ \\ \hline
$2.3\cdot10^{6}$ & 0.75 & $5.1\cdot10^{-2}$ & $4.0\cdot10^{-3}$ & $7.9\cdot10^{-4}$ & $2.5\cdot10^{-4}$ & $9.8\cdot10^{-5}$ & $5.0\cdot10^{-5}$ & $2.4\cdot10^{-5}$ & $1.5\cdot10^{-5}$ & $1.5\cdot10^{-5}$ \\ \hline
$3.8\cdot10^{6}$ & 0.66 & $2.4\cdot10^{-2}$ & $1.9\cdot10^{-3}$ & $3.0\cdot10^{-4}$ & $8.6\cdot10^{-5}$ & $2.9\cdot10^{-5}$ & $1.8\cdot10^{-5}$ & $6.4\cdot10^{-6}$ & $3.0\cdot10^{-6}$ & $4.9\cdot10^{-6}$ \\ \hline
$6.2\cdot10^{6}$ & 0.56 & $1.3\cdot10^{-2}$ & $8.1\cdot10^{-4}$ & $9.9\cdot10^{-5}$ & $2.2\cdot10^{-5}$ & $6.2\cdot10^{-6}$ & $4.0\cdot10^{-6}$ & $1.3\cdot10^{-6}$ & $3.7\cdot10^{-7}$ & $9.7\cdot10^{-7}$ \\ \hline
$1.0\cdot10^{7}$ & 0.47 & $7.5\cdot10^{-3}$ & $3.0\cdot10^{-4}$ & $3.4\cdot10^{-5}$ & $5.8\cdot10^{-6}$ & $1.6\cdot10^{-6}$ & $1.1\cdot10^{-6}$ & $4.2\cdot10^{-7}$ & $8.7\cdot10^{-8}$ & $2.6\cdot10^{-7}$ \\ \hline
\bottomrule
\end{tabular*}
\caption{\textbf{\textit{Conventional muon antineutrino}} passing fraction for $E_\nu = [10^3, \, 10^7]$~GeV and $\cos\theta_z = [0.1, \, 1.0]$.}
\label{tbl:conventional_antinumu}
\end{table}

\begin{table}[h]
\centering
\begin{tabular*}{\textwidth}{l @{\extracolsep{\fill}} r r r r r r r r r r}
\toprule
& \multicolumn{10}{c}{$\cos \theta_z$}\\
\cmidrule{2-11}
$E_\nu$ [GeV] & 0.1 & 0.2 & 0.3 & 0.4 & 0.5 & 0.6 & 0.7 & 0.8 & 0.9 & 1.0 \\ 
\midrule
$1.0\cdot10^{3}$ & 1.00 & 0.99 & 0.97 & 0.94 & 0.92 & 0.89 & 0.88 & 0.87 & 0.86 & 0.86 \\ \hline
$1.6\cdot10^{3}$ & 1.00 & 0.99 & 0.95 & 0.92 & 0.89 & 0.86 & 0.85 & 0.83 & 0.82 & 0.81 \\ \hline
$2.6\cdot10^{3}$ & 1.00 & 0.98 & 0.94 & 0.89 & 0.85 & 0.82 & 0.80 & 0.78 & 0.77 & 0.76 \\ \hline
$4.3\cdot10^{3}$ & 1.00 & 0.98 & 0.91 & 0.85 & 0.81 & 0.77 & 0.75 & 0.72 & 0.71 & 0.70 \\ \hline
$7.0\cdot10^{3}$ & 1.00 & 0.97 & 0.88 & 0.81 & 0.76 & 0.71 & 0.69 & 0.66 & 0.65 & 0.64 \\ \hline
$1.1\cdot10^{4}$ & 1.00 & 0.95 & 0.85 & 0.77 & 0.70 & 0.65 & 0.63 & 0.59 & 0.58 & 0.57 \\ \hline
$1.8\cdot10^{4}$ & 1.00 & 0.94 & 0.81 & 0.72 & 0.64 & 0.58 & 0.56 & 0.53 & 0.51 & 0.50 \\ \hline
$3.0\cdot10^{4}$ & 1.00 & 0.92 & 0.77 & 0.66 & 0.58 & 0.52 & 0.49 & 0.46 & 0.45 & 0.44 \\ \hline
$4.8\cdot10^{4}$ & 1.00 & 0.90 & 0.72 & 0.60 & 0.52 & 0.45 & 0.43 & 0.39 & 0.38 & 0.37 \\ \hline
$7.8\cdot10^{4}$ & 1.00 & 0.87 & 0.67 & 0.54 & 0.45 & 0.39 & 0.36 & 0.33 & 0.32 & 0.31 \\ \hline
$1.3\cdot10^{5}$ & 1.00 & 0.84 & 0.61 & 0.48 & 0.39 & 0.32 & 0.30 & 0.27 & 0.26 & 0.25 \\ \hline
$2.1\cdot10^{5}$ & 1.00 & 0.80 & 0.55 & 0.41 & 0.33 & 0.26 & 0.24 & 0.21 & 0.20 & 0.19 \\ \hline
$3.4\cdot10^{5}$ & 0.99 & 0.74 & 0.48 & 0.34 & 0.26 & 0.20 & 0.18 & 0.16 & 0.15 & 0.14 \\ \hline
$5.5\cdot10^{5}$ & 0.99 & 0.68 & 0.41 & 0.27 & 0.20 & 0.14 & 0.13 & 0.11 & 0.10 & $9.5\cdot10^{-2}$ \\ \hline
$8.9\cdot10^{5}$ & 0.98 & 0.60 & 0.31 & 0.19 & 0.13 & $8.9\cdot10^{-2}$ & $7.7\cdot10^{-2}$ & $6.2\cdot10^{-2}$ & $5.8\cdot10^{-2}$ & $5.4\cdot10^{-2}$ \\ \hline
$1.4\cdot10^{6}$ & 0.96 & 0.50 & 0.22 & 0.12 & $7.3\cdot10^{-2}$ & $4.6\cdot10^{-2}$ & $3.8\cdot10^{-2}$ & $3.0\cdot10^{-2}$ & $2.7\cdot10^{-2}$ & $2.5\cdot10^{-2}$ \\ \hline
$2.3\cdot10^{6}$ & 0.94 & 0.41 & 0.14 & $6.6\cdot10^{-2}$ & $3.5\cdot10^{-2}$ & $2.0\cdot10^{-2}$ & $1.6\cdot10^{-2}$ & $1.2\cdot10^{-2}$ & $1.1\cdot10^{-2}$ & $9.9\cdot10^{-3}$ \\ \hline
$3.8\cdot10^{6}$ & 0.92 & 0.32 & $8.9\cdot10^{-2}$ & $3.3\cdot10^{-2}$ & $1.5\cdot10^{-2}$ & $7.8\cdot10^{-3}$ & $6.0\cdot10^{-3}$ & $4.1\cdot10^{-3}$ & $3.6\cdot10^{-3}$ & $3.3\cdot10^{-3}$ \\ \hline
$6.2\cdot10^{6}$ & 0.88 & 0.22 & $4.6\cdot10^{-2}$ & $1.4\cdot10^{-2}$ & $5.3\cdot10^{-3}$ & $2.3\cdot10^{-3}$ & $1.6\cdot10^{-3}$ & $1.0\cdot10^{-3}$ & $8.6\cdot10^{-4}$ & $7.7\cdot10^{-4}$ \\ \hline
$1.0\cdot10^{7}$ & 0.84 & 0.14 & $2.0\cdot10^{-2}$ & $4.9\cdot10^{-3}$ & $1.6\cdot10^{-3}$ & $5.9\cdot10^{-4}$ & $4.0\cdot10^{-4}$ & $2.3\cdot10^{-4}$ & $1.9\cdot10^{-4}$ & $1.7\cdot10^{-4}$ \\ \hline
\bottomrule
\end{tabular*}
\caption{\textbf{\textit{Prompt electron neutrino}} passing fraction for $E_\nu = [10^3, \, 10^7]$~GeV and $\cos\theta_z = [0.1, \, 1.0]$.} \vspace{-6mm}
\label{tbl:prompt_nue}
\end{table}

\begin{table}[h!]
\centering
\begin{tabular*}{\textwidth}{l @{\extracolsep{\fill}} r r r r r r r r r r}
\toprule
& \multicolumn{10}{c}{$\cos \theta_z$}\\
\cmidrule{2-11}
$E_\nu$ [GeV] & 0.1 & 0.2 & 0.3 & 0.4 & 0.5 & 0.6 & 0.7 & 0.8 & 0.9 & 1.0 \\ 
\midrule
$1.0\cdot10^{3}$ & 1.00 & 0.99 & 0.97 & 0.94 & 0.92 & 0.90 & 0.89 & 0.87 & 0.87 & 0.86 \\ \hline
$1.6\cdot10^{3}$ & 1.00 & 0.99 & 0.95 & 0.92 & 0.89 & 0.86 & 0.85 & 0.83 & 0.82 & 0.81 \\ \hline
$2.6\cdot10^{3}$ & 1.00 & 0.98 & 0.94 & 0.89 & 0.85 & 0.82 & 0.80 & 0.78 & 0.77 & 0.76 \\ \hline
$4.3\cdot10^{3}$ & 1.00 & 0.98 & 0.91 & 0.86 & 0.81 & 0.77 & 0.75 & 0.72 & 0.71 & 0.70 \\ \hline
$7.0\cdot10^{3}$ & 1.00 & 0.97 & 0.89 & 0.81 & 0.76 & 0.71 & 0.69 & 0.66 & 0.65 & 0.64 \\ \hline
$1.1\cdot10^{4}$ & 1.00 & 0.95 & 0.85 & 0.77 & 0.70 & 0.65 & 0.63 & 0.60 & 0.59 & 0.57 \\ \hline
$1.8\cdot10^{4}$ & 1.00 & 0.94 & 0.81 & 0.72 & 0.65 & 0.59 & 0.56 & 0.53 & 0.52 & 0.51 \\ \hline
$3.0\cdot10^{4}$ & 1.00 & 0.92 & 0.77 & 0.66 & 0.59 & 0.52 & 0.50 & 0.46 & 0.45 & 0.44 \\ \hline
$4.8\cdot10^{4}$ & 1.00 & 0.90 & 0.72 & 0.60 & 0.52 & 0.45 & 0.43 & 0.40 & 0.38 & 0.37 \\ \hline
$7.8\cdot10^{4}$ & 1.00 & 0.87 & 0.67 & 0.54 & 0.46 & 0.39 & 0.37 & 0.33 & 0.32 & 0.31 \\ \hline
$1.3\cdot10^{5}$ & 1.00 & 0.84 & 0.62 & 0.48 & 0.39 & 0.33 & 0.30 & 0.27 & 0.26 & 0.25 \\ \hline
$2.1\cdot10^{5}$ & 1.00 & 0.80 & 0.56 & 0.42 & 0.33 & 0.27 & 0.24 & 0.21 & 0.20 & 0.20 \\ \hline
$3.4\cdot10^{5}$ & 0.99 & 0.75 & 0.49 & 0.35 & 0.26 & 0.20 & 0.18 & 0.16 & 0.15 & 0.14 \\ \hline
$5.5\cdot10^{5}$ & 0.99 & 0.68 & 0.41 & 0.27 & 0.20 & 0.15 & 0.13 & 0.11 & 0.10 & $9.7\cdot10^{-2}$ \\ \hline
$8.9\cdot10^{5}$ & 0.98 & 0.60 & 0.32 & 0.20 & 0.13 & $9.4\cdot10^{-2}$ & $8.1\cdot10^{-2}$ & $6.6\cdot10^{-2}$ & $6.2\cdot10^{-2}$ & $5.8\cdot10^{-2}$ \\ \hline
$1.4\cdot10^{6}$ & 0.96 & 0.51 & 0.23 & 0.12 & $7.6\cdot10^{-2}$ & $5.0\cdot10^{-2}$ & $4.2\cdot10^{-2}$ & $3.3\cdot10^{-2}$ & $3.0\cdot10^{-2}$ & $2.8\cdot10^{-2}$ \\ \hline
$2.3\cdot10^{6}$ & 0.95 & 0.41 & 0.15 & $6.8\cdot10^{-2}$ & $3.7\cdot10^{-2}$ & $2.2\cdot10^{-2}$ & $1.8\cdot10^{-2}$ & $1.3\cdot10^{-2}$ & $1.2\cdot10^{-2}$ & $1.1\cdot10^{-2}$ \\ \hline
$3.8\cdot10^{6}$ & 0.92 & 0.32 & $9.2\cdot10^{-2}$ & $3.6\cdot10^{-2}$ & $1.8\cdot10^{-2}$ & $9.8\cdot10^{-3}$ & $7.8\cdot10^{-3}$ & $5.6\cdot10^{-3}$ & $5.1\cdot10^{-3}$ & $4.7\cdot10^{-3}$ \\ \hline
$6.2\cdot10^{6}$ & 0.88 & 0.23 & $4.9\cdot10^{-2}$ & $1.6\cdot10^{-2}$ & $6.8\cdot10^{-3}$ & $3.4\cdot10^{-3}$ & $2.6\cdot10^{-3}$ & $1.9\cdot10^{-3}$ & $1.7\cdot10^{-3}$ & $1.6\cdot10^{-3}$ \\ \hline
$1.0\cdot10^{7}$ & 0.84 & 0.15 & $2.4\cdot10^{-2}$ & $6.5\cdot10^{-3}$ & $2.7\cdot10^{-3}$ & $1.4\cdot10^{-3}$ & $1.1\cdot10^{-3}$ & $8.6\cdot10^{-4}$ & $8.0\cdot10^{-4}$ & $7.6\cdot10^{-4}$ \\ \hline
\bottomrule
\end{tabular*}
\caption{\textbf{\textit{Prompt electron antineutrino}} passing fraction for $E_\nu = [10^3, \, 10^7]$~GeV and $\cos\theta_z = [0.1, \, 1.0]$.}
\label{tbl:prompt_antinue}
\end{table}

\begin{table}[h!]
\centering
\begin{tabular*}{\textwidth}{l @{\extracolsep{\fill}} r r r r r r r r r r}
\toprule
& \multicolumn{10}{c}{$\cos \theta_z$}\\
\cmidrule{2-11}
$E_\nu$ [GeV] & 0.1 & 0.2 & 0.3 & 0.4 & 0.5 & 0.6 & 0.7 & 0.8 & 0.9 & 1.0 \\ 
\midrule
$1.0\cdot10^{3}$ & 1.00 & 0.99 & 0.97 & 0.94 & 0.91 & 0.88 & 0.87 & 0.85 & 0.84 & 0.82 \\ \hline
$1.6\cdot10^{3}$ & 1.00 & 0.99 & 0.95 & 0.91 & 0.87 & 0.83 & 0.81 & 0.78 & 0.76 & 0.74 \\ \hline
$2.6\cdot10^{3}$ & 1.00 & 0.98 & 0.93 & 0.86 & 0.81 & 0.75 & 0.71 & 0.68 & 0.65 & 0.63 \\ \hline
$4.3\cdot10^{3}$ & 1.00 & 0.97 & 0.89 & 0.80 & 0.72 & 0.64 & 0.60 & 0.55 & 0.52 & 0.49 \\ \hline
$7.0\cdot10^{3}$ & 1.00 & 0.96 & 0.83 & 0.70 & 0.60 & 0.51 & 0.46 & 0.41 & 0.38 & 0.36 \\ \hline
$1.1\cdot10^{4}$ & 1.00 & 0.93 & 0.75 & 0.59 & 0.47 & 0.38 & 0.33 & 0.29 & 0.26 & 0.24 \\ \hline
$1.8\cdot10^{4}$ & 1.00 & 0.89 & 0.64 & 0.46 & 0.35 & 0.26 & 0.22 & 0.18 & 0.16 & 0.15 \\ \hline
$3.0\cdot10^{4}$ & 1.00 & 0.83 & 0.53 & 0.34 & 0.24 & 0.17 & 0.14 & 0.11 & $9.3\cdot10^{-2}$ & $8.0\cdot10^{-2}$ \\ \hline
$4.8\cdot10^{4}$ & 1.00 & 0.74 & 0.41 & 0.24 & 0.15 & $9.8\cdot10^{-2}$ & $7.6\cdot10^{-2}$ & $5.7\cdot10^{-2}$ & $4.8\cdot10^{-2}$ & $3.9\cdot10^{-2}$ \\ \hline
$7.8\cdot10^{4}$ & 1.00 & 0.65 & 0.30 & 0.15 & $8.9\cdot10^{-2}$ & $5.2\cdot10^{-2}$ & $3.8\cdot10^{-2}$ & $2.7\cdot10^{-2}$ & $2.2\cdot10^{-2}$ & $1.7\cdot10^{-2}$ \\ \hline
$1.3\cdot10^{5}$ & 0.99 & 0.54 & 0.21 & $9.0\cdot10^{-2}$ & $4.7\cdot10^{-2}$ & $2.4\cdot10^{-2}$ & $1.7\cdot10^{-2}$ & $1.2\cdot10^{-2}$ & $9.6\cdot10^{-3}$ & $7.4\cdot10^{-3}$ \\ \hline
$2.1\cdot10^{5}$ & 0.98 & 0.43 & 0.13 & $5.0\cdot10^{-2}$ & $2.4\cdot10^{-2}$ & $1.1\cdot10^{-2}$ & $7.9\cdot10^{-3}$ & $5.4\cdot10^{-3}$ & $4.4\cdot10^{-3}$ & $3.3\cdot10^{-3}$ \\ \hline
$3.4\cdot10^{5}$ & 0.96 & 0.32 & $7.8\cdot10^{-2}$ & $2.5\cdot10^{-2}$ & $1.1\cdot10^{-2}$ & $4.8\cdot10^{-3}$ & $3.3\cdot10^{-3}$ & $2.3\cdot10^{-3}$ & $1.8\cdot10^{-3}$ & $1.3\cdot10^{-3}$ \\ \hline
$5.5\cdot10^{5}$ & 0.93 & 0.23 & $4.2\cdot10^{-2}$ & $1.2\cdot10^{-2}$ & $5.0\cdot10^{-3}$ & $1.9\cdot10^{-3}$ & $1.3\cdot10^{-3}$ & $8.7\cdot10^{-4}$ & $6.7\cdot10^{-4}$ & $4.8\cdot10^{-4}$ \\ \hline
$8.9\cdot10^{5}$ & 0.88 & 0.14 & $2.1\cdot10^{-2}$ & $4.8\cdot10^{-3}$ & $1.8\cdot10^{-3}$ & $5.9\cdot10^{-4}$ & $3.7\cdot10^{-4}$ & $2.5\cdot10^{-4}$ & $1.9\cdot10^{-4}$ & $1.3\cdot10^{-4}$ \\ \hline
$1.4\cdot10^{6}$ & 0.81 & $8.4\cdot10^{-2}$ & $8.6\cdot10^{-3}$ & $1.5\cdot10^{-3}$ & $4.9\cdot10^{-4}$ & $1.6\cdot10^{-4}$ & $9.1\cdot10^{-5}$ & $5.1\cdot10^{-5}$ & $3.9\cdot10^{-5}$ & $2.7\cdot10^{-5}$ \\ \hline
$2.3\cdot10^{6}$ & 0.73 & $4.5\cdot10^{-2}$ & $3.1\cdot10^{-3}$ & $5.0\cdot10^{-4}$ & $1.4\cdot10^{-4}$ & $4.2\cdot10^{-5}$ & $2.2\cdot10^{-5}$ & $1.0\cdot10^{-5}$ & $7.1\cdot10^{-6}$ & $6.0\cdot10^{-6}$ \\ \hline
$3.8\cdot10^{6}$ & 0.64 & $2.3\cdot10^{-2}$ & $1.2\cdot10^{-3}$ & $1.5\cdot10^{-4}$ & $3.5\cdot10^{-5}$ & $1.0\cdot10^{-5}$ & $5.6\cdot10^{-6}$ & $2.3\cdot10^{-6}$ & $1.3\cdot10^{-6}$ & $1.5\cdot10^{-6}$ \\ \hline
$6.2\cdot10^{6}$ & 0.55 & $1.1\cdot10^{-2}$ & $4.4\cdot10^{-4}$ & $4.7\cdot10^{-5}$ & $9.4\cdot10^{-6}$ & $2.4\cdot10^{-6}$ & $1.2\cdot10^{-6}$ & $4.6\cdot10^{-7}$ & $2.1\cdot10^{-7}$ & $3.0\cdot10^{-7}$ \\ \hline
$1.0\cdot10^{7}$ & 0.45 & $4.6\cdot10^{-3}$ & $1.4\cdot10^{-4}$ & $1.1\cdot10^{-5}$ & $1.6\cdot10^{-6}$ & $3.5\cdot10^{-7}$ & $1.6\cdot10^{-7}$ & $6.1\cdot10^{-8}$ & $1.7\cdot10^{-8}$ & $3.1\cdot10^{-8}$ \\ \hline
\bottomrule
\end{tabular*}
\caption{\textbf{\textit{Prompt muon neutrino}} passing fraction for $E_\nu = [10^3, \, 10^7]$~GeV and $\cos\theta_z = [0.1, \, 1.0]$.} \vspace{-6mm}
\label{tbl:prompt_numu}
\end{table}

\begin{table}[h!]
\centering
\begin{tabular*}{\textwidth}{l @{\extracolsep{\fill}} r r r r r r r r r r}
\toprule
& \multicolumn{10}{c}{$\cos \theta_z$}\\
\cmidrule{2-11}
$E_\nu$ [GeV] & 0.1 & 0.2 & 0.3 & 0.4 & 0.5 & 0.6 & 0.7 & 0.8 & 0.9 & 1.0 \\ 
\midrule
$1.0\cdot10^{3}$ & 1.00 & 0.99 & 0.97 & 0.94 & 0.91 & 0.88 & 0.87 & 0.85 & 0.84 & 0.83 \\ \hline
$1.6\cdot10^{3}$ & 1.00 & 0.99 & 0.95 & 0.91 & 0.87 & 0.83 & 0.81 & 0.78 & 0.76 & 0.75 \\ \hline
$2.6\cdot10^{3}$ & 1.00 & 0.98 & 0.93 & 0.86 & 0.81 & 0.75 & 0.72 & 0.68 & 0.65 & 0.63 \\ \hline
$4.3\cdot10^{3}$ & 1.00 & 0.97 & 0.89 & 0.80 & 0.72 & 0.64 & 0.60 & 0.55 & 0.52 & 0.49 \\ \hline
$7.0\cdot10^{3}$ & 1.00 & 0.96 & 0.83 & 0.70 & 0.60 & 0.51 & 0.46 & 0.41 & 0.38 & 0.36 \\ \hline
$1.1\cdot10^{4}$ & 1.00 & 0.93 & 0.75 & 0.59 & 0.47 & 0.38 & 0.33 & 0.29 & 0.26 & 0.24 \\ \hline
$1.8\cdot10^{4}$ & 1.00 & 0.89 & 0.64 & 0.46 & 0.35 & 0.26 & 0.22 & 0.18 & 0.16 & 0.15 \\ \hline
$3.0\cdot10^{4}$ & 1.00 & 0.83 & 0.53 & 0.34 & 0.24 & 0.17 & 0.14 & 0.11 & $9.4\cdot10^{-2}$ & $8.1\cdot10^{-2}$ \\ \hline
$4.8\cdot10^{4}$ & 1.00 & 0.74 & 0.41 & 0.24 & 0.15 & $9.8\cdot10^{-2}$ & $7.6\cdot10^{-2}$ & $5.8\cdot10^{-2}$ & $4.8\cdot10^{-2}$ & $4.0\cdot10^{-2}$ \\ \hline
$7.8\cdot10^{4}$ & 1.00 & 0.65 & 0.30 & 0.15 & $8.9\cdot10^{-2}$ & $5.2\cdot10^{-2}$ & $3.8\cdot10^{-2}$ & $2.7\cdot10^{-2}$ & $2.2\cdot10^{-2}$ & $1.7\cdot10^{-2}$ \\ \hline
$1.3\cdot10^{5}$ & 0.99 & 0.54 & 0.21 & $9.2\cdot10^{-2}$ & $4.9\cdot10^{-2}$ & $2.5\cdot10^{-2}$ & $1.8\cdot10^{-2}$ & $1.2\cdot10^{-2}$ & $9.9\cdot10^{-3}$ & $7.6\cdot10^{-3}$ \\ \hline
$2.1\cdot10^{5}$ & 0.98 & 0.43 & 0.13 & $5.0\cdot10^{-2}$ & $2.4\cdot10^{-2}$ & $1.1\cdot10^{-2}$ & $7.9\cdot10^{-3}$ & $5.4\cdot10^{-3}$ & $4.4\cdot10^{-3}$ & $3.3\cdot10^{-3}$ \\ \hline
$3.4\cdot10^{5}$ & 0.96 & 0.32 & $7.8\cdot10^{-2}$ & $2.5\cdot10^{-2}$ & $1.2\cdot10^{-2}$ & $4.9\cdot10^{-3}$ & $3.4\cdot10^{-3}$ & $2.3\cdot10^{-3}$ & $1.9\cdot10^{-3}$ & $1.3\cdot10^{-3}$ \\ \hline
$5.5\cdot10^{5}$ & 0.93 & 0.23 & $4.2\cdot10^{-2}$ & $1.2\cdot10^{-2}$ & $5.1\cdot10^{-3}$ & $1.9\cdot10^{-3}$ & $1.3\cdot10^{-3}$ & $8.8\cdot10^{-4}$ & $6.7\cdot10^{-4}$ & $4.8\cdot10^{-4}$ \\ \hline
$8.9\cdot10^{5}$ & 0.88 & 0.15 & $2.1\cdot10^{-2}$ & $4.9\cdot10^{-3}$ & $1.9\cdot10^{-3}$ & $6.1\cdot10^{-4}$ & $3.8\cdot10^{-4}$ & $2.6\cdot10^{-4}$ & $1.9\cdot10^{-4}$ & $1.4\cdot10^{-4}$ \\ \hline
$1.4\cdot10^{6}$ & 0.82 & $8.4\cdot10^{-2}$ & $8.8\cdot10^{-3}$ & $1.6\cdot10^{-3}$ & $5.1\cdot10^{-4}$ & $1.7\cdot10^{-4}$ & $9.7\cdot10^{-5}$ & $5.5\cdot10^{-5}$ & $4.1\cdot10^{-5}$ & $3.0\cdot10^{-5}$ \\ \hline
$2.3\cdot10^{6}$ & 0.73 & $4.5\cdot10^{-2}$ & $3.2\cdot10^{-3}$ & $5.2\cdot10^{-4}$ & $1.4\cdot10^{-4}$ & $4.5\cdot10^{-5}$ & $2.4\cdot10^{-5}$ & $1.2\cdot10^{-5}$ & $8.0\cdot10^{-6}$ & $6.8\cdot10^{-6}$ \\ \hline
$3.8\cdot10^{6}$ & 0.64 & $2.3\cdot10^{-2}$ & $1.2\cdot10^{-3}$ & $1.6\cdot10^{-4}$ & $3.8\cdot10^{-5}$ & $1.2\cdot10^{-5}$ & $6.5\cdot10^{-6}$ & $2.7\cdot10^{-6}$ & $1.4\cdot10^{-6}$ & $1.8\cdot10^{-6}$ \\ \hline
$6.2\cdot10^{6}$ & 0.55 & $1.1\cdot10^{-2}$ & $4.4\cdot10^{-4}$ & $4.8\cdot10^{-5}$ & $9.7\cdot10^{-6}$ & $2.7\cdot10^{-6}$ & $1.4\cdot10^{-6}$ & $6.4\cdot10^{-7}$ & $2.8\cdot10^{-7}$ & $4.1\cdot10^{-7}$ \\ \hline
$1.0\cdot10^{7}$ & 0.45 & $4.9\cdot10^{-3}$ & $1.6\cdot10^{-4}$ & $1.5\cdot10^{-5}$ & $3.0\cdot10^{-6}$ & $9.2\cdot10^{-7}$ & $5.5\cdot10^{-7}$ & $2.5\cdot10^{-7}$ & $8.2\cdot10^{-8}$ & $1.7\cdot10^{-7}$ \\ \hline
\bottomrule
\end{tabular*}
\caption{\textbf{\textit{Prompt muon antineutrino}} passing fraction for $E_\nu = [10^3, \, 10^7]$~GeV and $\cos\theta_z = [0.1, \, 1.0]$.}
\label{tbl:prompt_antinumu}
\end{table}

\end{document}